\documentclass[traditabstract]{aa}
\pdfoutput=1
\usepackage{flushend}
\usepackage{savesym}
\usepackage[fleqn]{amsmath}
\savesymbol{iint}
\usepackage{txfonts}
\restoresymbol{TXF}{iint}
\usepackage{natbib}
\usepackage{graphicx}
\usepackage{afterpage}
\usepackage{lscape}
\usepackage{psfrag}
\usepackage[hyphens]{url}
\PassOptionsToPackage{hyphens}{url}
\usepackage{microtype}
\usepackage{float}
\usepackage{afterpage}
\usepackage{longtable}
\usepackage[figuresright]{rotating}
\usepackage{fixltx2e}
\usepackage[colorlinks=true, linkcolor=blue, citecolor=blue, urlcolor=blue]{hyperref}
\usepackage{breakurl} 
\usepackage{xcolor}
\bibpunct{(}{)}{;}{a}{}{,}

\def\apjl{ApJL}

\def\aap{A\&A}
\def\apjs{ApJS}

\newcommand{\footnoteremember}[2]{
\footnote{#2}
\newcounter{#1}
\setcounter{#1}{\value{footnote}}
}
\newcommand{\footnoterecall}[1]{
\footnotemark[\value{#1}]
}

\begin{document}

   \title{Insights into gas heating and cooling in the disc of NGC~891\\ from \textit{Herschel}\thanks{Based on observations from {\it Herschel}, an ESA space observatory with science instruments provided by European-led Principal Investigator consortia and with important participation from NASA.} far-infrared spectroscopy}

   \subtitle{ }

   \author{T. M. Hughes\inst{1},  K. Foyle\inst{2}, M. R. P. Schirm\inst{2}, T. J. Parkin\inst{2}, I. De Looze\inst{1}, C. D. Wilson\inst{2}, G. J. Bendo\inst{3}, \\ M. Baes\inst{1}, J. Fritz\inst{1}, A. Boselli\inst{4}, A. Cooray\inst{5}, D. Cormier\inst{6}, O. \L. Karczewski\inst{7}, \\ V. Lebouteiller\inst{8}, N. Lu\inst{9}, S. C. Madden\inst{8}, L. Spinoglio\inst{10}, S. Viaene\inst{1}}
	  
\institute{Sterrenkundig Observatorium, Universiteit Gent, Krijgslaan 281-S9, Gent 9000, Belgium
               \\ \email{thomas.hughes@ugent.be}
\and Department of Physics \& Astronomy, McMaster University, Hamilton, Ontario L8S 4M1, Canada 
\and UK ALMA Regional Centre Node, Jodrell Bank Centre for Astrophysics, School of Physics and Astronomy, \\ University of Manchester, Oxford Road, Manchester M13 9PL, UK
\and Laboratoire d'Astrophysique de Marseille, Université d'Aix-Marseille and CNRS, UMR7326, F-13388 Marseille Cedex 13, France
\and Center for Cosmology, Department of Physics  and  Astronomy, University  of California, Irvine, CA 92697, USA
\and Institut für theoretische Astrophysik, Zentrum für Astronomie der Universität Heidelberg, Albert-Ueberle Str. 2, D-69120 Heidelberg, Germany
\and Department of Physics \& Astronomy, University of Sussex, Brighton, BN1 9QH, UK
\and CEA, Laboratoire AIM, Université Paris VII, IRFU/Service d'Astrophysique, Bat. 709, Orme des Merisiers, F-91191 Gif-sur-Yvette, France
\and Infrared Processing and Analysis Center, California Institute of Technology, MS 100-22, Pasadena, CA 91125, USA
\and Istituto di Astrofisica e Planetologia Spaziali, INAF-IAPS, Via Fosso del Cavaliere 100, I-00133 Roma, Italy
  } 

   \date{Accepted for publication in A\&A.}

\newcommand{\hi}{H{\sc i}} 
\newcommand{\hii}{H{\sc ii}\ }
\newcommand{\cii}{[C{\sc ii}]}
\newcommand{\oi}{[O{\sc i}]}
\newcommand{\oii}{[O{\sc ii}]}
\newcommand{\oiii}{[O{\sc iii}]}
\newcommand{\oiv}{[O{\sc iv}]}
\newcommand{\nii}{[N{\sc ii}]}
\newcommand{\niii}{N{\sc iii}}
\newcommand{\tir}{$F_{\mathrm{TIR}}$}
\newcommand{\fir}{$F_{\mathrm{FIR}}$}
\newcommand{\ciitir}{[C{\sc ii}]/$F_{\mathrm{TIR}}$}
\newcommand{\ciifir}{[C{\sc ii}]/$F_{\mathrm{FIR}}$}
\newcommand{\ciioitir}{([C{\sc ii}]+[O{\sc i}]63)/$F_{\mathrm{TIR}}$}
\newcommand{\ciioipah}{([C{\sc ii}]+[O{\sc i}]63)/$F_{\mathrm{PAH}}$}
\newcommand{\ciioialttir}{([C{\sc ii}]+[O{\sc i}]145)/$F_{\mathrm{TIR}}$}
\newcommand{\ltir}{$L_{\mathrm{TIR}}$}
\newcommand{\lfir}{$L_{\mathrm{FIR}}$}
\newcommand{\ciiltir}{[C{\sc ii}]/$L_{\mathrm{TIR}}$}
\newcommand{\ciilfir}{[C{\sc ii}]/$L_{\mathrm{FIR}}$}
\newcommand{\ciioiltir}{([C{\sc ii}]+[O{\sc i}]63)/$L_{\mathrm{TIR}}$}
\newcommand{\ciioilpah}{([C{\sc ii}]+[O{\sc i}]63)/$L_{\mathrm{PAH}}$}
\newcommand{\ciioi}{[C{\sc ii}]/[O{\sc i}]63}
\newcommand{\ciioialt}{[C{\sc ii}]/[O{\sc i}]145}

\newcommand{\ciiion}{${\mathrm{[C\textsc{ii}]}}^{\,}_{\tiny\textsc{ionised}}$} 
\newcommand{\ciisynion}{${\mathrm{[C\textsc{ii}]}}^{\tiny 24\,\mu\mathrm{m}}_{\tiny\textsc{ionised}}$} \newcommand{\ciiobsion}{${\mathrm{[C\textsc{ii}]}}^{\mathrm{\tiny[N\textsc{ii}]205}}_{\tiny\textsc{ionised}}$} 

\newcommand{\ha}{H{\sc $\alpha$}}
\newcommand{\hd}{H{\sc $\delta$}}
\newcommand{\hg}{H{\sc $\gamma$}}
\newcommand{\hb}{H{\sc $\beta$}}
\newcommand{\kms}{km~s$^{-1}$\ }
\newcommand{\sdust}{$\Sigma_{\mathrm{dust}}$}
\newcommand{\sgas}{$\Sigma_{\mathrm{gas}}$}
\newcommand{\shii}{$\Sigma_{\mathrm{H}_{2}}$} 
\newcommand{\shi}{$\Sigma_{\mathrm{H}\tiny{\textsc{i}}}$} 
\newcommand{\ssfr}{$\Sigma_{\mathrm{SFR}}$}

  \abstract{We present \textit{Herschel} PACS and SPIRE spectroscopy of the most important far-infrared cooling lines in the nearby, edge-on spiral galaxy, NGC~891: \cii ~158~$\mu$m,  \nii ~122,~205~$\mu$m, \oi ~63,~145~$\mu$m, and \oiii~88~$\mu$m. We find that the photoelectric heating efficiency of the gas, traced via the \ciioitir \ ratio, varies from a mean of $3.5\times$10$^{-3}$ in the centre up to $8\times$10$^{-3}$ at increasing radial and vertical distances in the disc. A decrease in \ciioitir \ but constant \ciioipah \ with increasing FIR colour suggests that polycyclic aromatic hydrocarbons (PAHs) may become important for gas heating in the central regions. We compare the observed flux of the FIR cooling lines and total IR emission with the predicted flux from a PDR model to determine the gas density, surface temperature and the strength of the incident far-ultraviolet (FUV) radiation field, $G_{0}$. Resolving details on physical scales of $\sim$0.6~kpc, a pixel-by-pixel analysis reveals that the majority of the PDRs in NGC~891's disc have hydrogen densities of $1 < \log (n/\mathrm{cm}^{-3}) < 3.5$ experiencing an incident FUV radiation field with strengths of $1.7 < \log G_0 < 3$. Although these values we derive for most of the disc are consistent with the gas properties found in PDRs in the spiral arms and inter-arm regions of M51, observed radial trends in $n$ and $G_0$ are shown to be sensitive to varying optical thickness in the lines, demonstrating the importance of accurately accounting for optical depth effects when interpreting observations of high inclination systems. Increasing the coverage of our analysis by using an empirical relationship between the MIPS 24~$\mu$m and \nii ~205~$\mu$m emission, we estimate an enhancement of the FUV radiation field strength in the far north-eastern side of the disc relative to the rest of the disc that coincides with the above-average star formation rate surface densities and gas-to-dust ratios. However, an accurate interpretation remains difficult due to optical depth effects, confusion along the line-of-sight and observational uncertainties.} 


	\authorrunning{T. M. Hughes et al.}
	\titlerunning{Gas heating and cooling in NGC~891}
   \maketitle
 
\section{Introduction}

Star formation in galaxies converts gas into stars, which in turn produce the heavy elements via nucleosynthesis. Upon their demise, stars expel the heavy elements along with any unprocessed gas back into the interstellar medium (ISM), where the metals either mix with the gas phase or condense to form dust grains in enriched, cooling gas (see e.g. \citealp*{nozawa2013}). Dust grains not only aid the synthesis of molecular hydrogen from atomic hydrogen gas (\citealp*{gould1963}), but also act as the dominant heating mechanism of the neutral interstellar gas, via photoelectrons that are ejected by incident UV photons originating from young stars (e.g. \citealp{watson1972}; \citealp{hollenbach1991}), in addition to other heating sources (cosmic rays, X-rays, mechanical heating, etc.). Polycyclic aromatic hydrocarbons (PAHs) are considered to be a key source of photoelectrons (e.g. \citealp*{bakes1994}; \citealp{draine2007}). For molecular clouds to collapse to form stars, the gas must be able to cool sufficiently to enable gravity to overcome random motion and remove the increasing thermal energy in the contracting clouds. The primary cooling mechanism of neutral gas is the collisional excitation of forbidden transitions of heavy elements followed by radiative decay. The efficiency of these processes that heat and cool the gas therefore affect the global star formation process and the overall evolution of the ISM components. 

The far-infrared fine-structure cooling lines, such as the \cii \ 158~$\mu$m, \nii \ 122 and 205~$\mu$m, \oi \ 63 and 145~$\mu$m, and \oiii \ 88~$\mu$m lines, play a crucial role in the thermal balance of the gas. Photons emitted through de-exciting forbidden transitions from collisionally-excited atoms cool the gas by removing thermal energy. The low ionization potential of atomic carbon means that far-ultraviolet (FUV) photons with energies greater than 11.26~eV can produce C$^{+}$, and so both neutral and ionized gas are traced by \cii \ emission. The \cii \ line luminosity is typically 0.1-1\% of the far-infrared luminosity in normal star-forming galaxies, making it one of the dominant cooling lines (e.g. \citealp{crawford1985}; \citealp{stacey1991}). The \oi \ lines originate in the neutral gas of photon dominated regions (PDRs), as atomic oxygen has an ionisation potential greater than hydrogen (13.6~eV). A harder radiation field is required to ionise N and O$^{+}$ due to their ionization potentials of 14.5 and 35.1~eV, respectively. The \nii \ and \oiii \ lines therefore trace only ionised gas predominantly found in \hii \ regions. Thus, observations of these lines can tell us important characteristics about the gas in the cold neutral and ionized regimes of the ISM. 

The fraction of far-ultraviolet (FUV) photons heating the dust via absorption compared to the fraction responsible for ejecting electrons from dust grains or PAHs that heat the gas, provides a measure of the photo-electric heating efficiency of the FUV radiation field. Since warm dust is traced via the re-emission of absorbed UV and optical photons that peak at FIR wavelengths, and gas heated from photoelectrons ejected from small dust grains may be traced during cooling via the fine-structure lines, both dust heating and gas cooling can be investigated via FIR observations. Previous studies have thus probed the photo-electric heating efficiency via the observed \ciiltir \ or \ciioiltir \ ratios using, for example, the \emph{Kuiper} Airborne Observatory (KAO, e.g. \citealp{stacey1991}; \citealp{madden1993}) and the Infrared Space Observatory (ISO, e.g. \citealp{hunter2001}; \citealp{contursi2002}). An ISO LWS spectrometer survey of 60 normal, star-forming galaxies spanning a range in various global properties, such as morphology and FIR colour, found a decreasing ratio of \ciiltir \ towards warmer IR colours (\citealp{malhotra2001}; see also \citealp{luhman1998}; \citealp{brauher2008}). One interpretation of this result is that warmer dust becomes more positively charged in stronger FUV radiation fields, lowering the efficiency of the photoelectric effect.

Most recently, the \textit{Herschel} Space Observatory \citep{pilbratt2010} with two of its instruments, the Photodetector Array Camera and Spectrometer \citep[PACS,][]{poglitsch2010} and the Spectral and Photometric Imaging REceiver \citep[SPIRE,][]{griffin2010}, was capable of observing both the FIR cooling lines and FIR/submm spectral energy distribution at unprecedented resolution, enabling the study of gas heating and cooling on galactic and spatially-resolved, sub-kiloparsec scales. Whilst studies using \textit{Herschel} observations of nearby galaxies (e.g. \citealp{cormier2010}; \citealp{mookerjea2011}; \citealp{beirao2012}) and the LMC-N11 \hii region \citep{lebouteiller2012} find that \ciitir \ varies on local scales, the \ciioitir \ ratio has also been found to vary as a function of FIR colour on small scales (\citealp{croxall2012}; \citealp{lebouteiller2012}; \citealp{parkin2013}). In addition, \citet{croxall2012} and \citet{lebouteiller2012} report even tighter correlations between the heating efficiency of PAHs, \ciioipah , versus the FIR colour, which suggests that PAHs may also trace the gas heating. In M51, the warmer dust showed a stronger decrease in heating efficiency when traced by \ciioitir \ than with the \ciioipah \ ratio \citep{parkin2013}. Whilst there remains a possibility that PAHs are responsible for the majority of gas heating, their true role is still unknown.

These diagnostic ratios may be used to determine the physical properties of the gaseous components of the ISM by comparing the observed values to the predictions of a PDR model. Several PDR models for determining the gas density, temperature and strength of the FUV radiation field are available (see \citealp{rollig2007} for a discussion, and references within). One of the most commonly used PDR models is that of \citet{tielens1985}, which characterises the physical conditions in a semi-infinite, plane-parallel slab PDR by two free variables: the hydrogen nucleus density, $n$, and the strength of the FUV radiation field in units of the Habing Field, $G_0 = 1.6 \times 10^{-3}$~erg~cm$^{-2}$~s$^{-1}$ \citep{habing1968}. The model has since been updated by \citet{wolfire1990}, \cite{hollenbach1991} and \citet{kaufman1999,kaufman2006}. Predictions from PDR models have been compared to \textit{Herschel} observations of both Galactic PDRs and nearby galaxies. For example, \citet{croxall2012} studied a late-type spiral, NGC~1097, and a Seyfert 1 galaxy, NGC~4559, finding $50 \le G_{0} \le 1000$ varying with $10^{2.5}\,\mathrm{cm}^{-3} \le n \le 10^{3}\,\mathrm{cm}^{-3}$ across both discs. Most recently, \citet{parkin2013} examined the $n$ and $G_0$ in various regions of M51; the hydrogen density and FUV radiation peak in the nucleus and similarly decline in both the spiral arm and interarm regions, suggesting similar physical conditions in clouds in these environments (see also \citealp{parkin2014}).

\begin{figure}
\begin{center}
\includegraphics[width=0.76\columnwidth]{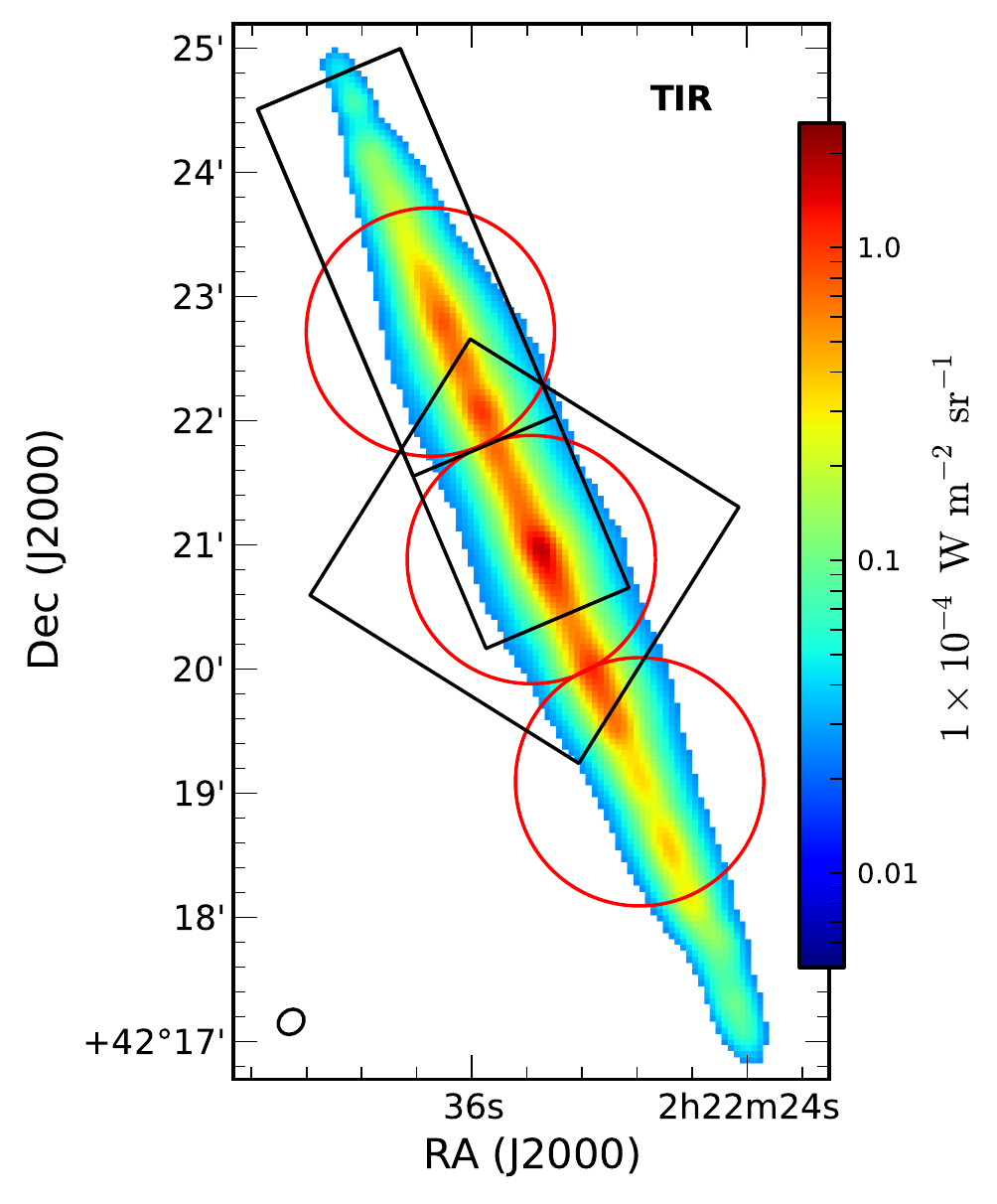}
\end{center}
\vspace{-0.3cm}
\caption[Total infrared flux map]{The total infrared flux, $F_{\mathrm{TIR}}$, derived from the MIPS 24~$\mu$m, PACS 70 and 160~$\mu$m maps using Equation~\ref{eqn:tirflux}. The map is presented in the 12$\arcsec$ resolution of the PACS 160~$\mu$m image with a pixel scale of 4$\arcsec$ and units of $\mathrm{W}\,\mathrm{m}^{-2}\,\mathrm{sr}^{-1}$. The black square, black rectangle and red circles demarcate the coverage of the PACS spectroscopic maps, PACS strips, and the SPIRE FTS observations, respectively. North is up, east is to the left.}\label{fig:tirmap}
\end{figure}

To complement these recent studies of typical face-on systems, we can use observations of edge-on galaxies to study the vertical variations in the physical conditions of the interstellar gas, particularly important for understanding possible vertical gas outflows driven by star formation and the structure of the multiphase ISM (see e.g. \citealp*{shapiro1976}; \citealp{bregman1980}; \citealp*{norman1989}). NGC~891 is a prototypical example of a near perfect edge-on ($ i >$89$^{\circ}$, e.g. \citealp{xilouris1998}), non-interacting spiral galaxy (SA(s)b, \citealp{RC21976}) located right in our neighbourhood ($D = $9.6 Mpc, e.g. \citealp{strickland2004}), and which exhibits many properties similar to our own Milky Way. These characteristics make NGC~891 an ideal target for studying the interstellar material in a star-forming disc, and so it has already been extensively observed at a range of wavelengths (see \citealp{hughes2014} and references therein). \citet{madden1994} mapped the [C\,\textsc{ii}](158~$\mu$m) line emission over an 8$\arcmin$ region of the galaxy with the Far-Infrared Fabry-Perot Imaging Spectrometer (FIFI) on the KAO at $55 \arcsec$ resolution (i.e. $\sim$2.6~kpc), finding extraplanar \cii \ emission near the nucleus. More recently, \citet{stacey2010} investigated the radial profiles of the \cii~158~$\mu$m, \oi~63~$\mu$m and \nii~122~$\mu$m fine-structure lines in NGC 891 from reprocessed observations made with the ISO LWS spectrometer at $\sim$75$\arcsec$ resolution (see \citealp{brauher2008}, and references therein). A comparison to PDR models found $G_{0}=200-400$ and $n \sim 10^{4}$~cm$^{-3}$ across the disc. However, the low resolution of these datasets, typically $\sim$ 1$\arcmin$, meant a spatially-resolved pixel-by-pixel analysis of the gas heating and cooling efficiencies determined from the FIR cooling lines and the FIR/submm SED was not previously feasible.

In this paper, we present new \textit{Herschel} far-infrared spectroscopy of NGC~891 obtained as part of the \textit{Herschel} Guaranteed Time Key Project, the Very Nearby Galaxies Survey (VNGS; P.~I.~:~C.~D.~Wilson), which aims to study the gas and dust in the ISM of a diverse sample of 13 nearby galaxies using \textit{Herschel}. We focus on the \cii \ 158~$\mu$m, \nii \ 122~$\mu$m, \oi \ 63 and 145~$\mu$m and \oiii \ 88~$\mu$m fine structure lines observed at unprecedented resolution (better than $\sim 12\arcsec$, or roughly 0.5~kpc) with the PACS instrument. We also present observations of the \nii \ 205~$\mu$m line from the SPIRE Fourier Transform Spectrometer (FTS) at $\sim$17$\arcsec$ resolution.  We use these spectra combined with the multi-wavelength photometry presented in \citet{hughes2014} to investigate the physical properties of the interstellar gas in the galaxy by using the PDR model of \citet{kaufman1999, kaufman2006}. In particular, we are interested in comparing the gas heating and cooling mechanisms observed in the disc from a near-perfect edge-on orientation, as in NGC 891, to those of the face-on spiral galaxy, M51. Our paper is structured as follows. In Sec.~\ref{sec:observations}, we describe our observations and data reduction methodology. In Sec.~\ref{sec:results}~and~\ref{sec:pdrmodelling}, we describe the characteristics of the gas and compare our observations to theoretical PDR models, respectively. Finally, Sec.~\ref{sec:discussion} and ~\ref{sec:conclusions} present our discussion and conclusions.

\begin{figure}
\begin{center}
\includegraphics[width=0.76\columnwidth]{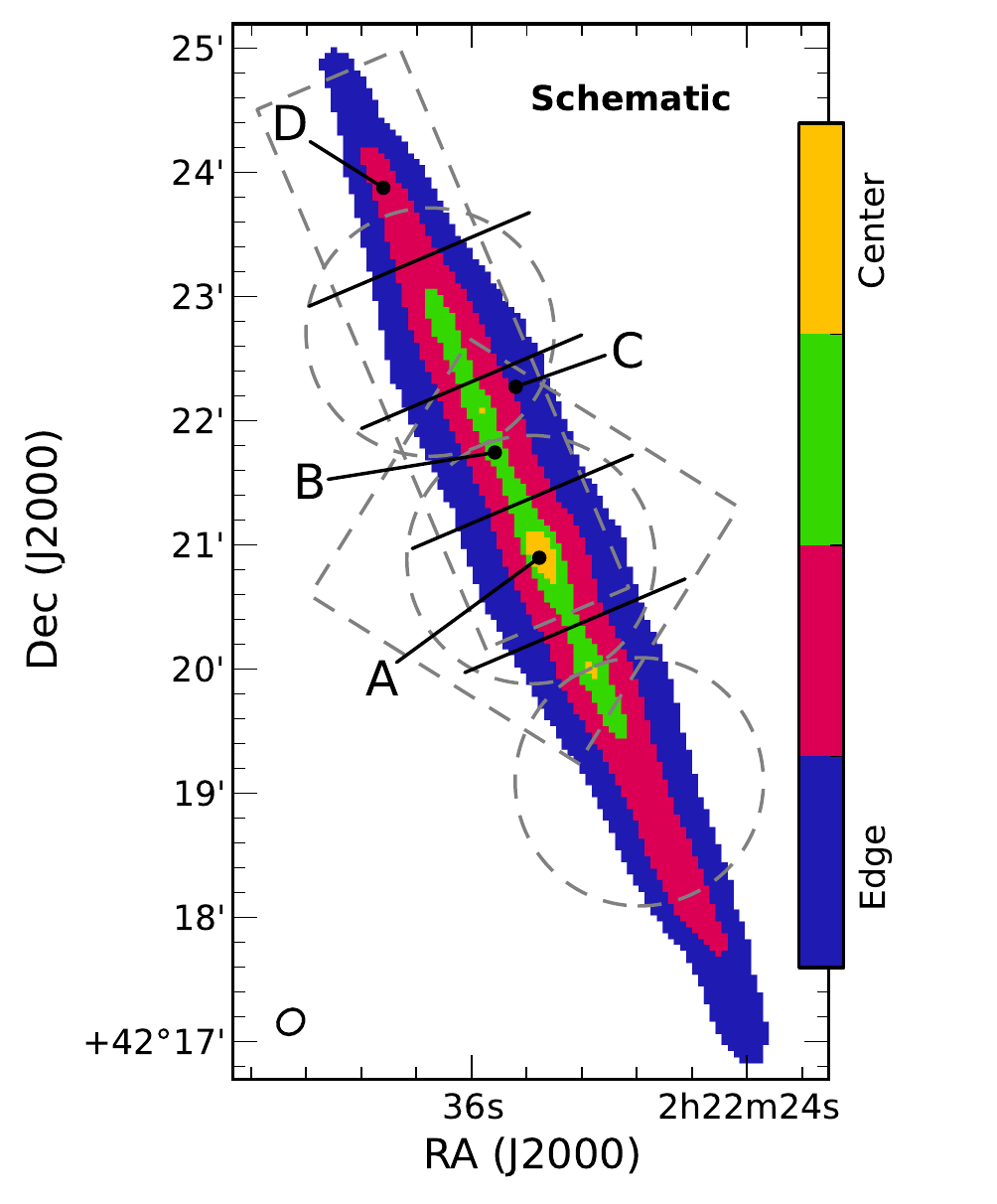}
\end{center}
\vspace{-0.3cm}
\caption[Total infrared flux map]{A schematic diagram of the radial bins (divided by solid black lines) and the four regions based on the TIR flux density (see Sec.~\ref{sec:ancillary} and Fig.~\ref{fig:tirmap}) that we use in our analysis to search for regional variations in the gas properties. Labels A-D mark positions of the typical line spectra shown in Fig.~\ref{fig:spectra}. The dashed grey lines demarcate the coverage of the PACS and SPIRE spectroscopic maps for reference.}\label{fig:tirmapmask}
\end{figure}

\begin{table*}
 \begin{center}
\begin{minipage}{\textwidth}
  \caption{Properties of our \textit{Herschel} PACS and SPIRE spectroscopic observations of NGC~891.}
  \label{tab:obsproperties}
   \begin{center}
  \begin{tabular}{l l c c c  c c c c}
\hline 
\hline
Instrument& Line  & $\lambda_{\mathrm{rest}}$& Transition &  OBSID     &  Date of     & Map size     & FWHM          &  Duration  \\
          &       & ($\mu$m) &            &           &  Observation & ($\arcmin\,\times\,\arcmin$) &  ($\arcsec$)   &   (s)   \\      
\hline
PACS      & \oi   &  63.184  & $^{3}P_{1}\, ^{\vector(1,0){+10}\, 3}P_{2} $ & 1342214876 &  01 Mar 2011 & $2.5\,\times\,2.5$    &  9.3 & 11182  \\
          &       &          & & 1342214881 &  01 Mar 2011 &  $0.72\,\times\,2.25$  & 9.3 &1462\\ 
          & \oiii &  88.356  & $^{3}P_{1}\, ^{\vector(1,0){+10}\, 3}P_{0} $ & 1342214877 &  01 Mar 2011 & $0.72\,\times\,3.25$  & 9.3 & 3575\\ 
          & \nii  &  121.898 & $^{3}P_{2}\, ^{\vector(1,0){+10}\, 3}P_{1} $ & 1342214875 &  28 Feb 2011 & $2.5\,\times\,2.5$  & 10 & 10266 \\ 
          &       &          & & 1342214878 &  01 Mar 2011 & $0.72\,\times\,2.25$ & 10 & 2400 \\
          & \oi   &  145.525 & $^{3}P_{0}\, ^{\vector(1,0){+10}\, 3}P_{1} $ & 1342214880 &  01 Mar 2011 & $0.72\,\times\,3.25$  & 11 & 6809\\ 
          & \cii  &  157.741 & $^{3}P_{3/2}\, ^{\vector(1,0){+10}\, 3}P_{1/2} $ & 1342214874 &  28 Feb 2011 & $2.5\,\times\,2.5$   & 11.5 & 5588\\ 
          &       &          & &  1342214879 &  01 Mar 2011 & $0.72\,\times\,2.25$  & 11.5 & 1462\\
SPIRE FTS & \nii  &  205.178 &  $^{3}P_{0}\, ^{\vector(1,0){+10}\, 3}P_{1} $ & 1342213376 &  28 Jan 2011 & $\sim$2$\arcmin$ diameter circle  & 17 & 17603\\
          &       &          & & 1342224765 &  26 Jul 2011 & $\sim$2$\arcmin$ diameter circle  & 17 & 7668\\
          &       &          & & 1342224766 &  26 Jul 2011 & $\sim$2$\arcmin$ diameter circle  & 17 & 7668\\    
\hline
\end{tabular}
\end{center}
\end{minipage}
\end{center}
\end{table*}

\section{Observations}\label{sec:observations}

We first present our \textit{Herschel} PACS and SPIRE spectroscopic observations, summarised in Table~\ref{tab:obsproperties}, and describe the data reduction steps for producing maps of the far-infrared cooling line emission.

\subsection{\textit{Herschel} PACS spectroscopy}

Covering a wavelength range of 51 to 220$\mu$m, the PACS spectrometer comprises 25 spaxels each with a 9.4$\arcsec$ field of view and arranged in a 5$\times$5 grid with an approximately square field of view of 47$\arcsec$ on each side. The spectral resolution ranges between 75 and 300 km s$^{-1}$ and the beam FWHM varies from approximately 9 to 13$\arcsec$. Our VNGS PACS spectroscopic observations were taken on the 28th February and 1st March, 2011, using the unchopped grating scan mode. They consist of raster maps and strips of the \cii, \nii \ 122 and \oi \ 63 $\mu$m line emission that cover the central 2.5$\arcmin\times$2.5$\arcmin$ and 0.72$\arcmin\times$2.25$\arcmin$ along the north-eastern side of the disc, and 0.72$\arcmin\times$3.25$\arcmin$ raster strips of the \oi \ 145~$\mu$m and \oiii \ 88~$\mu$m emission that also cover the northern disc. In Fig.~\ref{fig:tirmap}, we superimpose the outlines of our observations onto a map of the total infrared flux (see Section \ref{sec:ancillary}).  

\begin{figure*}
\begin{center}
\includegraphics[width=0.98\textwidth]{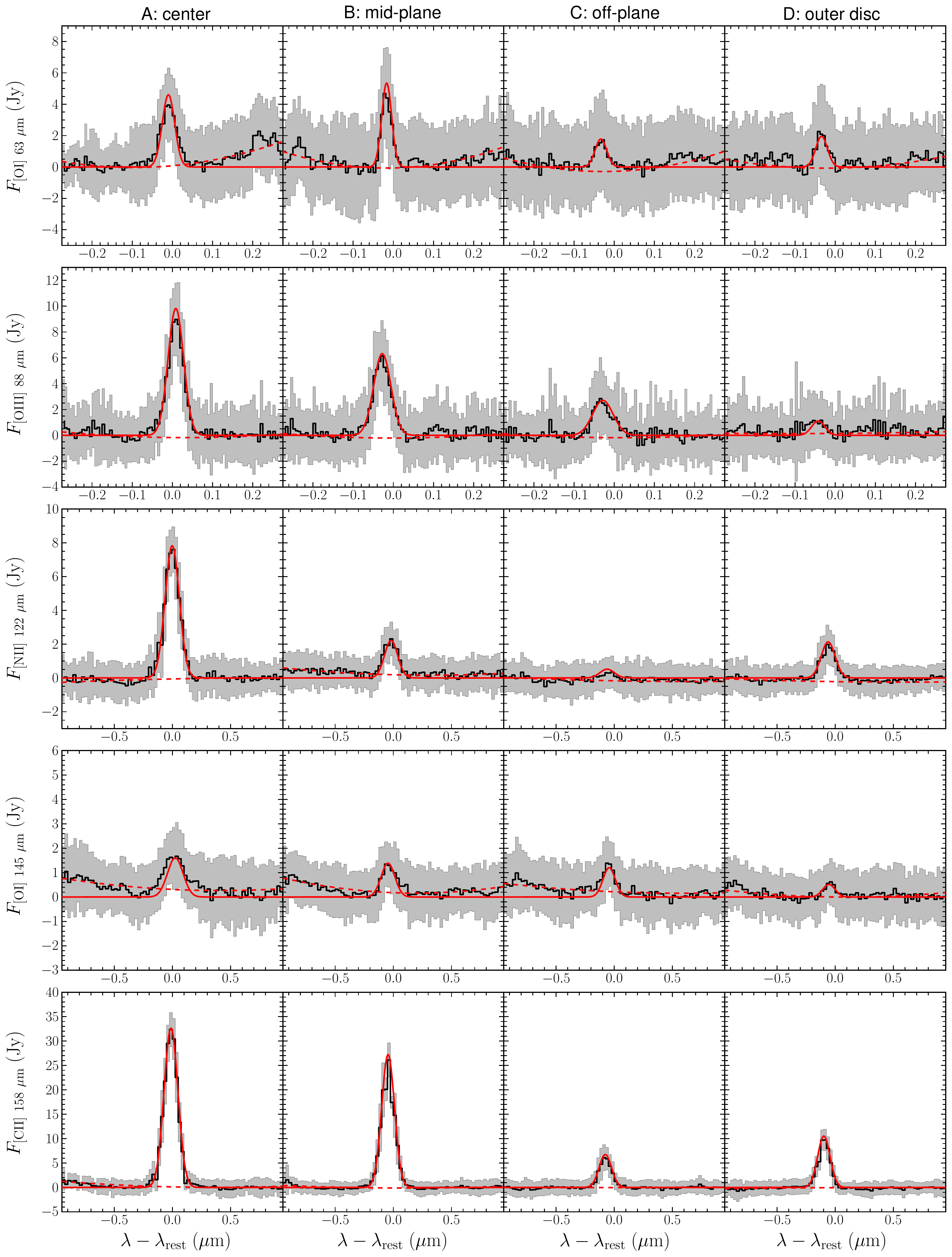}
\end{center}
\vspace{-0.4cm}
\caption[Line spectra]{The \oi ~63, \oiii ~88, \nii ~122, \oi ~145 and \cii ~158~$\mu$m line spectra (\textit{top} to \textit{bottom}) found in locations A-D (see Fig.~\ref{fig:tirmapmask}) representative of the central, mid-plane, off-plane and outer regions (\textit{left} to \textit{right}). We plot the $rms$ after 5$\sigma$ clipping (black line) and corresponding $\pm$1$\sigma$ values in each bin (grey area), and our best fit Gaussian profile and baseline (solid and dashed red lines). }\label{fig:spectra}
\end{figure*}

These observations were processed from Level 0 to Level 1 using the \textit{Herschel} Interactive Processing Environment \citep[HIPE, v.9.2][]{ott2010} with calibration files FM,41, following the standard pipeline reduction steps for the unchopped observing mode. Further details may be found in \citet{parkin2013}. Level 1 cubes were exported to PACSman v.3.5.2 \citep{lebouteiller2012}, where each individual spaxel's spectrum is fit with a Gaussian function and second order polynomial for the line and continuum baseline, respectively. Representative spectra observed from different locations in the galaxy, as indicated in Fig.~\ref{fig:tirmapmask}, are shown with the best fitting functions in Fig.~\ref{fig:spectra}. Intensity maps of the integrated flux are created from the individual rasters, also using PACSman, by projecting the rasters onto a common, over-sampled grid with a 3.133$\arcsec$ pixel size. Table~\ref{tab:obsproperties} lists the resolution and sizes of the final mosaicked maps of the \cii \ 158~$\mu$m, \nii \ 122~$\mu$m, \oi \ 63 and 145~$\mu$m, and \oiii \ 88~$\mu$m emission.

\subsection{\textit{Herschel} SPIRE spectroscopy}

The \textit{Herschel} SPIRE FTS instrument \citep{griffin2010} consists of two bolometer arrays, the SPIRE Short Wavelength (SSW) array and the SPIRE Long Wavelength (SLW) array, covering wavelength ranges of 194 to 313 $\mu$m and 303 to 671 $\mu$m, respectively, with a 2$\arcmin$ diameter field of view. Using the SPIRE FTS, we observed the center of NGC891 ($\alpha = 2^\mathrm{h}\ 22^\mathrm{m}\ 33\fs41$, 
$\delta = +42\degr\ 20\arcmin\ 56\farcs9$) in high spectral resolution (FWHM $=0.048$ cm$^{-1}$), intermediate-sampling mode. These observations were taken on 28th January, 2011 (OD 625) with observation ID 1342213376 and a total integration time of $17603$ seconds ($\sim$5 hours) for the entire map. In addition, we supplemented these data with publicly available open time observations (P.~I.:~G.~Stacey) of the upper (ID 1342224766, $\alpha = 2^\mathrm{h}\ 22^\mathrm{m}\ 37\fs29$, $\delta = +42\degr\ 22\arcmin\ 41\farcs8$) and lower (ID 1342224765, $\alpha = 2^\mathrm{h}\ 22^\mathrm{m}\ 28\fs69$, $\delta = +42\degr\ 19\arcmin\ 05\farcs6$) portions of the disc of NGC~891. These observations were both performed in high spectral resolution, sparse-sampling mode with a total integration time of $7668$ seconds ($\sim$2 hours) each. 

We reduce the FTS data for our intermediate-sampling observations using a modified version of the standard Spectrometer Mapping user pipeline from HIPE v.11.0.1 along with SPIRE calibration context v.11.0. The standard pipeline assumes that the source is extended and uniformly fills the beam; however, NGC 891 only partially fills the beam. As such, we apply the same point-source correction to each jiggle position as done in \citet{schirm2014}. For reducing the sparse-sampling FTS data, we use the standard Spectrometer Single Pointing user pipeline along with HIPE v.11.1 and SPIRE calibration context v11.1. The standard pipeline outputs both extended source calibrated and point-source calibrated data. Once again, since the beam is not uniformly filled, we opt to use the point-source calibrated data. We combine all 4 jiggle positions from the intermediate-sampled observations and the 2 sparse-sampling observations into 2 data cubes using the \emph{spireProjection} task in HIPE: one cube for the SLW and one for the SSW. The flux calibration uncertainty is 7\%, although this does not include uncertainty from source-beam coupling. Finally, the \nii \ 205 $\mu$m line was fit with a Sinc function across the entire map using the same technique as described in \citet{schirm2014}. The final emission map has a resolution of $\sim$17$\arcsec$ with a 15$\arcsec$ pixel size.

\subsection{Ancillary data}\label{sec:ancillary}

We use the VNGS \textit{Herschel} PACS photometric maps at 70 and 160~$\mu$m originally presented in \citet{hughes2014}. The maps have pixel sizes of $1\farcs4$ and $2\farcs85$, which correspond to one quarter of the point spread function (PSF) full width half maximum (FWHM) for the scan speed used for these observations, for the 70 and 160~$\mu$m maps respectively. The calibration uncertainty is 5\%. We also use the Multiband Imaging Photometer for {\it Spitzer} (MIPS; \citealp{rieke2004}) 24~$\mu$m data, which were reprocessed by \citet{bendo2012} using the MIPS Data Analysis Tools \citep{gordon2005} along with additional processing steps. The image has a pixel scale of $1\farcs5$, the PSF FWHM is $6\arcsec$, and the calibration uncertainty is 4\% \citep{engelbracht2007}.

We estimate the total infrared flux emitting from 3 to 1100~$\mu$m, $F_{\mathrm{TIR}}$, using these MIPS 24~$\mu$m, PACS 70 and 160~$\mu$m maps. The maps were first convolved to a common 12$\arcsec$ resolution of the PACS 160~$\mu$m image using the common-resolution convolution kernels\footnoteremember{aniano}{PSFs, convolution kernels and the IDL task \textsc{convolve\_image.pro} from Aniano et al. are available from \url{http://www.astro.princeton.edu/~ganiano/Kernels.html}.} of \citet{aniano2011}, and rescaled to a 4$\arcsec$ pixel scale. The total infrared flux is then calculated from these images via the empirical equation from \citet{galametz2013}, given by 
\begin{align}\label{eqn:tirflux}
F_{\mathrm{TIR}}\,=\,c_{24}\nu_{24}F_{24}\,+\,c_{70}\nu_{70}F_{70}\,+\,c_{160}\nu_{160}F_{160}
\end{align}
where the coefficients $\left[c_{24},\,c_{70},\,c_{160}\right]\,=\,\left[2.133,\,0.681,\,1.125\right]$ are taken from \citet[][see their Table 3]{galametz2013}. We eschew the techniques that include the SPIRE photometric maps in estimating the TIR emission (\citealp{galametz2013}) out of our desire to preserve the relatively high (12$\arcsec$) common spatial resolution attained with the PACS 160~$\mu$m map compared to the SPIRE maps ($>$18$\arcsec$; see \citealp{hughes2014}). The calibration is shown to as reliably reproduce estimates of the properly modelled TIR emission (within $\sim$20\%) as when using the complete sampling of the FIR/submm emission, i.e. data at 24, 70, 100, 160 and 250~$\mu$m (see Fig.~10 in \citealp{galametz2013}). Whilst the resulting $F_{\mathrm{TIR}}$ map, presented in Fig.~\ref{fig:tirmap}, covers the entire disc of NGC~891, we only use those regions that overlap with the spectroscopic maps in our analysis. Furthermore, we use the contours of the TIR map as a means of crudely dissecting the galaxy into different morphological regions: flux densities of $F_{\mathrm{TIR}} \geq$ 1.2$\times10^{-4}$, 0.5$\times10^{-4} \leq F_{\mathrm{TIR}} < 1.2 \times10^{-4}$, 1.5$\times10^{-5} \leq F_{\mathrm{TIR}} <  5\times10^{-5}$, and 0.4$\times10^{-5} \leq F_{\mathrm{TIR}} < 1.5\times10^{-5}$ $\mathrm{W}\,\mathrm{m}^{-2}\,\mathrm{sr}^{-1}$ correspond to the galaxy center, the mid-plane of the disc, and regions at intermediate and higher radial and vertical heights above the plane, respectively. A schematic is presented in Fig.~\ref{fig:tirmapmask}.

Finally, we use the \textit{Spitzer} Infrared Array Camery (IRAC; \citealp{fazio2004}) 3.6 $\mu$m map presented in \citet{hughes2014} to trace the stellar continuum emission, and a new IRAC 8~$\mu$m map as a proxy for the PAH emission. The latter data were obtained in astronomical observation requests 3631872. Individual corrected \clearpage
\clearpage 
\noindent
basic calibration data frames were processed with version 18.25.0 of the IRAC pipeline and remosaicked using the standard IRAC pipeline within the MOsaicker and Point source EXtractor \citep{makovoz2005}. The final image has a pixel scale of $0\farcs75$ and PSF FWHM of $1\farcs9$. Calibration uncertainties are 4\% (IRAC Instrument and Instrument Support Teams, 2013, IRAC Instrument Handbook, Version 2.0.3, JPL, Pasadena). To estimate the total PAH power with the IRAC~8~$\mu$m map, we first apply a colour correction following the method in the \textit{Spitzer} Data Analysis Cookbook\footnote{See \burl{http://irsa.ipac.caltech.edu/data/SPITZER/docs/dataanalysistools/cookbook/}.} and then subtract the stellar contribution estimated from the available IRAC~3.6~$\mu$m map via the \citet{marble2010} correction (see Eq.~2 in \citealp{croxall2012}). We discuss in detail the uncertainty in the PAH power in Sec.~\ref{sec:heatingcooling}, yet note a 6\% uncertainty in the aromatic fraction of the IRAC~8~$\mu$m flux reported by \citet{marble2010}.

\subsection{Image processing}

All spectroscopic images were first convolved to the resolution of the 160~$\mu$m image, since this band has a PSF with the largest FWHM, using the appropriate Gaussian common-resolution convolution kernels\footnoterecall{aniano} and the IDL task \textsc{convolve\_image.pro} \citep{aniano2011}. For the results presented here, the images were regridded to the pixel size of the 160~$\mu$m map using the {\sc Montage} software package. We note that since the pixel size ($4\arcsec$) is smaller than the 160~$\mu$m beam size ($12\arcsec$), adjacent pixels are not independent. However, we also performed the following analysis in its entirety using maps with a pixel scale matching the 160~$\mu$m beam size ($12\arcsec$) and, despite having far fewer pixels, the analysis reproduces the same trends and conclusions as found when oversampling the maps. Errors on each pixel were calculated by summing the flux calibration uncertainty, instrumental noise and sky background measurement in quadrature. For the pixels covering the galaxy, the flux errors are dominated by the calibration uncertainty. We use flux calibration uncertainties of 30 \% and 7 \% for the PACS and SPIRE FTS observations, respectively. Finally, we only consider pixels with a S/N ratio $>$ 3$\sigma$, excluding the calibration uncertainties, in our analysis.

\begin{figure*}
\begin{center}
\includegraphics[width=0.99\textwidth]{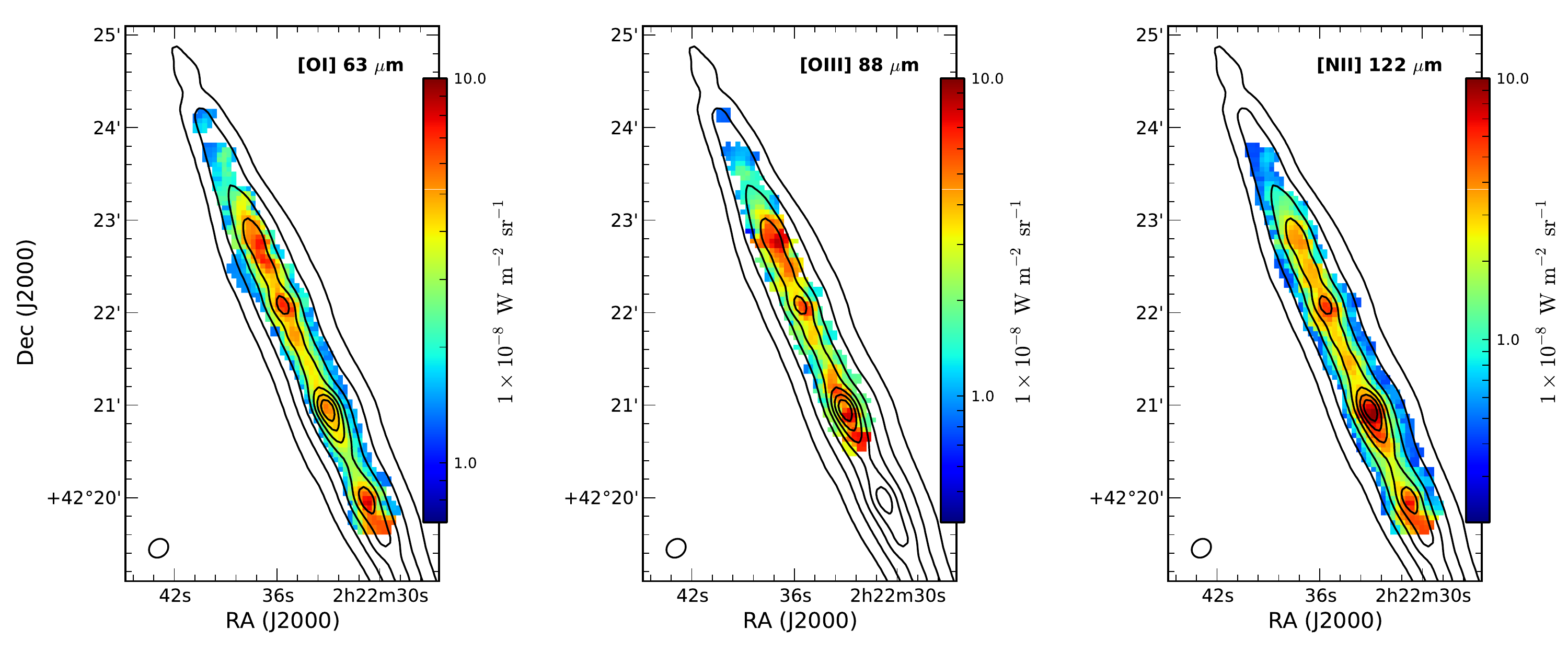}
\includegraphics[width=0.99\textwidth]{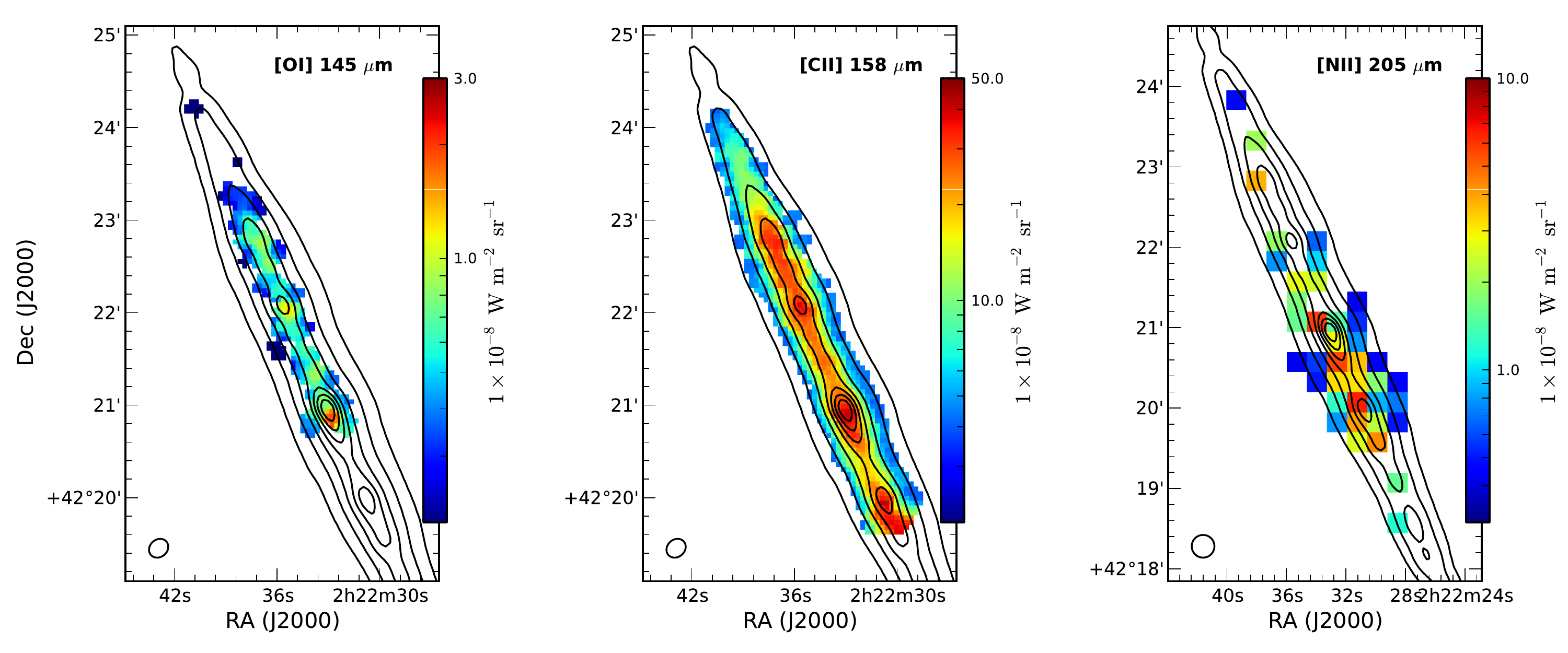}
\end{center}
\vspace{-0.3cm}
\caption[PACS far-infrared spectroscopic maps]{\textit{Herschel} PACS and SPIRE spectroscopic maps of the the most important far-infrared cooling lines in NGC~891: \oi ~$\lambda$~63~$\mu$m (\textit{upper left}), and \oiii~$\lambda$~88~$\mu$m (\textit{upper middle}), \nii ~$\lambda$~122~$\mu$m (\textit{upper right}), \oi ~$\lambda$~145~$\mu$m (\textit{lower left}), \cii ~$\lambda$~158~$\mu$m (\textit{lower middle}), and \nii ~$\lambda$~205~$\mu$m (\textit{lower right}). The maps are centred on $\alpha = 2^\mathrm{h}\ 22^\mathrm{m}\ 35\fs7$, $\delta = +42\degr\ 22\arcmin\ 05\farcs9$ (J2000.0), except for the \nii \ 205~$\mu$m image that is centred as in Fig.~\ref{fig:tirmap}, and are presented in their native resolution and pixel size. We only show pixels with S/N $>$ 3$\sigma$, yet note that only raster strips are available for the \oiii~$\lambda$~88 and \oi ~$\lambda$~145~$\mu$m line emission. Contours from the $F_{\mathrm{TIR}}$ map (see Fig.~\ref{fig:tirmap}) are superimposed on each image as a visual aid, with the levels corresponding to 4$\times10^{-6}$, 1$\times10^{-5}$, 2.5$\times10^{-5}$, 5$\times10^{-5}$, 7.5$\times10^{-5}$, 1$\times10^{-4}$, and 1.25$\times10^{-4}$ $\mathrm{W}\,\mathrm{m}^{-2}\,\mathrm{sr}^{-1}$. North is up, east is to the left.}\label{fig:obsmaps3sig}
\end{figure*}

\section{Physical properties of the gas}\label{sec:results}

\subsection{Morphology of line emission}

Our final \cii \ 158~$\mu$m, \nii \ 122~and~205~$\mu$m, \oi \ 63 and 145~$\mu$m, and \oiii \ 88~$\mu$m emission maps are presented in Fig.~\ref{fig:obsmaps3sig} at their native resolution and with an applied 3$\sigma$ cut. The \cii ~158~$\mu$m, \nii ~122~$\mu$m, and \oi ~63~$\mu$m all show remarkable spatial correlation with the main morphological features of NGC 891 evident in most observations at FIR/submm wavelengths (see e.g. Fig. 1 of \citealp{hughes2014}) and highlighted here with the contours of the \tir \ emission map: a peak in line emission in the galaxy centre, with two smaller maxima located either side of the centre at radial distances approximately 4 to 6 kpc along the semimajor axis. The \cii , \oi \ 63~$\mu$m and \oiii \ lines also appear to display an enhancement in their emission relative to the TIR contours on the far north eastern side of the disc, a location infamous for its higher luminosity at various wavelengths (e.g. \ha , \citealp{kamphuis2007}) compared to the opposite region on the southern side of the disc. Unfortunately, the lack of PACS spectroscopic observations towards the south means we cannot investigate whether such asymmetry also exists in the line emission from the disc. However, our new maps provide the ideal opportunity to study the gas properties in this region further and so we shall return to discuss this topic shortly.

\begin{figure}
\begin{center}
\includegraphics[width=\columnwidth]{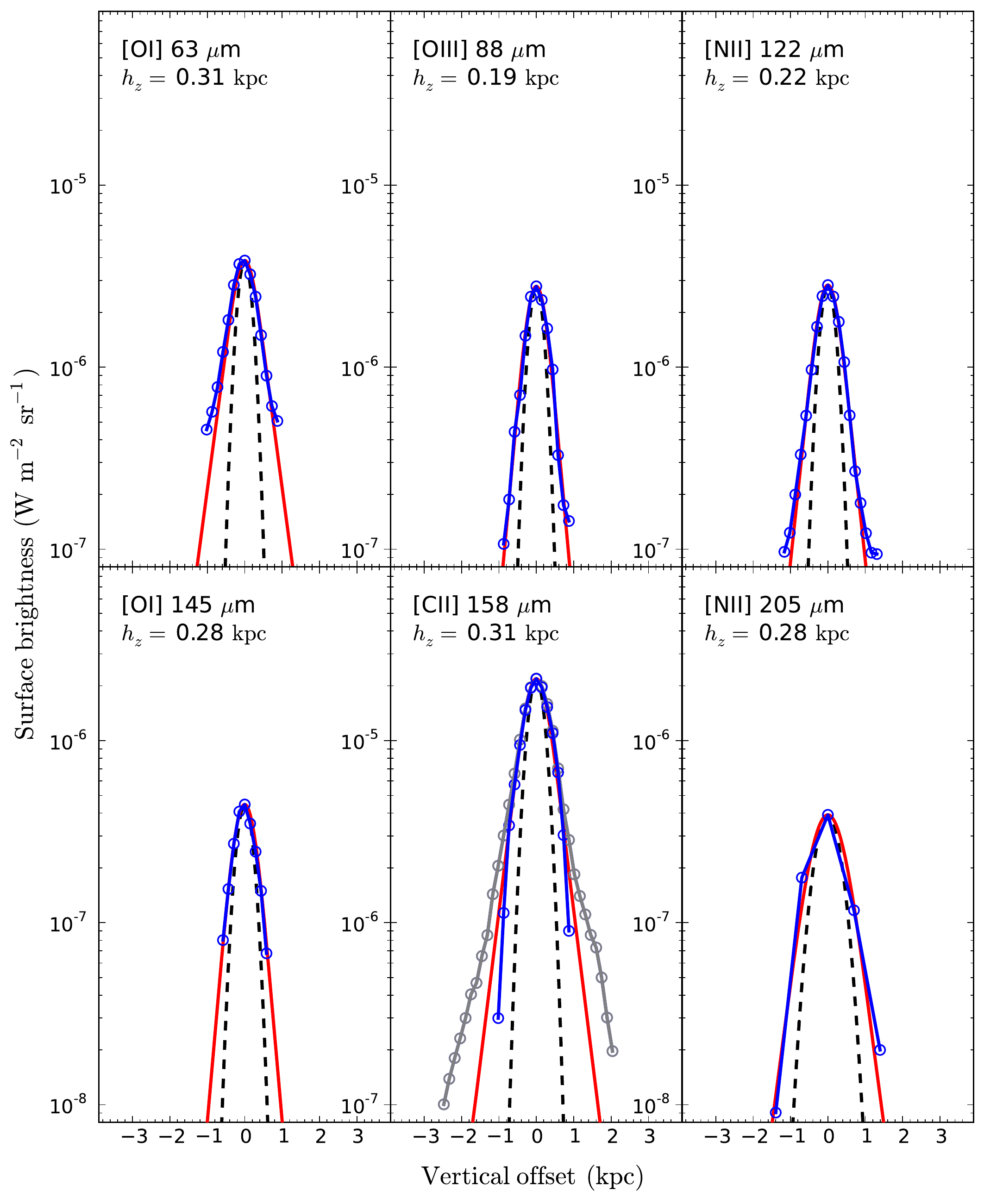}
\end{center}
\vspace{-0.3cm}
\caption[FIR line vertical profiles]{The vertical profiles from the emission maps (S/N $>$ 3$\sigma$) of the \cii \ 158~$\mu$m, \nii \ 122~and~205~$\mu$m, \oi \ 63 and 145~$\mu$m, and \oiii \ 88~$\mu$m fine-structure lines are plotted as blue circles connected by a blue line. We fit with an exponential profile (solid red line), which has been convolved with a Gaussian of FWHM similar to the \textit{Herschel} beam at the corresponding wavelength (black dashed line). For the \cii \ profile, we also show the profile including the extraplanar emission detected at the $>$1$\sigma$ level (grey circles and grey line).}\label{fig:vertiprofs}
\end{figure}

In Table~\ref{tab:measproperties}, we present the integrated line emission from each of the maps in Fig.~\ref{fig:obsmaps3sig} yet caution, however, that our integrated line fluxes are certainly an underestimate of the true global emission since our PACS observations do not cover the full extent of NGC~891's disc, thus missing any contribution from the south-west side. We measure a total \cii \ emission of (3.0$\pm$0.6)$\times$10$^{-14}$~W~m$^{-2}$ across a mapped area of $\sim$ 2.6$\times$10$^{-7}$ sr. NGC 891's \cii \ emission was previously mapped at 55$\arcsec$ resolution with the Far-Infrared Fabry-Perot Imaging Spectrometer on the KAO by \citet[][see their Fig.~5]{madden1994}, finding a peak intensity of 1$\times$10$^{-4}$~ergs~s$^{-1}$~cm$^{-2}$~sr$^{-1}$ in the center. Consistent with these results, our \cii \ map has an integrated intensity of (1.2$\pm$0.1)$\times$10$^{-4}$~ergs~s$^{-1}$~cm$^{-2}$~sr$^{-1}$ within a central 1$\arcmin$-diameter aperture. We also note that the integrated \cii \ line intensity contours from the \citet{madden1994} KAO observations correlate qualitatively with the spatial distribution of \cii \ seen in our \textit{Herschel} maps, with both sets of observations displaying a central peak and two secondary peaks along the semimajor axis. Extraplanar \cii \ emission matching the \citeauthor{madden1994} contours is evident in the original maps, but is not detected above the 3$\sigma$ level in the PACS map (see also Fig.~\ref{fig:vertiprofs}). In an 84$\arcsec$-diameter aperture, \citet{brauher2008} find a \cii \ flux of 7.79$\times$10$^{-15}$~W~m$^{-2}$ that is slightly lower than our value of $(9.2\pm 0.9)\times$10$^{-15}$~W~m$^{-2}$ found using the same aperture and smoothing our data to the $\sim$75$\arcsec$ resolution of the ISO LWS beam. More recently, \citet{stacey2010} calculated the luminosities of several fine-structure lines in NGC 891 from reprocessed observations made with the \textit{ISO} LWS spectrometer (see \citealp{brauher2008}, and references therein). Smoothing our data to match the ISO LWS beam resolution, we find an integrated \cii \ line luminosity of (1.20$\pm$0.20)$\times10^{8}\,L_{\odot}$ in agreement with value of 1.40$\times10^{8}\,L_{\odot}$ reported by \citet{stacey2010}.   

\begin{table}
 \begin{center}
\begin{minipage}{\columnwidth}
  \caption{Measured properties of fine-structure line emission.}
  \label{tab:measproperties}
   \begin{center}
   {\renewcommand{\arraystretch}{1.25}%
  \begin{tabular}{l c c c  c}
\hline 
\hline
Line  & $\lambda_{\mathrm{rest}}$ & Detected area &  Flux   &  Scale height   \\
      & ($\mu$m)  & ($10^{-7}$ sr) & (W m$^{-2}$) & (kpc)   \\      
\hline
\oi   &  63.184  & 1.8 & (4.6$\pm$0.5)$\times10^{-15}$ & 0.31$^{+0.09}_{-0.07}$ \\
\oiii &  88.356  & 1.2 & (3.0$\pm$0.4)$\times10^{-15}$ & 0.19$^{+0.07}_{-0.06}$\\ 
\nii  &  121.898 & 2.1 & (3.2$\pm$0.5)$\times10^{-15}$ & 0.22$^{+0.08}_{-0.07}$\\ 
\oi   &  145.525 & 1.0 & (4.3$\pm$1.3)$\times10^{-16}$ & 0.28$^{+0.08}_{-0.08}$\\ 
\cii  &  157.741 & 2.6 & (3.0$\pm$0.6)$\times10^{-14}$ & 0.31$^{+0.06}_{-0.05}$ \\ 
\nii  &  205.178 & 1.4 & (2.8$\pm$0.9)$\times10^{-16}$ & 0.27$^{+0.04}_{-0.03}$\\
\hline
\end{tabular}}
\end{center}
\end{minipage}
\end{center}
\end{table}

From the \nii \ 122~$\mu$m emission in our maps, we calculate a total intensity of (3.2$\pm$0.5)$\times$10$^{-15}$~W~m$^{-2}$ across our mapped area of $\sim$ 2.1 $\times$10$^{-7}$ sr.
The nuclear \nii \ 122~$\mu$m emission measured with the ISO LWS is 1.11$\times$10$^{-15}$~W~m$^{-2}$ \citep{brauher2008}, which agrees with the flux of (1.2$\pm$0.2)$\times$10$^{-15}$~W~m$^{-2}$ found from our data when matching the aperture properties and resolution of the LWS data. Similarly, our total \oi \ 63~$\mu$m emission is (4.6$\pm$ 0.5)$\times$10$^{-15}$~W~m$^{-2}$ across our mapped area of $\sim$1.8$\times$10$^{-7}$ sr. As in the case of the \cii \ emission, the integrated \oi \ line luminosity in our smoothed $\sim$75$\arcsec$ resolution map of (1.81$\pm$0.25)$\times10^{7}\,L_{\odot}$, which we stress is a lower limit due to differences in spatial coverage, appears somewhat consistent with the measurement of 4.67$\times10^{7}\,L_{\odot}$ from the ISO LWS observations within the 50\% error margin (\citealp{stacey2010}).  

The VNGS observations of the \oi \ 145~$\mu$m and \oiii \ emission consist of only strips that cover a smaller area than the combination of strips plus maps available for the other lines, and are less sensitive due to the raster spacing. Furthermore, the two lines are intrinsically weak. The \oi \ 145~$\mu$m map displays a spatial distribution similar to that of the 63~$\mu$m emission, albeit with much weaker emission. We calculate a total \oi \ 145~$\mu$m intensity of (4.3$\pm$1.3)$\times$10$^{-16}$~W~m$^{-2}$ across our mapped area of $\sim$1.0$\times$10$^{-7}$ sr. The corresponding integrated \oi \ 145~$\mu$m line luminosity is (1.25$\pm$0.5)$\times10^{6}\,L_{\odot}$ in the $\sim$75$\arcsec$ resolution map, which is only consistent with the \citet{stacey2010} measurement of 4.74$\times10^{6}\,L_{\odot}$ when taking into account our 30\% calibration error, the \citet{stacey2010} 50\% error, and the differences in spatial coverage. Finally, we measure a total \oiii \ 88~$\mu$m line intensity of (3.0$\pm$0.4)$\times$10$^{-15}$~W~m$^{-2}$ across our mapped area of $\sim$1.2$\times$10$^{-7}$ sr, and note that the spatial distribution is similar to that of the other FIR line emission and appears to follow the \tir \ contours. \citet{brauher2008} report an \oiii \ flux of 1.52$\times$10$^{-15}$~W~m$^{-2}$ from the central 84$\arcsec$-diameter aperture, in agreement with our value of $(1.65\pm 0.24)\times$10$^{-15}$~W~m$^{-2}$. 

Thus, we conclude that our observations appear consistent to previous measurements from both KAO and ISO observations. 

\subsection{Line scale heights}

\begin{figure*}
\begin{center}
\includegraphics[width=0.99\textwidth]{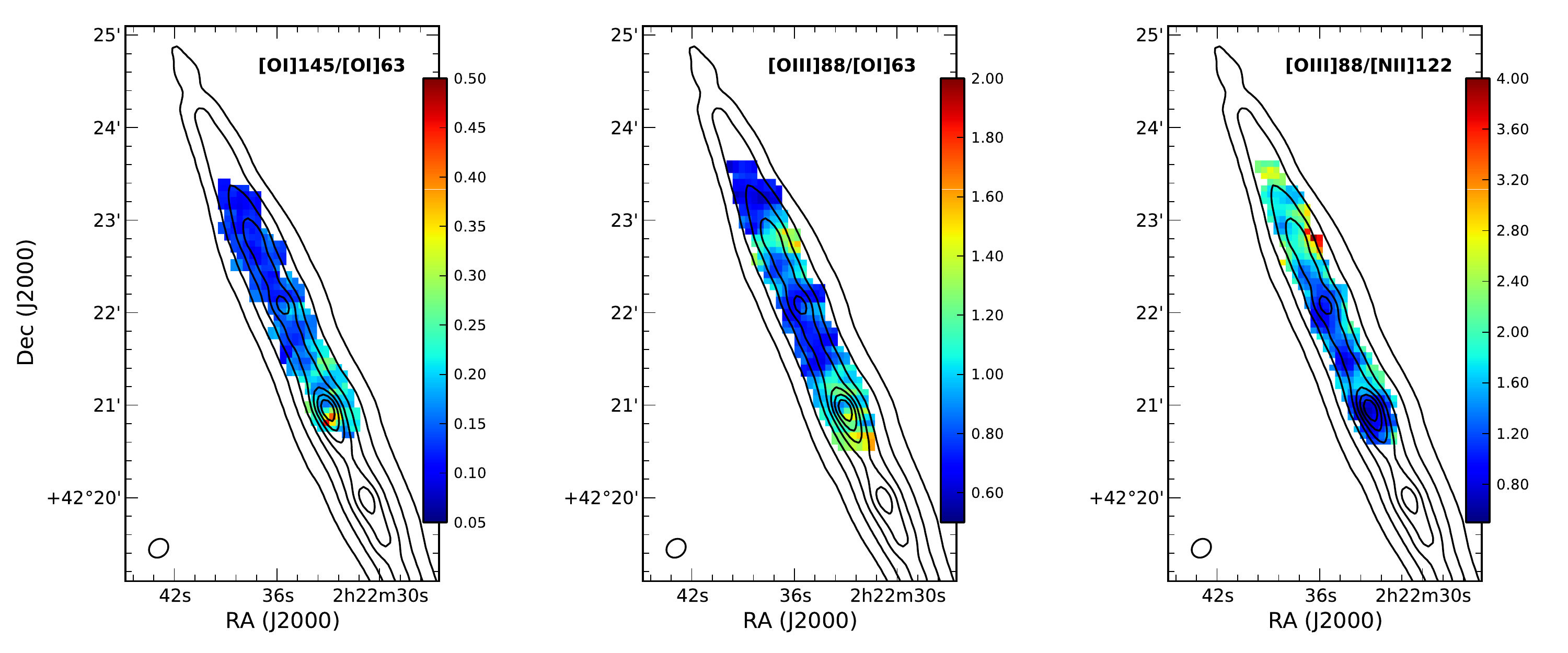}
\end{center}
\vspace{-0.3cm}
\caption[PACS far-infrared spectroscopic maps]{Maps of indicators of the interstellar gas properties: the \oi 145/\oi 63 line ratio for tracing the gas temperature (\textit{left}), the \oiii 88/\oi 63 line ratio as a tracer of the ionised gas fraction (\textit{middle}), and the \oiii 88/\nii 122 ratio for deriving the stellar effective temperatures of the ionising sources (\textit{right}). The maps are centred on $\alpha = 2^\mathrm{h}\ 22^\mathrm{m}\ 35\fs7$, $\delta = +42\degr\ 22\arcmin\ 05\farcs9$ (J2000.0) and are presented in the resolution and pixel size of the PACS 160~$\mu$m map. Contours from the $F_{\mathrm{TIR}}$ map (see Fig.~\ref{fig:tirmap}) are superimposed on each image as a visual aid with levels as listed in Fig.~\ref{fig:obsmaps3sig}.}\label{fig:mapsdiags}
\end{figure*}

Crucial for our later discussion about the FIR line emission above the mid-plane, we now investigate whether or not the observations are spatially-resolved in the vertical direction. We extracted vertical profiles of the six FIR fine-structure lines from the emission maps (see Fig.~\ref{fig:obsmaps3sig}). Following the same methodology as \citet{verstappen2013} and \citet{hughes2014}, we rotated our maps to a horizontal orientation before summing all the values of pixels with S/N $>$ 3$\sigma$ along the major axis to generate the profiles. We model these profiles with an exponential function appropriate for an exactly edge-on, double-exponential disc, given by
\begin{equation}\label{eqn:vertiprof}
\Sigma_{\mathrm{ver},\lambda}(z)=\frac{1}{2 h_{z,\lambda}}\exp\left(-\frac{|z|}{h_{z,\lambda}}\right)
\end{equation}
where $h_{z,\lambda}$ is the scale height of each line. We first convolve the vertical profile model with the \textit{Herschel} beams at each corresponding wavelength, using the appropriate Gaussian PSF images\footnoterecall{aniano} available from \citet{aniano2011}. In order to obtain one-dimensional beams, the two-dimensional PSFs are averaged along one direction in the same manner as we obtain the vertical profiles. The optimal value of $h_z$ that reconciles the observed and model profiles was found by using a $\chi^2$ minimisation technique with uncertainties also derived from the $\chi^2$ probability distribution. We adopt position angles of 22.9$^{\circ}$ for our analysis (see \citealp*{bianchi2011}; \citealp{hughes2014}).

The resulting vertical profiles are shown in Fig.~\ref{fig:vertiprofs}. For NGC 891, we derive scale heights of 0.31$^{+0.06}_{-0.05}$, 0.22$^{+0.08}_{-0.07}$, 0.27$^{+0.04}_{-0.03}$, 0.31$^{+0.09}_{-0.07}$, 0.28$^{+0.08}_{-0.08}$ and 0.19$^{+0.07}_{-0.06}$~kpc for the \cii \ 158~$\mu$m, \nii \ ~122~and~205~$\mu$m, \oi \ ~63 and 145~$\mu$m, and \oiii \ ~88~$\mu$m emission, respectively. Interestingly, the scale heights of the ionized gas tracers, i.e. the \nii ~122 and \oiii ~88 lines, are consistent with scale heights found in previous studies using optical and mid-IR emission lines to trace the more diffuse ionised gas. From \textit{Spitzer} Infrared Spectrograph observations of NGC~891, \citet{rand2008,rand2011} found scale heights of between approximately 0.25--0.5~kpc for the [Ne{\sc ii}]~12.81, [Ne{\sc iii}]~15.56 and [S{\sc iii}]~18.71~$\mu$m line emission (see Fig.~8 in \citealp{rand2011}). Following the reasoning of \citet{verstappen2013}, as our profiles are not dominated by the telescope beam, as evident in Fig.~{\ref{fig:vertiprofs}}, and the deconvolved scale height values we derive from the profile fitting are \textit{not} consistent with zero at the 5$\sigma$ level, we conclude that our vertical profiles of the FIR lines are spatially resolved. 

Previous observations have uncovered significant amounts of extended extraplanar emission from dust (e.g., \citealp*{howk1999}) as well as PAHs and small dust grains (e.g., \citealp{burgdorf2007}; \citealp{rand2008}; \citealp{whaley2009}) in NGC 891, and also other edge-on spirals (e.g. \citealp{thompson2004}; \citealp{rand2011}; \citealp{holwerda2012}; \citealp{verstappen2013}). Since PAHs appear to dominate the photoelectic heating of the gas (see Sec.~\ref{sec:heatingcooling} for a discussion), we would expect to see cooling from the fine-structure lines at least up to the same scale heights as the PAH features. Such extraplanar emission would be most evident from the primary gas coolants, the \cii~158~$\mu$m and \oi~63~$\mu$m lines (e.g. \citealp{wolfire1995}; \citealp{kaufman1999}). In fact, some extraplanar \cii \ emission matching the \citet{madden1994} contours is evident in the original maps albeit not detected in the PACS map above the 3$\sigma$ level and, in the lower middle panel of Fig.~\ref{fig:vertiprofs}, we present the \cii \ vertical profile including the extraplanar emission detected at the 1 to $<$3$\sigma$ level. Furthermore, we find our \cii \ scale height is in rough agreement with the scale heights of the PAH features ($\sim$ 0.4--0.5 kpc) derived by \citet{rand2011}.

\subsection{Consideration of optical depth effects}\label{sec:opticaldepth}

With edge-on galaxies like NGC~891, we face the possibility that variations in the optical depth along the line-of-sight may affect certain line ratios and thereby modify any trends found in the analysis, which becomes particularly important when encountering higher column densities as we observe towards the center of the galaxy. We can check whether such effects pose an issue in this work using our \oi \ line observations. The \oi ~63 and 145~$\mu$m lines have respective upper level energies, $\Delta E/k$, of 228~K and 327~K above the ground state (see e.g. \citealp*{tielens1985}; \citealp{liseau2006}), meaning the ratio of \oi 145/\oi 63 can probe optically thin neutral gas with temperatures of $\sim 300$~K. The \oi \ 63~$\mu$m line can become optically thick at lower column densities faster than the 145~$\mu$m line, leading to an apparent increase in the ratio at gas temperatures lower than $\sim 1000$~K \citep*{tielens1985}. We thus examine the optical thickness of the neutral gas by comparing our observed values to the theoretical values expected for gas with varying temperatures. 

In the left panel of Fig.~\ref{fig:mapsdiags}, we present our map of the \oi 145/\oi 63 ratio. Even though the \oi ~145~$\mu$m line was only mapped along a radial strip (c.f.~Fig.~\ref{fig:obsmaps3sig}), our measurements of the ratio cover the central region of the galaxy where we would expect optical depth effects to become most important. In fact, towards the center region, the \oi 145/\oi 63 ratio is typically $>$0.15 with uncertainties of $\sim 10$\%. Comparing the inverse of this value with Fig.~4 of \citet{liseau2006}, we find that the \oi \ 63~$\mu$m line is either optically thick with $T \gtrsim 200$~K and $n \gtrsim 10^{3}$~cm$^{-3}$, or optically thin and hot with $T \gtrsim 1000$~K and a density of approximately $10^{3}$~cm$^{-3}$. For the central peak of \oi 145/\oi 63 $\sim$ 0.41, the gas is likely to be completely optical thick. Such a high ratio could also indicate optical depth effects in the continuum emission at 63~$\mu$m and/or foreground absorption by low-density, diffuse gas, especially since NGC 891's almost perfect edge-on inclination could significantly increase $\tau_{\mathrm{dust}}$. Radiative transfer modelling of the disc, beyond the scope of this work, would be required to accurately investigate such effects on the continuum. The remainder of the disc, however, exhibits \oi 145 /\oi 63 ratios $<$0.15 that correspond to optically thin neutral gas at temperatures $\sim 100-300$~K. 

We attempt to derive a rough constraint on the extinction $A_V$ from the dust mass surface density map, derived from VNGS Herschel PACS and SPIRE photometry\footnote{In brief, we fit the six \textit{Herschel} PACS/SPIRE photometric bands with a one-component modified blackbody originally presented by \citep{hildebrand1983}, assuming a power-law dust emissivity with  $\kappa_{\nu}$ = 0.192 m$^{2}$ kg$^{-1}$ at 350~$\mu$m \citep{draine2007} and fixing the spectral index $\beta$ $=$ 1.8 (e.g. \citealp{galametz2012}).} (see \citealp{hughes2014}), using Eq. 4 from \citet{kreckel2013} for a simple geometry that assumes the dust is distributed in a uniform screen between the emitter and the observer, and which adopts the observed Milky Way ratio of visual extinction to hydrogen column density ($A_V$/N$_H$ = $5.34 \times 10^{-22}$ mag cm$^2$/H), and a fixed dust-to-gas ratio ($\Sigma_{\mathrm{dust}}/(N_H m_H) = 0.010$) from the \citet*{draineli2007} model prescription (see their Table 3). The dust mass surface densities from pixels where both \oi \ lines are detected result in extinctions of 5 $\lesssim$ $A_V$ $\lesssim$ 17 mag, where we note the uniform dust screen geometry yields an upper limit to the extinction. Although the absence/weakness of a cold diffuse dust component suggests a one-component greybody (as adopted here) is most appropriate to derive a reasonable estimate of the dust mass in NGC 891 (\citealp{hughes2014}), using just a single thermal component to fit cases where the FIR SEDs are clearly divisible into separate thermal components may underestimate the dust mass by a factor of two \citep{bendo2014}. Should this be the case for our target galaxy, the derived $A_V$ would simply shift to higher values. 

The resulting $A_V$ map is then regridded to match the pixel size of the \oi ~63 and 145~$\mu$m emission maps ($4\arcsec$) to facilitate a pixel-by-pixel comparison between the \oi 63/\oi 145 ratio and the extinction (see Fig.~\ref{fig:oilines_av}). For reference, we compare our observations to the predicted \oi 63/\oi 145 ratio along the line-of-sight as a function of $A_V$ (the red line in Fig.~\ref{fig:oilines_av}) from the open geometry PDR model of \citet[][see their Fig.~3]{abel2007}, which represents a lower limit. We find the \oi 63/\oi 145 ratio tends to decrease with increasing $A_V$, particularly evident in the mid-plane and central (i.e. on-axis) pixels, further suggesting that optical depth effects become important for the \oi \ 63 line \citep{abel2007}. We stress that the $A_V$ derived here is a global, beam-averaged measurement that effectively probes the global ISM opacity arising not only from PDRs. Furthermore, whilst we adopt a dust screen for simplicity, in reality we expect a mixing of stars and dust within the disc, of which the overall distribution of local star-dust geometries dictates the shape of the global SED and thus the effective $A_V$ (see e.g. \citealp{karczewski2013}). However, an empirical measure of \oi 63/\oi 145 -- $A_V$ is difficult to constrain. In the absence of better constraints, we keep in mind that the central and dust lane regions likely suffer from the effects of increasing optical thickness as we proceed with our analysis.

\subsection{Ionised gas characteristics}\label{sec:ioncharac}

Whilst the \oi \ transition arises from the neutral gas in PDRs, the high excitation potential of 35 eV required to further ionise O$^{+}$ means the \oiii ~88~$\mu$m transition predominantly originates in low-density \hii \ regions and diffuse ionised gas, and so the ratio of these lines can give some indication of the relative distributions of ionised and neutral media. From our \oiii 88/\oi 63 line ratio map, presented in the middle panel of Fig.~\ref{fig:mapsdiags}, we find that most of the disc is dominated by neutral gas. The emission from ionised gas appears to peak either side of the center, although this may be due to the \oi \ 63~$\mu$m line becoming increasingly optically thick towards the nucleus (see the previous section) and boosting the \oiii 88/\oi 63 ratio. Further along the disc, however, there is a second peak where the emission arising from ionised gas is stronger relative to the neutral gas, which coincides with the region that often demonstrates an asymmetry at numerous wavebands compared to the region diametrically opposite. Since the gas here is optically thin, we are likely integrating along a line-of-sight through the \hii \ regions of a spiral arm (see Fig.~3 in \citealp{kamphuis2007}). 

We can further probe the ionized gas via the \oiii/\nii 122 ratio. Since the ionization potentials of N and O$^{+}$ are 14.5 and 35~eV and the \nii ~122 and \oiii \ lines have critical densities of 310 and 510~cm$^{-3}$, the ratio of these two lines is relatively insensitive to the gas density. If the emission arises from \hii \ regions\footnote{Within the narrow line region of an AGN, the \oiii/\nii 122 ratio can also probe the strength of the ionization parameter, $U$, as described in e.g. \citealp{abel2009}.}, then the ratio gives an indication of the effective stellar temperature of the ionizing source (\citealp{ferkinhoff2011}) and thus can be used to constrain the stellar classification of the youngest stars in a \hii \ region. We find that for the majority of the mapped region of NGC 891 (see Fig.~\ref{fig:mapsdiags}, right panel), the \oiii/\nii 122 ratio is $\sim$ 1.0 but shows an increase in the area where the \oiii 88/\oi 63 peaks and at larger radii. Following the method of \citet{ferkinhoff2011} also adopted by \citet{parkin2014}, we compare our observed ratios to the model predictions. In Fig.~7 of \citet{parkin2014}, the theoretical line ratios are plotted as a function of stellar temperature for various gas densities as predicted from the \hii \ region models of \citet{rubin1985}. The \oiii/\nii 122 line ratios we measure across the disc correspond to a range of stellar effective temperatures of approximately 3.43$\times$10$^{4}$ and 3.65$\times$10$^{4}$~K, which in turn correspond to stellar classifications of O9 to O9.5 for the most luminous stars (see Fig.~1 of \citealp{vacca1996}), suggesting that young stars are present in the disc of NGC~891.

\subsection{Ionised gas contribution to \cii \ emission}\label{sec:ionisedcontrib}

\begin{figure}
\begin{center}
\includegraphics[width=0.98\columnwidth]{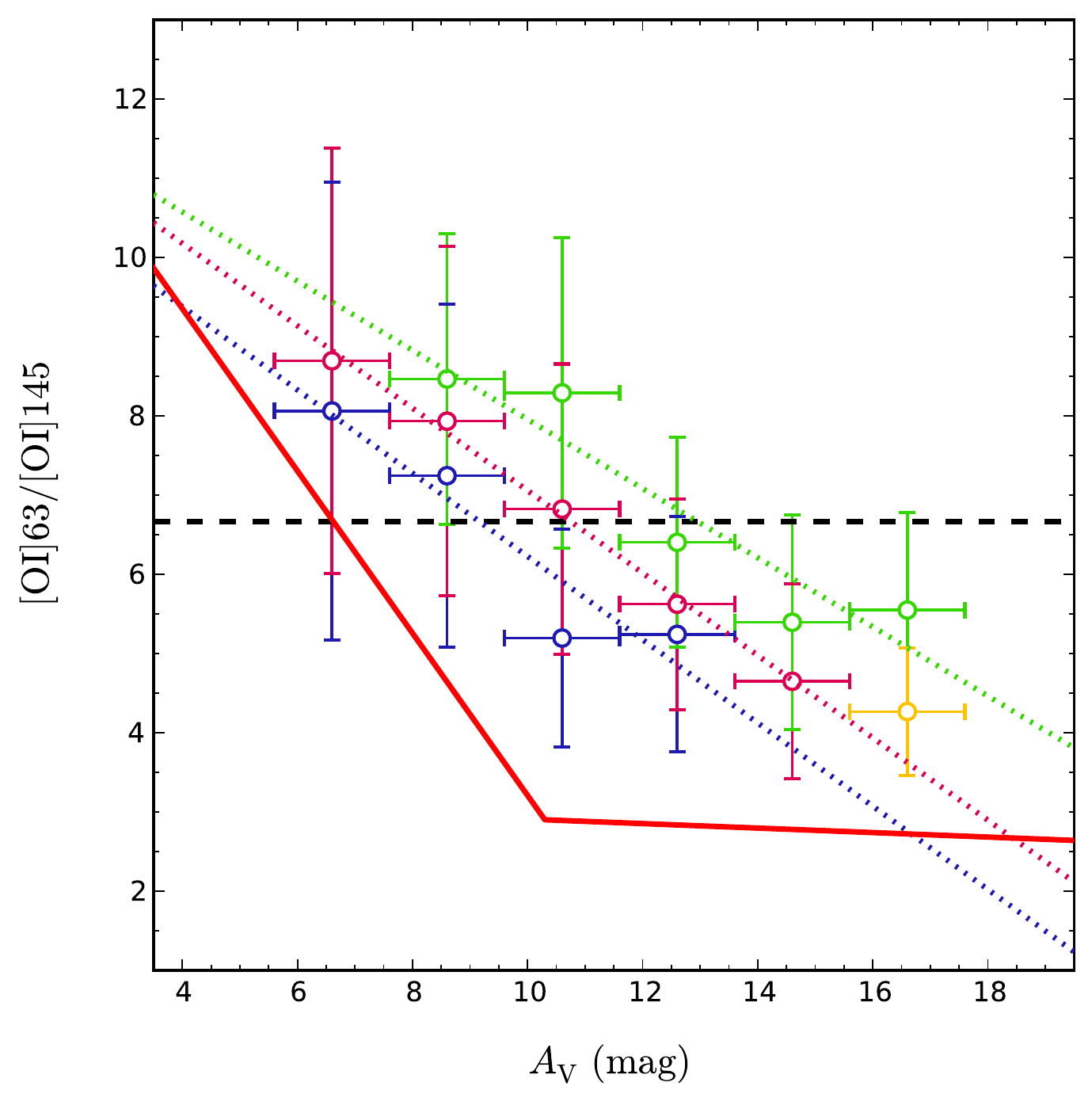}
\end{center}
\vspace{-0.3cm}
\caption[]{The \oi 63/\oi 145 line ratio plotted as a function of the attenuation. We distinguish the median line ratio derived for pixels from different regions of the disc in $A_V$ bins with widths of 2 mag (open circles) according to the colour scheme as depicted in Fig.~\ref{fig:tirmapmask}, and plot the corresponding best linear fits to the binned data (coloured dotted lines). We note the limited coverage of the \oi \ 145~$\mu$m map means these colours correspond mainly to varying vertical height from the mid-plane of the disc. The black dashed line corresponds to the ratio that \textit{approximately} divides the optically thin and optically thick regime as in Fig.~4 of \citet{liseau2006}, and the red solid line demonstrates the \oi 63/\oi 145 -- $A_V$ relationship predicted by the PDR model of \citet[][see their Fig. 3]{abel2007}.}\label{fig:oilines_av}
\end{figure}

\begin{figure*}
\begin{center}
\includegraphics[width=0.96\columnwidth]{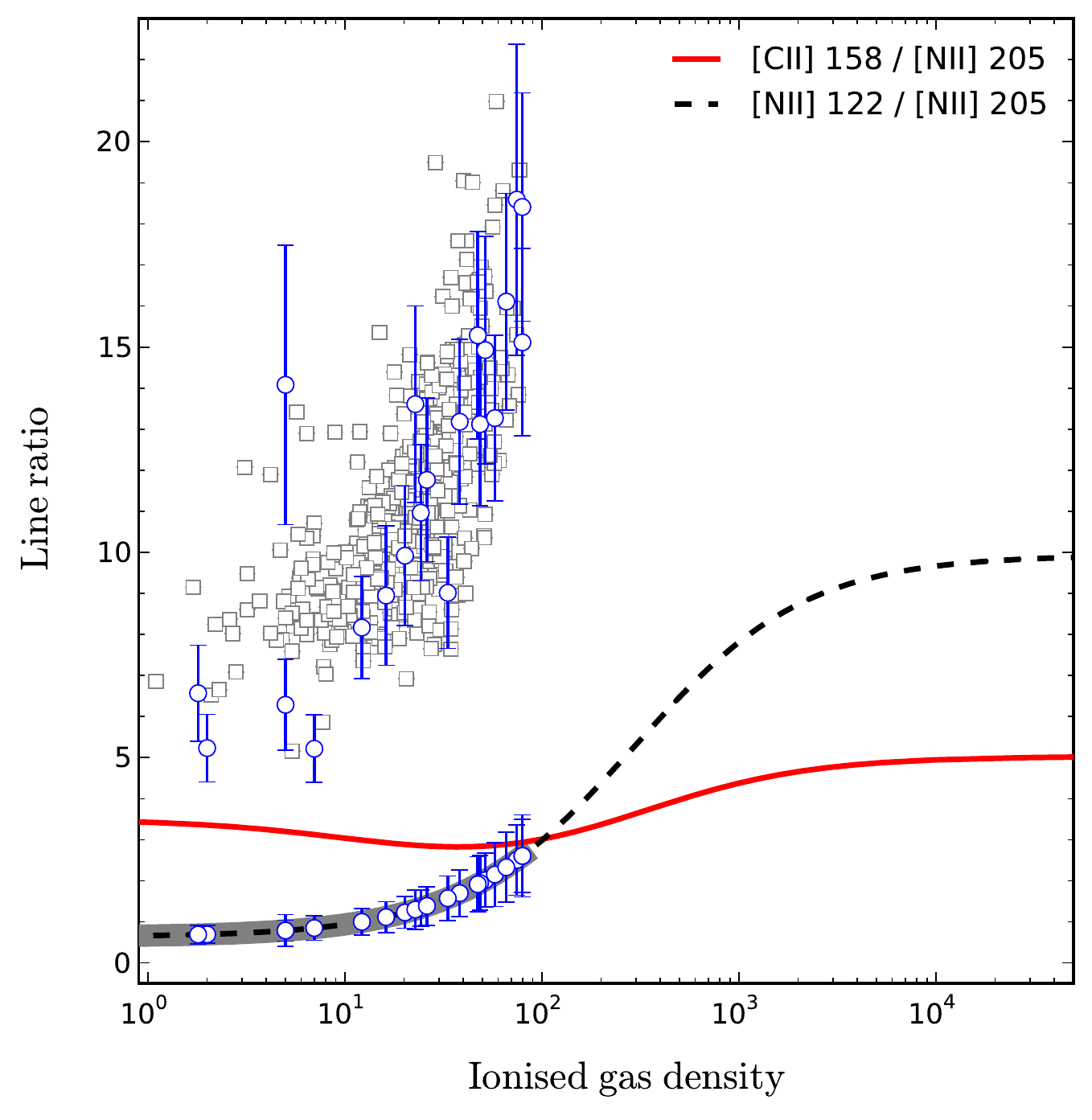}
\includegraphics[width=0.99\columnwidth]{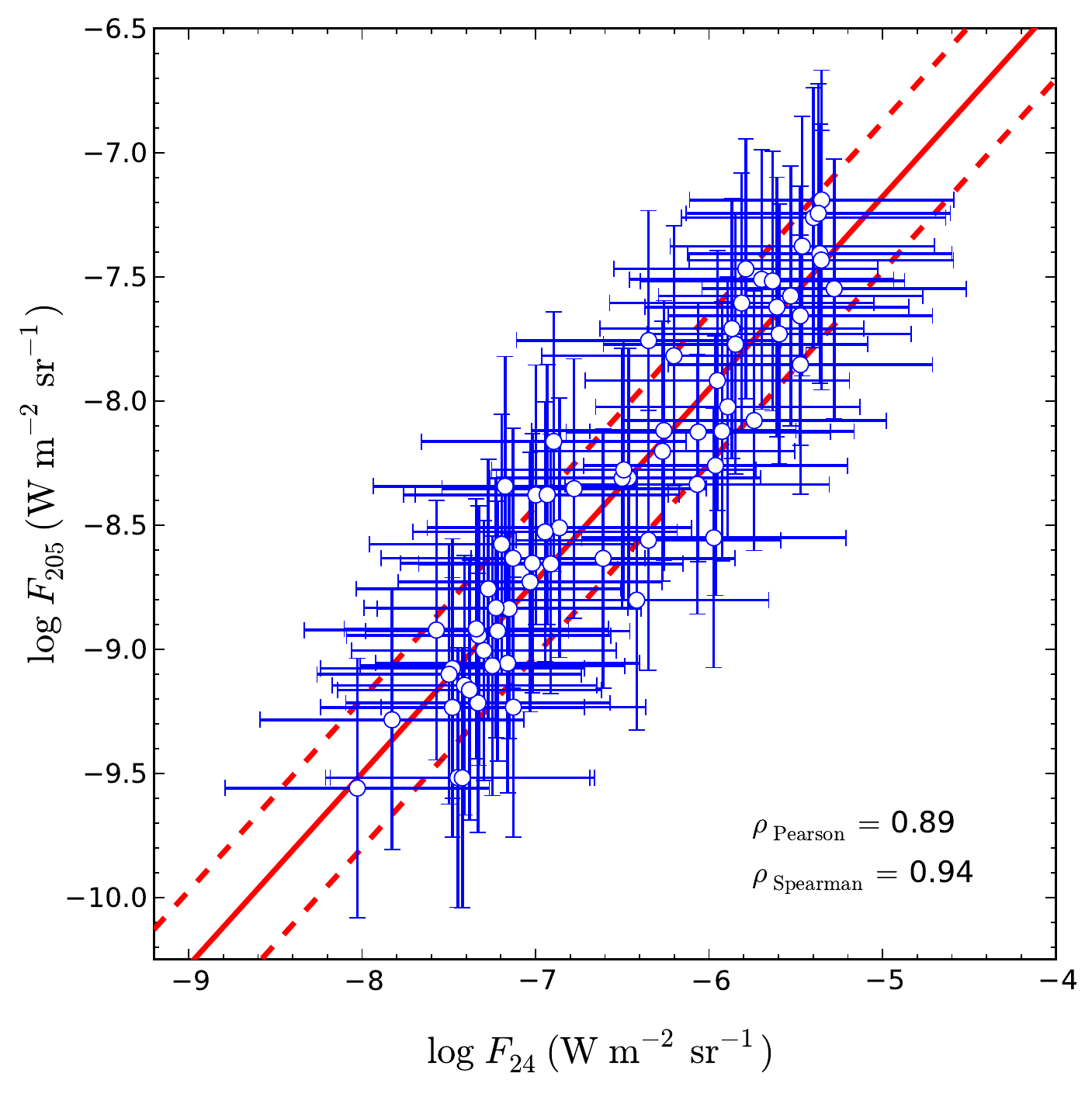}
\end{center}
\vspace{-0.3cm}
\caption[]{\textit{Left}: The theoretical line ratios for the \cii \ 158, \nii \ 122 and 205~$\mu$m transitions from diffuse ionised gas, plotted as a function of the hydrogen density. We interpolate the ionised gas density from the observed \nii \ 122 / \nii \ 205 ratio (blue circles following the black dashed line), which is subsequently used to determine the theoretical \cii \ 158 / \nii \ 205 emission for comparison with the observed \cii 158/\nii 205 emission (open blue circles), and hence calculate the fraction of \cii \ emission attributed to PDRs. The synthetic \nii \ 122 / \nii \ 205 line ratios (occupying the grey shaded region) estimated via a correlation with the MIPS 24~$\mu$m emission (see the right hand panel) are used to interpolate the ionised gas density to compare with the estimated \cii \ 158 / \nii \ 205 ratio (open grey squares). \textit{Right}: The correlation between the MIPS 24~$\mu$m emission and the \nii~205 ~$\mu$m emission. The best linear fit including 1$\sigma$ limits are represented by the solid and dashed red lines, respectively. In both panels, each data point represents one pixel.}\label{fig:ionisedgasfraction}\label{fig:24to205correlation}
\end{figure*}

The \cii \ emission originates from both ionised and neutral gas and thus, for accurate comparison to PDR models that consider the emission arising purely from the neutral gas, we must take into account the fraction of \cii \ emission originating from the ionised gas that we investigated in the previous section. We estimate the fraction of the \cii \ emission originating from ionised gas, \ciiion , following the method of \citet{oberst2006,oberst2011}  via the \cii /\nii 205 and \nii 122/\nii 205 ratios. The latter ratio is a sensitive probe of the ionised gas density in \hii \ regions, with the \nii \ emission arising entirely from ionised gas, due to the N ionization potential (14.5 eV) being greater than that of neutral hydrogen (13.6 eV). Since the \cii \ and \nii \ 205 $\mu$m lines have very similar critical densities for collisional excitation by electrons (46 and 44 cm$^{-3}$ at $T_{e}$ = 8000 K, respectively), the \cii /\nii 205 line ratios are mainly dependent on the relative abundances of C and N in the \hii regions. Comparing our observed \nii 122/\nii 205 ratios to the theoretical  ratio will allow us to infer the ionised gas density, from which we can predict the theoretical \cii /\nii 205 ratio arising from the ionised gas and subsequently estimate the neutral gas contribution to the \cii \ emission. Adopting solar gas phase abundances of $n(\mathrm{C}^{+})/n_{e} = $ 1.4$\times$10$^{-4}$ and $n(\mathrm{N}^{+})/n_{e} = $ 7.9$\times$10$^{-5}$ \citep*{savage1996}, respective \cii \ and \nii \ collision strengths from \citet*{blum1992} and \citet*{hudson2004}, and Einstein coefficients for \cii \ and \nii \ from \citet{galavis1997} and \citet{galavis1998}, i.e. the same values as \citet{parkin2013}, we calculate the theoretical \cii /\nii 205 and \nii 122/\nii 205 ratios as a function of the ionised gas density (see Fig.~\ref{fig:ionisedgasfraction}). For determining the ionised gas contribution to our observed \cii \ emission, we compare these curves to our observations using two different approaches. 

In our first approach, we base our calculations solely on pixels where the \nii \ 205~$\mu$m line is detected (see Fig.~\ref{fig:obsmaps3sig}). We convolve and rebin the \cii \ and \nii~122~$\mu$m maps to the resolution ($\sim$17$\arcsec$) and pixel size (15$\arcsec$) of the \nii \ 205~$\mu$m map. From the 30 pixels with a 3$\sigma$ detection of the \nii \ 205~$\mu$m line, we find the \nii 122/\nii 205 ratios range from 0.7 to 2.5, from which we infer ionised gas densities ranging from 1.9 to 80 cm$^{-3}$ with a mean $n_{e}=$ 21.9 cm$^{-3}$ (Fig.~\ref{fig:ionisedgasfraction}). Thus, the emission of these lines stems from diffuse gas. We interpolate the theoretical \cii /\nii 205 line ratio in each pixel from our inferred $n_{e}$ values, then compare these to our observed \cii /\nii 205 line ratios. In the disc of NGC 891, we calculate that the fraction of the \cii \ emission originating from ionised gas, \ciiobsion , varies from 0.15 up to 0.65 with a median and standard deviation of 0.22 and 0.15, respectively. Whilst both the number of pixels and their spatial resolution are low, providing only coarse estimates of the local ionised gas contribution to the \cii \ emission, our values across the disc appear consistent with previous results from a variety of astronomical sources (see e.g. \citealp{oberst2006,oberst2011}; \citealp{croxall2012}; \citealp{parkin2013}; \citealp{farrah2013}; \citealp{parkin2014}).

The small number of low resolution pixels available when using the \nii \ 205~$\mu$m map introduces a problematic limitation to our analysis, particularly evident when we attempt to use these results to correct our observed \cii \ emission for the contribution arising from ionised gas in order to facilitate a comparison with the \citet{kaufman1999,kaufman2006} PDR model (Sec.~\ref{sec:pdrmodelling}). To circumvent this issue, we experiment using a second approach in which we exploit a strong relationship we observe between the \nii \ 205~$\mu$m line emission and the 24~$\mu$m emission. Both these emission sources have been found to correlate with SFR on global and local scales. The 24~$\mu$m emission may be used to trace the obscured star formation (e.g. \citealp{calzetti2007}). In a recent study of 70 galaxies in the \textit{Herschel} Spectroscopic Survey of Warm Molecular Gas in Local Luminous Infrared Galaxies, \citet{zhao2013} found that the SFR determined from the TIR luminosity via the relationship in \citet*{kennicutt2012} correlated with the \nii \ 205~$\mu$m line luminosity. More recently, \citet{wu2014} found a spatially-resolved correlation between the surface densities of the SFR and \nii \ 205~$\mu$m line in the M83 galaxy, and that intersection of this local relationship and the global relationship of \citet{zhao2013} at high $\Sigma_{\mathrm{SFR}}$ indicates the latter correlation is dominated by active star-forming regions. Since both the 24~$\mu$m and the \nii \ 205~$\mu$m line emission both seem to trace the SFR on spatially-resolved scales, one might expect to find a relationship between these two quantities that may subsequently be used to predict the \nii \ 205~$\mu$m line emission from the higher resolution 24~$\mu$m images.  

\begin{figure}
\begin{center}
\includegraphics[width=\columnwidth]{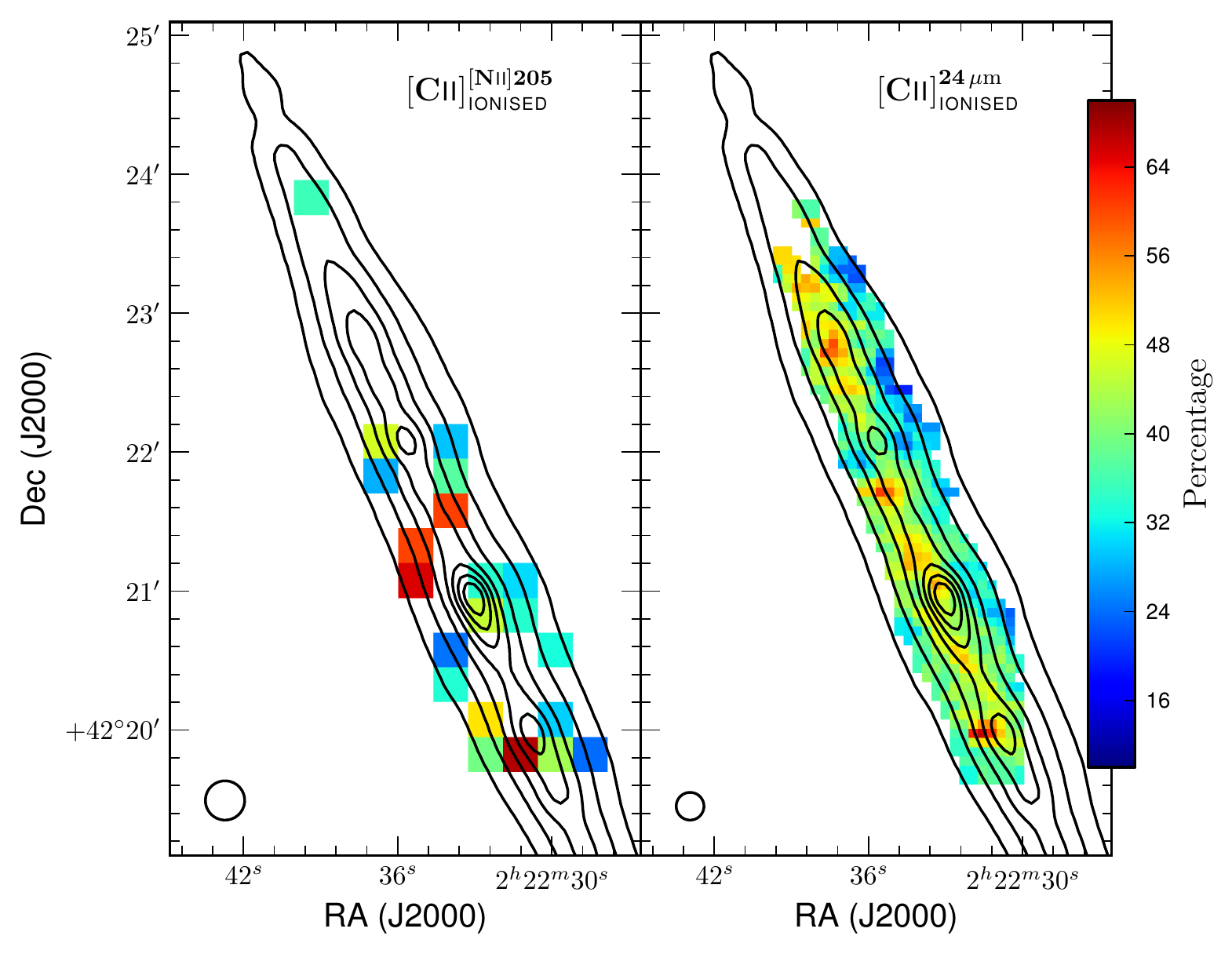}
\end{center}
\vspace{-0.3cm}
\caption[Total infrared flux map]{A comparison of the fraction of the \cii \ emission originating from ionised gas as determined from the observed map of the \nii~205~$\mu$m line emission at 17$\arcsec$ resolution (\textit{left panel}), and that as determined from a synthetic map of the \nii~205~$\mu$m line emission estimated using the 24~$\mu$m emission at 12$\arcsec$ resolution (\textit{right panel}) via the correlation presented in Fig.~\ref{fig:24to205correlation}. The maps are centred on $\alpha = 2^\mathrm{h}\ 22^\mathrm{m}\ 35\fs7$, $\delta = +42\degr\ 22\arcmin\ 05\farcs9$ (J2000.0) and the black contours from the $F_{\mathrm{TIR}}$ map (see Fig.~\ref{fig:tirmap}) represent TIR flux levels as listed in the caption of Fig.~\ref{fig:obsmaps3sig}.}\label{fig:iongasmap}
\end{figure}

We first convolve and regrid our 24~$\mu$m map to the resolution and pixel size of the \nii \ 205~$\mu$m line emission map, and in Fig.~\ref{fig:24to205correlation} we examine the relationship between their logarithmic fluxes on a pixel-by-pixel basis. We observe a strong correlation with Spearman and Pearson coefficients of rank correlation of 0.94 and 0.88, respectively, where a value of 1 represents a perfect correlation. The best linear fit to the data is given by
\begin{eqnarray}
\log\, F_{205} = (0.77\pm 0.01)\,\log\, F_{24} - (3.31\pm 0.04)
\end{eqnarray}
\noindent where both flux densities are in units of W m$^{-2}$ sr$^{-1}$. By applying this relation to the original 24~$\mu$m map, we can therefore estimate the \nii \ 205~$\mu$m line emission for all pixels at the resolution ($12\arcsec$) and pixel size ($4\arcsec$) of the \cii \ 158 and \nii ~122 $\mu$m maps. We set the error bars on these flux estimates at 50\%. Finally, we perform the same calculation as described above to estimate the fraction of the \cii \ emission originating from ionised gas, \ciisynion , using flux ratios based on this synthetic \nii \ 205~$\mu$m line emission map. From the synthetic map, we again estimate the \nii 122/\nii 205 ratios range from 0.7 to 2.5, from which we infer ionised gas densities ranging from 1.9 to 80 cm$^{-3}$ with a mean $n_{e}=$ 21.9 cm$^{-3}$ (Fig.~\ref{fig:ionisedgasfraction}). We thus find similar fractional contributions as before: the fraction of \cii \ emission from ionised gas varies from 0.13 to 0.61 with an average and standard deviation of 0.27 and 0.07, respectively. In Fig.~\ref{fig:iongasmap}, we present the maps of the fraction of the \cii \ emission arising from ionised gas, \ciiion , estimated from our two methods. Features in the estimated higher resolution map appear qualitatively consistent with those measured at lower resolution. We find \ciiion \ decreases with increasing height, implying the diffuse neutral component dominates the \cii \ emission in extraplanar regions, and that some regions have up to 50\% of the \cii \ emission originating from ionised gas. One clear peak corresponds to the region of enhancement in the \oiii \ emission relative to the TIR contours on the far north eastern side of the disc. The peaks also show a remarkably similar distribution to the 24/850~$\mu$m ratio map (Fig. 7 in \citealp{whaley2009}) and the \ha \ emission map (see \citealp{kamphuis2007}), indicating the presence of star-forming regions.

\subsection{Distribution of \cii /FIR ratio}\label{sec:ciideficit}

Before focussing our analysis on our main diagnostics of the gas heating and cooling mechanisms, we briefly consider the emission of the \cii \ line compared to the \tir \ emission. Compared to normal galaxies, low-redshift ultraluminous infrared galaxies (ULIRGS) exhibit much lower global \ciitir \ ratios of $\leq 5\times10^{-4}$ (see e.g. \citealp{luhman1998,luhman2003}). Although typically not seen in higher-redshift ULIRGS (e.g. \citealp{rigopoulou2014}; a notable exception is HFLS 3, \citealp{riechers2013}), this apparent deficit in the \cii \ line emission from normal star-forming galaxies has been extensively studied using surveys (e.g. \citealp{crawford1985}; \citealp{malhotra2001}) and observations of individual objects (e.g. \citealp{contursi2002,contursi2013}). In NGC 891, we calculate an integrated \ciitir \ value of $(4.6\pm 0.9)\times 10^{-3}$ across $\sim$2.5 $\times 10^{-7}$ sr, which, if we adopt \tir $=$ 1.3\fir \ \citep{graciacarpio2008}, corresponds to a \ciifir \ ratio of $(5.9\pm 1.2)\times 10^{-3}$ that is consistent with numerous other studies of the \ciifir \ ratio\footnote{We note that the relation of \tir $=$ 2\fir \ found by \citealp{dale2001} yields a \ciifir \ ratio of $(9.2\pm 1.8)\times 10^{-3}$.}. For example, the \citet{malhotra2001} ISO survey of 60 star-forming galaxies found global \ciifir \ value greater than $2\times 10^{-3}$ in a large fraction of the sample (see also \citealp{graciacarpio2011}). For reference, the Milky Way has a measured global \ciifir \ ratio of $3\times10^{-3}$ \citep{stacey1985}.         

\begin{figure*}
\begin{center}
\includegraphics[width=0.99\textwidth]{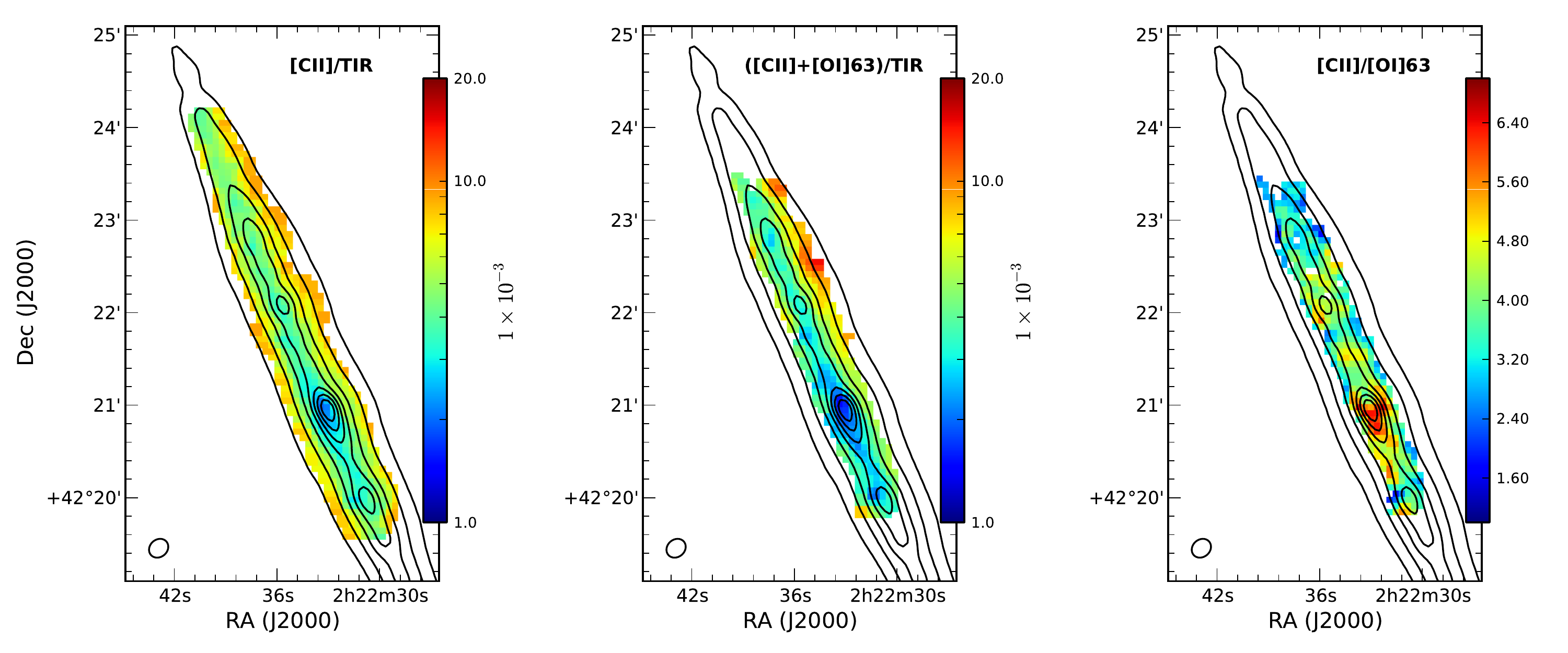}
\end{center}
\vspace{-0.3cm}
\caption[PACS far-infrared spectroscopic maps]{Maps of the main indicators of the gas heating and cooling: the \textit{uncorrected} \cii \ emission divided by the total infrared flux, \ciitir \ (\textit{left}), the sum of the \cii \ emission, \textit{corrected} for the contribution from ionised gas, and \oi \ 63 emission divided by the total infrared flux, \ciioitir \ (\textit{middle}), and the \ciioi \ ratio, where the \cii \ emission is also just the component from neutral gas (\textit{right}). The maps are centred on $\alpha = 2^\mathrm{h}\ 22^\mathrm{m}\ 35\fs7$, $\delta = +42\degr\ 22\arcmin\ 05\farcs9$ (J2000.0) and are presented in the resolution and pixel size of the PACS 160~$\mu$m map. Contours from the $F_{\mathrm{TIR}}$ map (see Fig.~\ref{fig:tirmap}) are superimposed on each image as a visual aid with levels as listed in Fig.~\ref{fig:obsmaps3sig}.}\label{fig:ciimapsdiags}
\end{figure*}

We examine the \ciitir \ distribution in Fig.~\ref{fig:ciimapsdiags} (left panel), finding that the ratio varies from $1.5 \times 10^{-3}$ in the nucleus to $13 \times 10^{-3}$ along the disc, corresponding to $ 2 \times 10^{-3}$ $\leq$ \ciifir \ $\leq$ $16.9 \times 10^{-3}$. In their spatially-resolved study of M51, \citet{parkin2013} found \ciifir \ ranging from $\sim1.3\times10^{-3}$ in the galaxy nucleus to values up to ten times higher in the spiral arms. Similar behaviour has been observed by \textit{Herschel} in M33, with \ciifir \ radially increasing from $8\times 10^{-3}$ up to $30\times 10^{-3}$ at $\sim$4.5 kpc from the center \citep{kramer2013}. Lower \ciitir \ values in the galaxy centre may be due to a higher fraction of UV photons being absorbed by dust instead of neutral hydrogen, contributing more to the TIR emission and less to the photoionisation heating of the gas in \hii \ regions (see e.g. \citealp{farrah2013}, and references therein). However, this is an unlikely explanation for non-starburst galaxies such as NGC~891, since the compactness of SF regions in the center may not be as high as in ULIRGS. For this galaxy, lower \ciitir \ values are most likely due to other lines becoming more important for gas cooling in the central regions, the photoelectric heating efficiency decreasing due to grain charging, and/or a varying contribution of PDRs to the TIR emission across the disc. In the following section, we examine in detail the gas heating and cooling mechanisms.

\subsection{Gas heating and cooling}\label{sec:heatingcooling}

The photoelectric heating efficiency of the interstellar gas, $\epsilon_{\mathrm{PE}}$, is defined as the ratio of the gas heating from photoelectrons to the dust heating from UV photons, i.e. the fraction of energy from the interstellar FUV radiation field that heats the gas via the photoelectric effect versus the fraction of the energy transferred to dust grains (see e.g. \citealp*{tielens1985}; \citealp{mochizuki2004}). Both dust heating and gas cooling can be investigated via FIR observations: warm dust is traced via the re-emission of absorbed UV and optical photons that peaks at FIR wavelengths, and gas heated from photoelectrons ejected from small dust grains may be traced during cooling via the collisionally-excited FIR fine-structure lines. The photoelectric heating efficiency maybe therefore be traceable with the observed FIR line-to-continuum ratio, \ciioitir, only if we assume the neutral gas cooling is dominated by the \cii~158~$\mu$m and \oi~63~$\mu$m lines (e.g. \citealp{wolfire1995}; \citealp{kaufman1999}) with negligible contributions from alternative cooling lines, such as e.g. [C{\sc i}], and that the TIR emission traces the gas heating in the same regions where these two lines originate with a negligible contribution from non-PDR emission. Since the photoelectric heating efficiency is only important in the neutral gas, in this section we consider only the \cii \ component originating from the neutral gas and thus correct our \cii \ emission for the fraction arising from the ionised gas.
 
In the middle panel of Fig.~\ref{fig:ciimapsdiags}, we present our map of \ciioitir \ as a proxy of the photoelectric heating efficiency, $\epsilon_{\mathrm{PE}}$. Our values of $\epsilon_{\mathrm{PE}}$ range from $\sim$1$\times$10$^{-3}$ to 2$\times$10$^{-2}$, consistent with the majority of studies on photoelectric heating. An \textit{ISO} LWS survey of 60 normal, star-forming galaxies spanning a range in various properties, such as morphology and FIR colour, found \ciioitir \ ranging from $\sim$10$^{-3}$ to 10$^{-2}$ \citep{malhotra2001}. More recent studies of $\epsilon_{\mathrm{PE}}$ using \textit{Herschel} observations have yielded similar ranges; NGC 1097 and NGC 4559 have heating efficiencies ranging from $\sim$2$\times$10$^{-3}$ to 10$^{-2}$ \citep{croxall2012}, and the \hii \ region LMC-N11B exhibits $\epsilon_{\mathrm{PE}}$ from $\sim$1$\times$10$^{-3}$ to 8$\times$10$^{-3}$ \citep{lebouteiller2012}. In the different regions of M51 defined by \citet{parkin2013}, the nucleus and central regions typically have \ciioitir \ ratios of $\sim$3$\times$10$^{-3}$ to 5$\times$10$^{-3}$, whereas the spiral arm and interarm regions show a broader range of values up to $\sim$10$^{-2}$. Interestingly, we observe similar behaviour of $\epsilon_{\mathrm{PE}}$ in the different regions of NGC~891 (Fig.~\ref{fig:ciimapsdiags}), which has a mean \ciioitir \ of $3.5\times$10$^{-3}$ in the centre, $5\times$10$^{-3}$ in the plane of the disc and $9\times$10$^{-3}$ towards higher vertical distances from the mid-plane. 

There is mounting evidence that the photoelectric heating efficiency correlates with the far-infrared colour, observed as a decrease in \ciioitir \ with increasing FIR colours, such as \textit{IRAS} 60~$\mu$m/100~$\mu$m or \textit{Herschel} 70~$\mu$m/160~$\mu$m colours (\citealp{malhotra2001}; \citealp{croxall2012}; \citealp{parkin2013}). One interpretation of this result is that warmer dust becomes more positively charged in stronger FUV radiation fields, lowering the efficiency of the photoelectric effect. However, \citet{croxall2012} and \citet{lebouteiller2012} report even tighter correlations between the heating efficiency traced by the PAH emission and the FIR colour, which suggests that PAHs rather than dust grains dominate the gas heating. Yet in M51, the warmer dust showed a stronger decrease in heating efficiency when traced by \ciioitir \ than with the \ciioipah \ ratio \citep{parkin2013}. 

\begin{figure}
\begin{center}
\includegraphics[width=0.99\columnwidth]{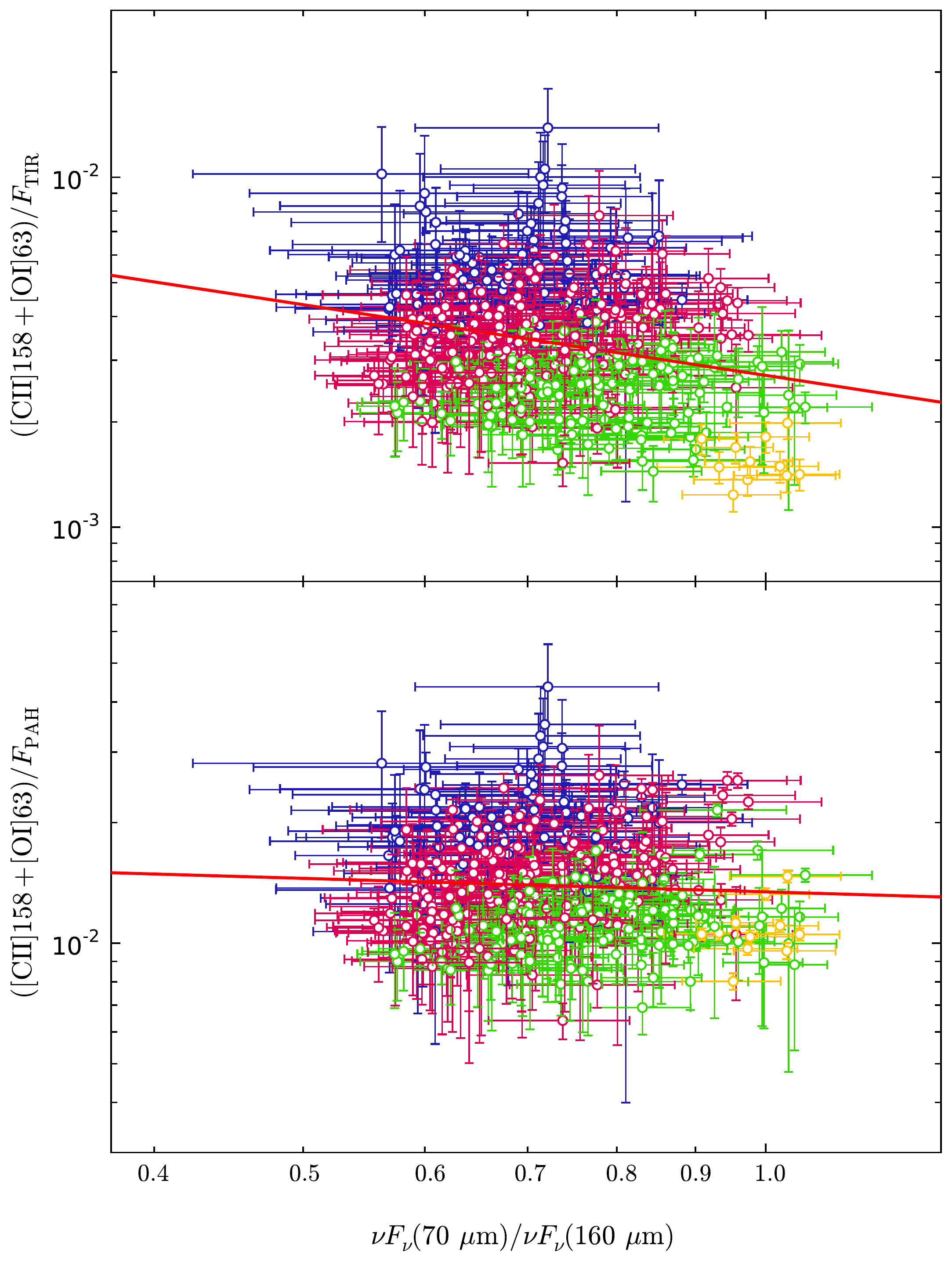}
\end{center}
\vspace{-0.3cm}
\caption[Comparison of the TIR versus PAH as a function of FIR colour]{A comparison of the \ciioitir \ and \ciioipah \ as a function of the $\nu F_{\nu}(70~\mu\mathrm{m})/\nu F_{\nu}(160~\mu\mathrm{m})$ FIR colour. The \cii \ emission has been corrected to remove the contribution originating from the ionised gas (see Sec.~\ref{sec:ionisedcontrib}). Each point represents a pixel, where pixels are coloured as depicted in Fig.~\ref{fig:tirmapmask}, according to their location in either the nucleus (yellow), mid-plane (green), or at increasing radial distances across in the disc (red to blue). The red solid lines represent the best linear fits to the data.}\label{fig:tirpahcomp}
\end{figure}

In Fig.~\ref{fig:tirpahcomp}, we investigate the \ciioitir \ and the \ciioipah \ ratios as a function of the  \textit{Herschel} $\nu F_{\nu}(70~\mu\mathrm{m})/\nu F_{\nu}(160~\mu\mathrm{m})$ FIR colour. As in the previous studies mentioned above, we find a decrease in \ciioitir \ of a factor $\sim$2 with increasing FIR colour, although the anti-correlation is very weak (with a Pearson correlation coefficient of -0.3). Our crude division of the galaxy into various regions based on the \tir \ emission indicates that this decrease in \ciioitir \ with increasing FIR colour corresponds to a decrease in the heating efficiency in the nucleus and inner plane regions than compared to regions at higher radial and vertical distances along the disc. The \ciioipah \ ratio varies across the galaxy from approximately 0.008 to 0.04, on average higher than the value of 0.01 found in M51 by \citealp{parkin2013}) yet less than the \ciioipah \ ratios found in the cases of NGC 1097 and NGC 4559 (0.03--0.06, \citealp{croxall2012}), and the LMC-N11B complex (0.07, \citealp{lebouteiller2012}). However, comparisons of the total PAH intensity estimated from IRAC 8~$\mu$m maps to the total PAH intensity derived from spectra from the \textit{Spitzer} Infrared Spectrograph via PAHfit \citep{smith2007} have demonstrated that the former method overestimates the total PAH intensity by 10\% \citep{croxall2012} up to 70\% \citep{parkin2013}. Applying such corrections to our PAH intensity therefore increases our \ciioipah \ ratio values to roughly coincide with those of \citet{croxall2012} and \citet[][see also \citealp{beirao2010}]{lebouteiller2012}. In contrast to the \ciioitir \ ratio, there is less variation with increasing FIR colour. This result may suggest that in the central regions the gas heating becomes dominated by PAHs rather than dust grains. However, the true role of the PAHs in the gas heating is still unclear. Given the plethora of studies indicating that star forming regions tend to destroy PAHs (e.g. \citealp{helou2004}; \citealp{calzetti2005,calzetti2007}; \citealp{lebouteiller2007}; \citealp{bendo2008}; \citealp{gordon2008}) and considering that NGC 891's FIR colours appear related to its star forming regions \citep{hughes2014}, PAH emission should be inhibited in locations with warmer colour temperatures. Thus, destruction of PAHs in star forming discs could affect the shape and interpretation of the \ciioipah \ - FIR colour relationship. 

Finally, we examine the gas cooling via our \cii/\oi 63 line ratio map (see Fig.~\ref{fig:ciimapsdiags}). Whereas the \cii \ line is more efficient at cooling PDRs at lower densities and cooler temperatures, the \oi \ 63~$\mu$m line is the predominant coolant mechanism of gas at higher densities and warmer temperatures \citep{tielens1985}. Focussing first on the central region, the ratio is higher than the rest of the observed region by a factor of $\sim\,2 - 3$, indicating the \oi \ 63~$\mu$m line is relatively weaker. This behaviour is contrary to the case of M51, where the ratio is lower towards the center, thus corresponding to a stronger \oi \ 63~$\mu$m emission \citep{parkin2013}. Our stronger central ratio is probably due to either the \oi \ 63~$\mu$m line becoming more optically thick towards the center (see Sec.\ref{sec:opticaldepth}), an apparent increase in the \cii \ emission due to conflating central and disc emission along the line-of-sight, or, most likely, a combination of these two effects. Along the northeastern disc of NGC 891, the ratio is always greater than 1, with several peaks in the ratio to upwards of 5 that are spatially coincident with lower values of the photoelectric heating efficiency (Fig.~\ref{fig:ciimapsdiags}, middle panel). The ratio is typically lower at higher altitudes from the disc, possibly indicating the \oi \ 63~$\mu$m line gains importance for gas cooling.

\section{PDR modelling}\label{sec:pdrmodelling}

We compare our observed line ratios to the PDR model of \citet{kaufman1999,kaufman2006}, based on the original model by \citet*{tielens1985}, which models PDR regions as homogeneous infinite plane slabs of hydrogen and characterised via two free parameters: the hydrogen nuclei density, $n$, and the strength of the incident FUV radiation field, $G_{0}$, normalised to the Habing field ($G_{0} = 1.6\times 10^{-6}$~W~m$^{-2}$; \citealp{habing1968}). In the model, the gas is collisionally heated via the ejection of photoelectrons from dust grains and PAH molecules by FUV photons. The FIR fine-structure line emission responsible for cooling the gas is predicted by simultaneously solving the chemical and energy equillibrium in the slab. The models cover a density range of $10^{1}\, \le n \le 10^{7}\,\mathrm{cm}^{-3}$ and a FUV radiation field range of $10^{-0.5} \le G_{0} \le 10^{6.5}$. For a given a set of observations of spectral line intensities, the corresponding best-fit $G_{0}$ and $n$ values from the PDR model are available online\footnote{The PDR Toolbox is available online at \url{http://dustem.astro.umd.edu}} via the `Photo Dissociation Region Toolbox' \citep*[PDRT,][]{pound2008}. We perform our comparison between our \cii , \oi ~63~$\mu$m and \tir \ observations and the PDR model on a pixel-by-pixel basis, but note that the pixel scale of 4$\arcsec$ in our maps means each pixel is not independent from its neighbours. 

\begin{figure*}
\begin{center}
\vspace{1.5cm}
\includegraphics[width=0.99\columnwidth]{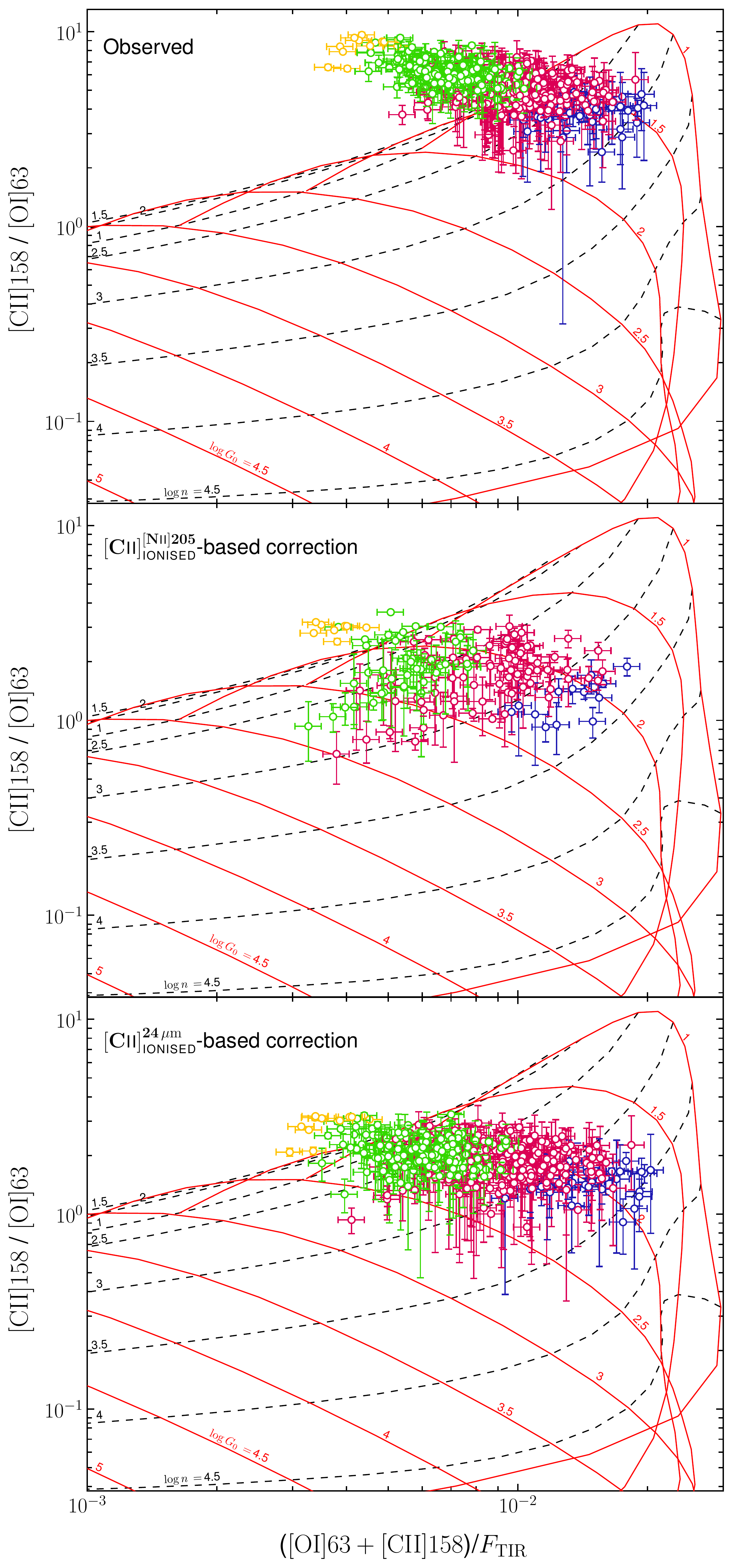}
\includegraphics[width=0.99\columnwidth]{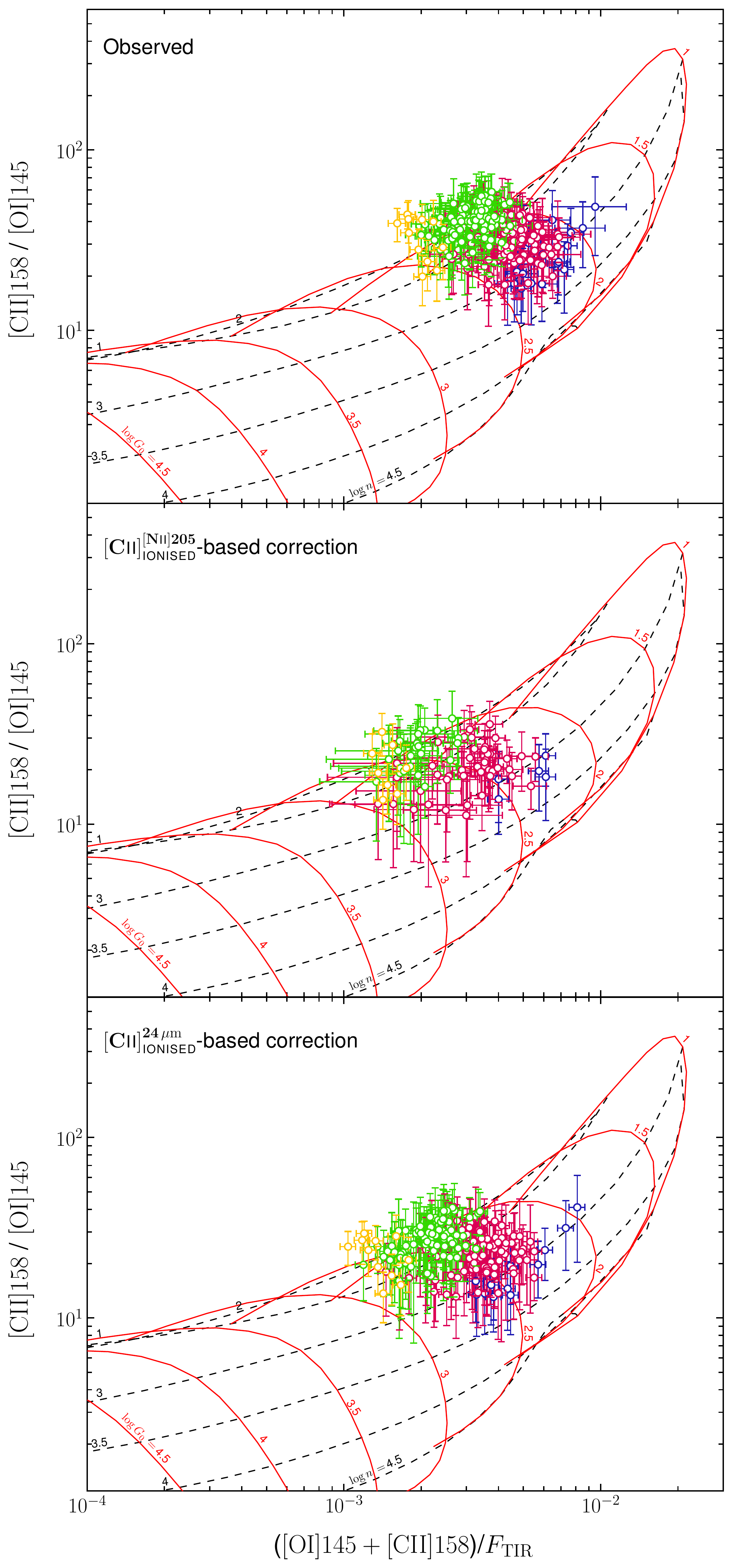}
\end{center}
\caption[Diagnostic diagrams]{Diagnostic diagrams of the \ciioi \ ratio plotted against the \ciioitir \ ratio (\textit{left panels}) and the \ciioialt \ versus \ciioialttir \ ratio (\textit{right panels}) for NGC 891. We superimpose our adjusted observations onto a grid of constant hydrogen nuclei density, log $n$ (black dashed lines), and FUV radiation field strength, log $G_{0}$ (red solid lines), determined from the \citet{kaufman1999,kaufman2006} PDR model. Each data point represents one pixel, with colours as described in Fig.~\ref{fig:tirpahcomp}. We present our unadjusted observations (\textit{upper panels}) and the observations including the adjustments applied to the \cii , \oi 63 and \tir \ emission as described in Sec.~\ref{sec:adjustments}. We compare our two approaches to estimate (and remove) the fraction of the \cii \ emission arising from ionised gas, whereby one method uses the reliable measurements of the \nii~205~$\mu$m line emission (\textit{middle panels}) and the alternative method uses the 24~$\mu$m emission as a proxy for the \nii~205~$\mu$m line emission (\textit{lower panels}) via the correlation presented in Fig.~\ref{fig:24to205correlation}. Error bars do not account for the uncertainties in these corrections to the \cii \ emission.}\label{fig:pdrdiag}
\end{figure*}

We first compare the observed \ciioi \ ratio versus the \ciioitir \ ratio for NGC 891 superimposed on the PDR model grid lines of constant log ($n/\mathrm{cm}^{-3}$) and log $G_{0}$ from the \citet{kaufman1999} diagnostic plots, as presented in Fig.~\ref{fig:pdrdiag}. We note that the parameter space formed via these two diagnostic ratios yields two possible model solutions - a high-$n$ ($\sim10^{3.5}$--$10^{4.25}$), low-$G_{0}$ ($\sim10^{0}$--$10^{0.75}$) regime and a moderate regime. For a face-on galaxy like M51, \citet{parkin2013} could eliminate one of the high-density solutions, following the reasoning of \citet{kramer2005}, by considering the number of clouds emitting within the beam. In the case of NGC~891, we estimate we would require several thousand PDR regions in our 17$\arcsec$ beam to reconcile our observed \cii \ emission with the \cii \ emission predicted by the model using the corresponding values of $n$ and $G_{0}$, which is not completely unrealistic given that our beam could contain between 60 and 75 thousand giant molecular clouds, assuming clouds with 50 pc diameters integrated along a $\sim$8--10 kpc line-of-sight through this edge-on galaxy. Thus, though we focus great part of the discussion on the moderate $n$ and $G_{0}$ solutions, we cannot entirely eliminate the high-$n$, low-$G_{0}$ solutions. It is clear from Fig.~\ref{fig:pdrdiag} (upper left panel) that the PDR model does not represent the observed quantities, as half of all pixels fall outside of the theoretical parameter space. In the next section, we describe the adjustments to our observations to facilitate a proper comparison with the PDR model.

\subsection{Adjustments to observed quantities}\label{sec:adjustments}

A proper comparison to the PDR model of \citet{kaufman1999,kaufman2006} requires us to make three adjustments to our observed quantities, for which we initially follow the strategy of \citet{parkin2013,parkin2014}. Firstly, the observed total infrared flux from extragalactic sources must be reduced by a factor of two in order to account for the optically thin infrared continuum flux emitting not just towards the observer but from both sides of the PDR slab. The model assumes the \tir \ and fine-structure line emission originates purely from the front side of the cloud. We apply this correction to the \tir \ emission for the entire map, which is equivalent to the bolometric far-infrared flux of the PDR model (see \citealp{kaufman1999}). Possible contamination arising from ionised gas remains the main uncertainty in the TIR emission. 

Our second adjustment is to remove the fraction of \cii \ emission arising from ionized gas, as the  \citet{kaufman1999,kaufman2006} PDR model only considers the contribution to the  \cii \ emission that originates from the neutral gas. We achieve this by \clearpage
\clearpage
\noindent
using the maps of the fractional contribution to the \cii \ emission from ionised gas (see Fig.~\ref{fig:iongasmap}), as discussed in Sec.~\ref{sec:ionisedcontrib}, to correct our \cii \ map. Both corrections to the \cii \ emission, derived from the observed \nii~205~$\mu$m line emission and the line emission predicted via the 24~$\mu$m data, are considered in the following analysis. We refer to these datasets as the `\ciiobsion -based' and `\ciisynion -based' corrections, respectively.

Finally, we must apply a correction to the \oi ~63~$\mu$m map to account for the likely case that the line becomes optically thick in regions of star formation much faster than the \cii \ line or the total infrared flux (\citealp{stacey1983}; \citealp*{tielens1985}). The PDR infinite plane slab experiences an incident radiation field from one side, but for extragalactic sources the ensemble of clouds in the PACS beam will not all be orientated with their irradiated side facing towards us. Thus, whilst we may observe all the emission from the optically thin \cii \ line and \tir , we may miss the emission from the more optically thick \oi ~63~$\mu$m line that escapes from those clouds with their irradiated sides orientated away from us. Under the assumption that we only observe about half of the total \oi \ ~63~$\mu$m emission from all PDRs within the PACS beam, \citet{parkin2013,parkin2014} multiply their observed \oi \ ~63~$\mu$m emission by a factor of two. This is a conservative correction; \citet{stacey2010} reason that the observed \oi \ line intensity should be corrected for these geometric issues by as much as a factor of four for high optical depth in a spherical cloud geometry. Furthermore, given that the optical depth of the ~63~$\mu$m emission may strongly vary across NGC~891's disc (Fig.~\ref{fig:oilines_av}), it may therefore be more appropriate to adopt different corrections for each region. We thus initially increase our observed \oi ~63~$\mu$m emission by a factor of two for the entire map and later, in Sec.~\ref{sec:oi145pdr}, investigate the effects of varying the correction factor.

\begin{figure*}
\begin{center}
\includegraphics[width=0.99\textwidth]{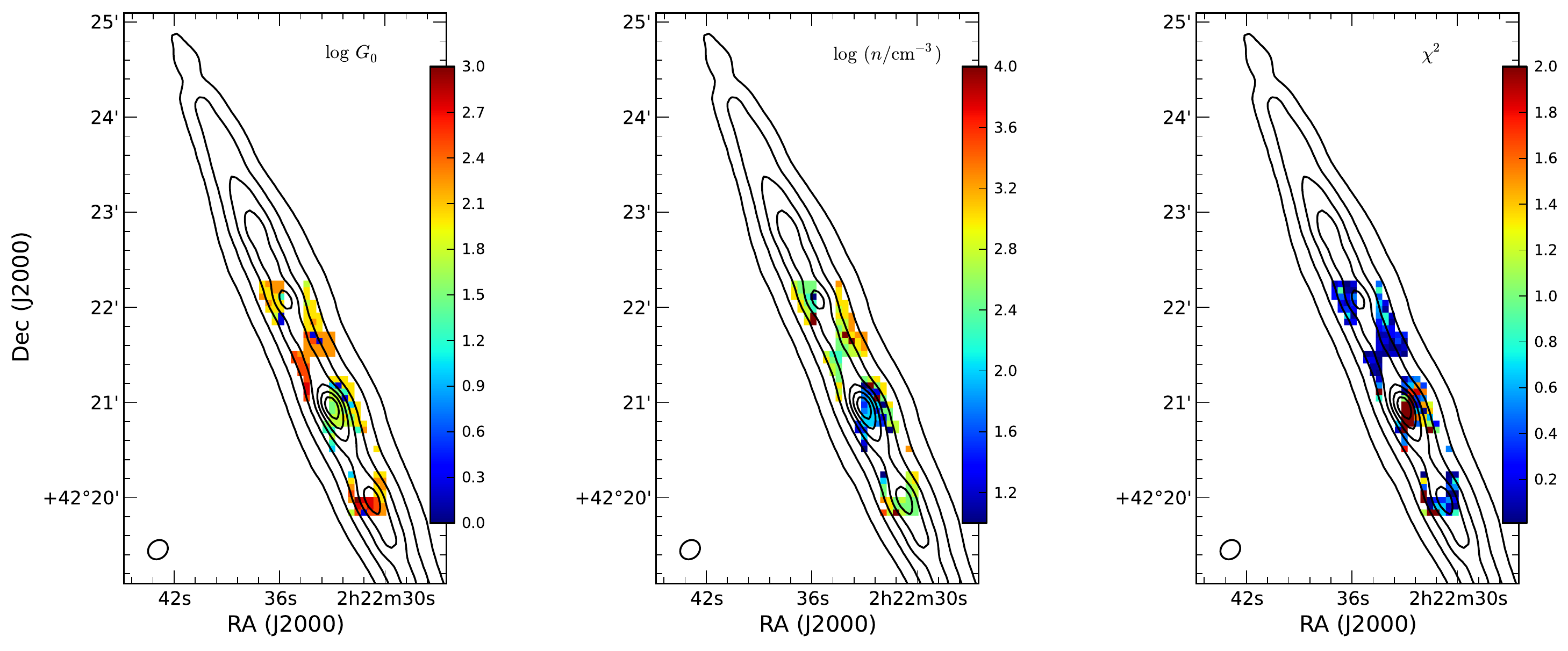}
\includegraphics[width=0.99\textwidth]{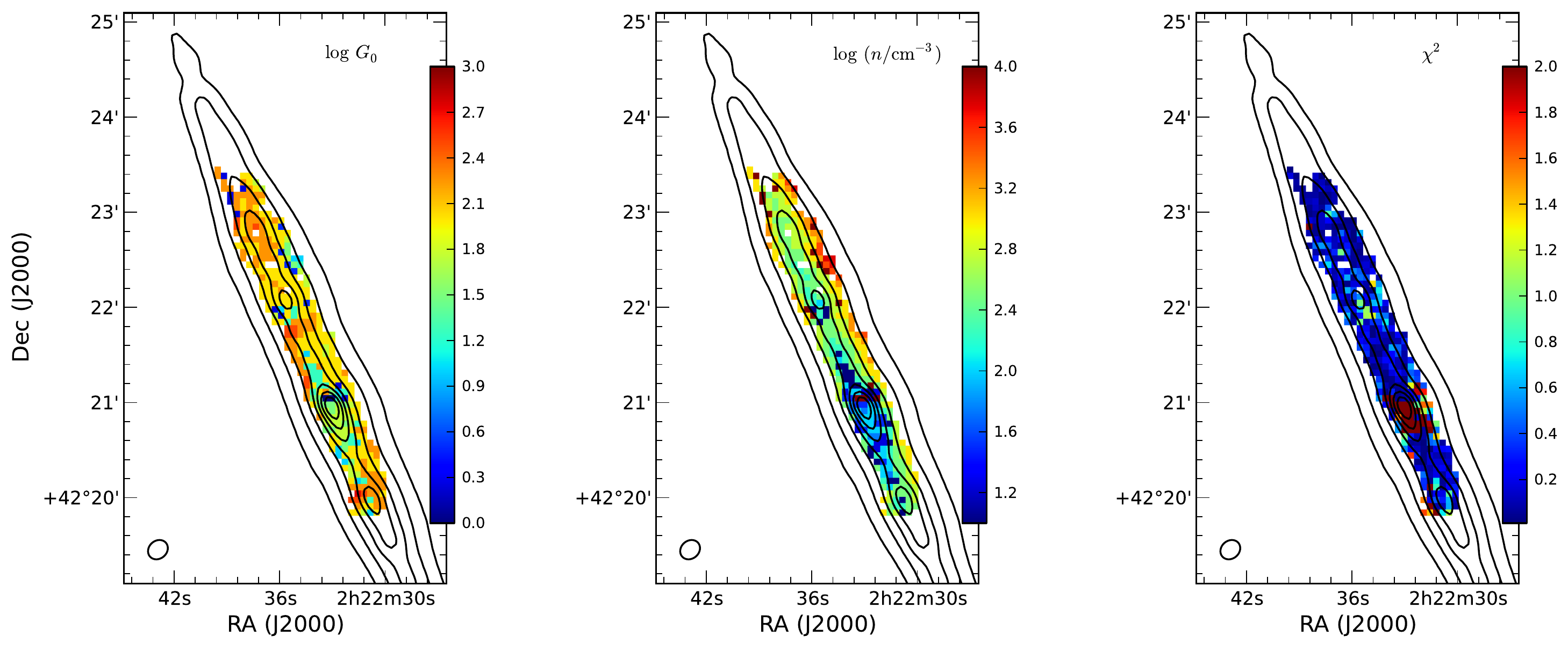}
\end{center}
\vspace{-0.3cm}
\caption[PACS far-infrared spectroscopic maps]{The maps of FUV radiation field strength, $G_{0}$ (\textit{left}), hydrogen nuclei density, $n$ (\textit{middle}), and $\chi^{2}$ (\textit{right}) determined from fitting the adjusted \cii , \oi 63, and \tir \ emission with the \citet{kaufman1999,kaufman2006} PDR model on a pixel-by-pixel basis. We compare the results from the observations adjusted via the \ciiobsion -based (\textit{upper panels}) and \ciisynion -based corrections (\textit{lower panels}) described in Sec.~\ref{sec:ionisedcontrib}. The maps are centred on $\alpha = 2^\mathrm{h}\ 22^\mathrm{m}\ 35\fs7$, $\delta = +42\degr\ 22\arcmin\ 05\farcs9$ (J2000.0) and are presented in the resolution and pixel size of the PACS 160~$\mu$m map. Contours from the $F_{\mathrm{TIR}}$ map (see Fig.~\ref{fig:tirmap}) are superimposed on each image as a visual aid with levels as listed in Fig.~\ref{fig:obsmaps3sig}. North is up, east is to the left.}\label{fig:mapsdiagparams}
\end{figure*}

\subsection{Insights from diagnostic diagrams}

Following these line adjustments, we now return to Fig.~\ref{fig:pdrdiag} to examine their effects on the observations in the \ciioi \ versus \ciioitir \ parameter space. Focussing first on our \ciiobsion -based correction (Fig.~\ref{fig:pdrdiag}, middle panel), we see that there is an expected overall shift of pixels to lower values of both \ciioitir \ and \ciioi \ after applying the appropriate corrections to the \cii , \oi ~63~$\mu$m and TIR emission. The distribution of these ratios on the diagnostic diagram indicates the majority of the disc has a density of hydrogen nuclei in the range of $1 < \log n < 3.5$ cm$^{-3}$, and typically experiences an incident FUV radiation field with a strength varying between $\log G_{0} \approx 2$ to 2.5, though this increases up to $\log G_{0} = 3$ in some regions. We find that the density appears to increase from the interior regions out to the extremities of the disc, whereas the strength of $G_{0}$ is relatively uniform except for a number of peaks along the galaxy mid-plane. However, despite the adjustments to the observations, the pixels covering the nuclear region still fall outside of the \ciioi \ versus \ciioitir \ parameter space defined by the \citet{kaufman1999,kaufman2006} PDR model - the closest contours of constant $\log n$ and $\log G_{0}$ are those of the lowest density and weakest field strength, respectively. We shall shortly return to discuss the various reasons for this behaviour (see Sec.~\ref{sec:oi145pdr}).

\begin{figure}
\begin{center}
\includegraphics[width=0.66\columnwidth]{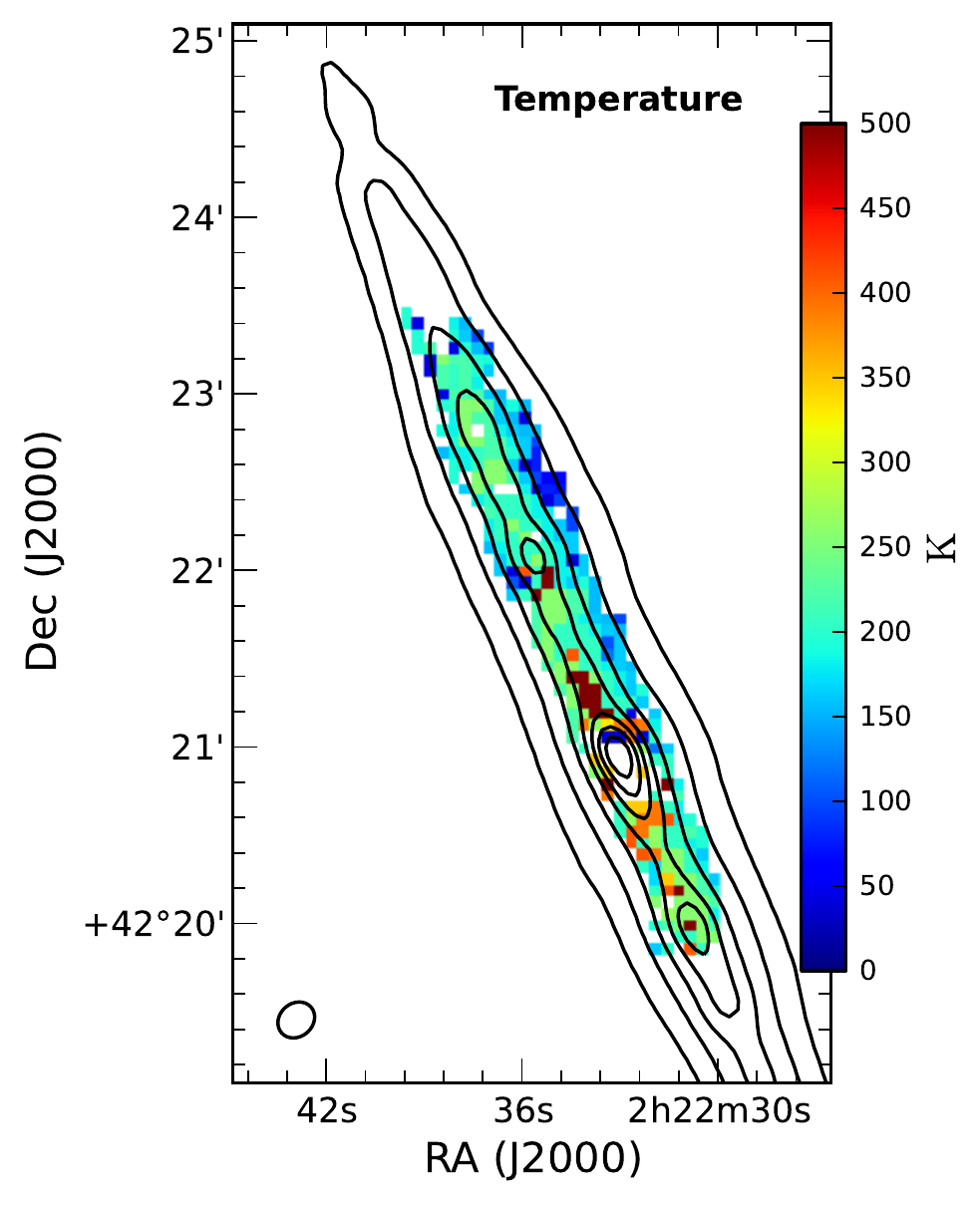}
\end{center}
\vspace{-0.3cm}
\caption[Surface temperature map]{The surface temperature of the atomic gas estimated from the corresponding FUV radiation field strength and hydrogen nuclei density, which were determined from fitting the \cii , \oi 63, and \tir \ emission, adjusted via the \ciiobsion -based correction, with the \citet{kaufman1999,kaufman2006} PDR model. The map centering, orientation, resolution, and contours of TIR emission are presented as in Fig.~\ref{fig:mapsdiagparams}.}\label{fig:temperature}
\end{figure}

We initially chose to constrain the best-fit $G_{0}$ and $n$ values from the PDR model using the \cii , \oi ~63~$\mu$m and TIR emission primarily because the \ciioi \ versus \ciioitir \ parameter space produces robust constraints to the parameters for the largest number of pixels and best facilitates a comparison with the literature. In Sec.~\ref{sec:opticaldepth}, however, we used a map of the \oi 145/\oi 63 line ratio to check where variations in the optical depth effects may become important (see Fig.~\ref{fig:mapsdiags}), finding that whilst most of the disc comprises optically thin neutral gas at temperatures $\sim 100-300$~K, the \oi ~63~$\mu$m line may become completely optical thick in the center. If the central regions indeed suffer from the effects of increasing optical thickness, then this may explain why the (adjusted) observations of the \cii , \oi ~63~$\mu$m and TIR emission are poorly described by the PDR model parameters towards the nucleus. We can test this hypothesis by performing our analysis using the \oi ~145~$\mu$m line map as an alternative observational constraint to the \oi ~63~$\mu$m line map, and comparing the results from the two lines.

In substituting the \oi ~63~$\mu$m emission map with that of the \oi ~145~$\mu$m line map, we correct the \cii \ and TIR emission as previously but do not apply a correction to the \oi ~145~$\mu$m map because, unlike the \oi ~63~$\mu$m line, the line is very optically thin and so we assume no emission will escape unobserved from those PDR clouds with their irradiated sides orientated away from us. We first construct a diagnostic diagram of the \ciioialt \ versus \ciioialttir \ parameter space and compare the observations superimposed on the PDR model grid lines of constant log ($n/\mathrm{cm}^{-3}$) and log $G_{0}$ from the \citet{kaufman1999} diagnostic plots (see right panels of Fig.~\ref{fig:pdrdiag}). The most interesting point to note from this exercise is that the distribution of the pixels, both in the parameter space and in relation to the model grid lines, are qualitatively similar to the results presented in Fig.~\ref{fig:pdrdiag} for all cases, i.e., the unadjusted observations and those adjusted with the \ciiobsion -based and \ciisynion -based corrections. However, unlike in the left panels of Fig.~\ref{fig:pdrdiag}, the central pixels now inhabit a region of the parameter space much closer to the PDR model grid lines such that the errorbars overlap with the parameter space defined by the model, where the closest contours of constant $\log n$ and $\log G_{0}$ remain those of the lowest density and weakest field strength, respectively.

\subsection{Results of model fitting}

For our set of observations of the \cii \ and \oi ~63~$\mu$m line intensities adjusted using the \nii ~205~$\mu$m line, we determine the corresponding best-fit $n$ and $G_{0}$ values from the model on a pixel-by-pixel basis via the online PDRT. We use these values to reconstruct maps of the hydrogen density and the incident FUV radiation field in NGC 891 (see Fig.~\ref{fig:mapsdiagparams}, upper panels). Here, one of the limitations of our analysis is clearly evident, as correcting the \cii ~158~$\mu$m emission for the contribution arising from ionised gas using the lower resolution and sparser coverage of the SPIRE FTS observations (compared to the PACS observations) results in only 140 pixels with adjusted \cii \ line intensities out of the original total of 416 pixels in which we detect both \cii \ and \oi ~63~$\mu$m at the 3$\sigma$ level. Furthermore, the nuclear regions yield low densities and field strengths with $\chi^{2} \gg 4$, indicating that the PDR model is unable to accurately describe the observations from the central pixels. 

Faced with these issues, making any quantitative statement on the variation of the properties of the photon dominated regions across NGC 891's disc becomes difficult. The density of hydrogen nuclei appears to become denser at greater radial distances and with increasing vertical height from the plane of the disc. In contrast, the average strength of $G_{0}$ is relatively uniform at the two secondary peaks in the TIR emission located on opposite sides of the nucleus. Within 20$\arcsec$ circular apertures centred on these two TIR peaks (traced by the contours in Fig.~\ref{fig:mapsdiagparams}), we find an average $\log G_{0}$ of $1.9 \pm 0.5$ from 38 pixels near the north-eastern TIR peak compared to $2.2 \pm 0.5$ from 29 pixels in the vicinity of the south-western TIR peak. However, with the observational dataset adjusted using the \nii ~205~$\mu$m emission line, we lack the necessary spatial coverage to investigate the region with enhanced \cii \ and \oi \ emission on the far north-eastern side of the disc, a location which, as we previously mentioned, exhibits higher luminosities at various wavelengths compared to the opposite location on the south-western side (e.g. \ha , \citealp{kamphuis2007}). 

These issues with the lack of spatial coverage and low resolution of the best-fit PDR model parameters may be resolved when using the 24~$\mu$m emission as a proxy for the \nii~205~$\mu$m line emission, enabling the estimation of the ionised gas for the majority ($\sim$91\%) of the pixels with \cii \ 158, \nii \ 122 and \oi ~63~$\mu$m measured at the 3$\sigma$ level. We now examine the effects of this \ciisynion -based correction on the distribution of our observations in the \ciioi \ versus \ciioitir \ diagnostic diagram (see Fig.~\ref{fig:pdrdiag}, lower panel). Our adjusted observations occupy very similar areas of the parameter space in diagnostic diagram as above, whereby most of the PDRs in the disc have hydrogen densities between $1 < \log n/\mathrm{cm}^{-3} < 3.5$ and an incident FUV radiation field strength varying between $\log G_{0} \approx 1.7$ to 2.5. The small peaks toward $\log G_{0} = 3$ in some inner regions are less evident. We also find similar behaviour in the variation of $n$ and $G_{0}$ across the disc, with $n$ increasing from the interior regions out to the extremities of the disc and $G_{0}$ typically following the $\log G_{0} =2$ contour.  

As in the observations adjusted with the \ciiobsion -based correction, pixels covering the nuclear region fall outside of the \ciioi \ versus \ciioitir \ parameter space defined by the \citet{kaufman1999,kaufman2006} PDR model. Thus, estimating the \nii \ 205~$\mu$m emission from the 24~$\mu$m flux density appears to reconstruct the same trends in the diagnostic diagrams as found when using the observed \nii \ 205~$\mu$m emission. Although this approach does not match the scatter in the distribution, this is expected when using such a best-fit linear relationship (Fig.~\ref{fig:ionisedgasfraction}) to estimate the \nii \ 205~$\mu$m emission. Within the adopted errorbars (see Fig.~\ref{fig:pdrdiag}), which do not account for the 30\% flux calibration errors nor the 50\% error assumed in our estimation of the ionised gas contribution to the \cii \ 158~$\mu$m emission, the two sets of results are consistent. 

In the lower panels of Fig.~\ref{fig:mapsdiagparams}, we map the best-fitting $n$ and $G_{0}$ values from the PDR model determined from our observations adjusted with the \ciisynion -based correction. Our results not only reproduce the \ciiobsion -based results, but also extend our estimate of $n$ and $G_{0}$ for great part of the northern section of NGC 891's disc. In addition to the aforementioned variations of $n$ and $G_{0}$ across the disc, i.e. where $n$ increases from the interior regions out to the extremities of the disc and the FUV field strength typically varies around $G_{0} \approx$ 10$^{2}$, we also find that the FUV radiation field in the far north-eastern side of the disc is on average slightly stronger ($\log G_{0} \sim 2.4$) than the rest of the disc albeit with a moderate hydrogen density ($\log n/\mathrm{cm}^{-3} \sim 2.5$). Furthermore, our estimate of the surface temperatures of the atomic gas, $T$, predicted from the best-fit $n$ and $G_{0}$ values of the PDR model (see Fig. 1 in \citealp{kaufman1999}), which, with the exclusion of the nucleus, ranges from $\sim$ 40 to 500 K with a mean of 210 K and standard deviation of 107 K (i.e. in agreement with the empirical results using the \oi 145/\oi 63 ratio), suggests that the gas surface temperature in this region ($\sim$ 230-260 K) is slightly warmer than the average gas temperature across the disc (see Fig.~\ref{fig:temperature}). 

\begin{figure}
\begin{center}
\includegraphics[width=0.99\columnwidth]{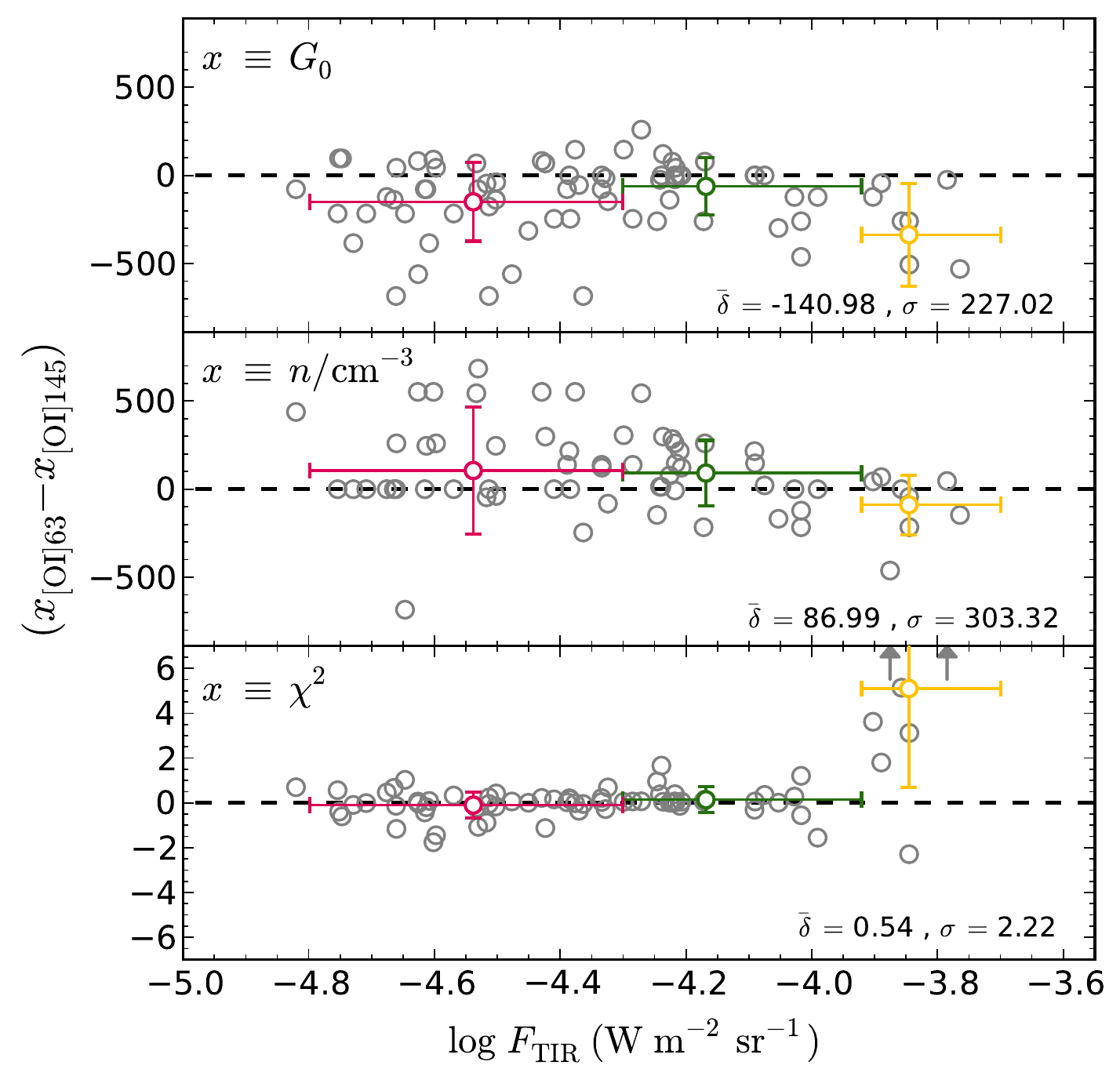}
\end{center}
\vspace{-0.3cm}
\caption[Surface temperature map]{A comparison of the FUV radiation field strength, $G_0$ (\textit{upper panel}), hydrogen nuclei density, $n$ (\textit{middle panel}), and the $\chi^{2}$ value (\textit{lower panel}) from fitting the \citet{kaufman1999,kaufman2006} PDR model to the \cii \ and \tir \ emission together with either the \oi ~63 or 145~$\mu$m line emission, adjusted via the \ciiobsion -based correction. The difference between the parameters are plotted against the logarithm of the TIR emission, and the coloured bins correspond to the schematic in Fig.~\ref{fig:tirmapmask}. The quoted values are the overall scatter ($\sigma$) defined as the standard deviation of the difference ($\delta$) between the PDR model parameters constrained with each \oi \ line.}\label{fig:oicomp}
\end{figure}

Finally, we exactly reproduce the analysis as described above, but now substitute the \oi ~63~$\mu$m emission map with that of the \oi ~145~$\mu$m line map. We perform our comparison between our adjusted \cii , \oi ~145~$\mu$m and \tir \ observations and the PDR model on a pixel-by-pixel basis via the PDRT. Again, we test both the \ciiobsion -based and \ciisynion -based corrections, yet focus on the former set of results for this discussion. In Fig.~\ref{fig:oicomp}, we present a comparison of the best-fit $G_{0}$ and $n$ parameters and corresponding $\chi^{2}$ value determined from the \oi ~63 and 145~$\mu$m lines for the 65 pixels covered by both maps (see Fig.~\ref{fig:obsmaps3sig}) and with valid estimates of the \cii \ emission arising from ionised gas. Defining the scatter ($\sigma$) as the standard deviation of the difference ($\delta$) between the parameters constrained with each line, e.g. $n_{\mathrm{[O\textsc{i}]} 63} - n_{\mathrm{[O\textsc{i}]} 145}$, we find there is significant scatter in both the $G_{0}$ and $n$ distribution. This large scatter is somewhat expected, considering the uncertainties on the observations, the various correction factors, and the model solutions, e.g. in some regions of the \ciioialt \ versus \ciioialttir \ parameter space, there are overlaps between several different solutions yielding $G_{0}$ and $n$ values spanning several orders of magnitude. However, the main result is that although the analysis with the \oi ~145~$\mu$m line frequently yields lower $G_0$ and higher $n$ values, it reproduces the overall trends in the variation of $G_{0}$ and $n$ with increasing radial and vertical differences found with the \oi ~63~$\mu$m line, particularly evident when we examine the different regional bins (see the coloured circles in Fig.~\ref{fig:oicomp}). Despite the consistencies between the two sets of results, the central pixels are typically better fit when we constrain the PDR model parameters with the \oi ~145~$\mu$m line compared to the 63~$\mu$m line, as indicated by the $\chi^{2}$ values (Fig.~\ref{fig:oicomp}).

\subsection{\cii \ column density and optical depth}

For each set of maps of best-fitting $n$, $G_0$ and $T$ results considered above, i.e. from the \ciiobsion \ and \ciisynion \ corrections, we can estimate the hydrogen nuclei column density associated with the \cii \ emission, $N_{\mathrm{C}^{+}}(\mathrm{H})$, and the corresponding optical depth, $\tau_{\mathrm{[CII]}}$. Following \citet{contursi2013}, we use the equation of \citet{crawford1985} that states
\begin{equation}\label{eqn:coldens}
N_{\mathrm{C}^{+}}(\mathrm{H})=\frac{4.25\times10^{20}}{\chi(C)}\left[\frac{1+2\mathrm{e}^{(-92/T)}+(n_{\mathrm{crit}}/n)}{2\mathrm{e}^{(-92/T)}}\right]\left(\frac{I_{\mathrm{[CII]}}}{\Phi_{\mathrm{[CII]}}}\right)
\end{equation}
where $N_{\mathrm{C}^{+}}(\mathrm{H})$ is in cm$^{-2}$, $\chi(C)$ is the gas-phase [C$^+$]/[H] abundance ratio of $1.4\times10^{-4}$, $n_\mathrm{crit}$ is the critical density for collisions of C$^{+}$ and H nuclei equal to $4\times10^{3}$ cm$^{-3}$, and $I_{\mathrm{[CII]}}$ is the \cii \ intensity from the PDR in units of erg s$^{-1}$ cm$^{-2}$ sr$^{-1}$. Finally, $\Phi_{\mathrm{[CII]}}$ is the fraction of the beam filled with \cii - emitting clouds, the beam area filling factor, estimated from the ratio of the \cii \ emission corresponding to the best fitting $G_{0}$ and $n$ values to the observed \cii \ emission. We find $\Phi_{\mathrm{[CII]}}$ ranges from 0.1 to 0.2 in the center, 0.6 to 1.5 all along the disc, and increases up to 3 at high vertical distances above the plane. To avoid circular reasoning, we make the assumption that $\Phi_{\mathrm{[CII]}}=1$, given the numerous clouds likely falling along the line-of-sight in NGC 891. Combining all our results from the PDR modelling, we derive $\mathrm{C}^{+}$ column densities ranging from $\sim1\times10^{20}$ to $6\times10^{22}$ cm$^{-2}$, where the values higher than the mean of $\sim10^{21}$ cm$^{-2}$ are primarily found in the galaxy center. We can relate the column density to the optical depth via
\begin{multline}\label{eqn:tau}
\tau_{\mathrm{[CII]}}=\frac{\lambda^{3}A_{ul}}{8\pi \Delta v}\left[\left(1+\frac{n_{\mathrm{crit}}}{n}\right)\mathrm{e}^{(92/T)}-1\right] \\ \times\left(\frac{2\mathrm{e}^{(-92/T)}}{1+2\mathrm{e}^{(-92/T)}+(n_{\mathrm{crit}}/n)}\right)N_{\mathrm{C}^{+}}(\mathrm{H})
\end{multline}
where $\Delta v$ is the line velocity width in units of 5 km s$^{-1}$ \citep{crawford1985}. Adopting the average velocity width from the \cii \ line fits of 200 km s$^{-1}$, an optical depth of $\tau_{\mathrm{[CII]}} = 1$ is reached for column densities of $N_{\mathrm{C}^{+}}(\mathrm{H})\approx 6\times10^{19}$~cm$^{-2}$. We find $\tau_{\mathrm{[CII]}}$ is of order unity for most of the disc. However, even considering the higher limit in column density required to reach $\tau_{\mathrm{[CII]}} = 1$ predicted by \citet*{tielens1985}, $N_{\mathrm{C}^{+}}(\mathrm{H})|_{\tau_{\mathrm{[CII]}}=1}$ equal to $1.2\times10^{21}$~cm$^{-2}$, the high central column densities and corresponding opacities indicate we cannot entirely rule out the effects of optical depth in the \cii \ line. Additionally, by considering the above equation for the case of the \oi ~63~$\mu$m line, substituting a gas-phase [O]/[H] abundance ratio of $3\times10^{-4}$ and a critical density for collisions of O and H atoms equal to $4.7\times10^{5}$ into Eq.~\ref{eqn:coldens}, and further assuming the O and C$^{+}$ are coexistent in the gas phase such that $N_{\mathrm{O}}(\mathrm{H})=2N_{\mathrm{C}^{+}}(\mathrm{H})$ (), we estimate that $\tau_{\mathrm{[OI]}}/\tau_{\mathrm{[CII]}}$ ranges from 2.5 to 8.0 for the range of PDR densities ($1 < \log n/\mathrm{cm}^{-3} < 3.5$) and temperatures ($\sim 40< T < 500$ K) seen in NGC 891's disc. In other words, whilst $\tau_{\mathrm{[OI]}}$ is likely $\gg$10 in the center, the optical depths of the two lines are expected to be of order unity in the rest of the galaxy.

\subsection{Attempts to counter \oi \ optical depth effects}\label{sec:oi145pdr}

We initially chose to constrain the best-fit $G_{0}$ and $n$ values from the PDR model using the \cii , \oi ~63~$\mu$m and TIR emission primarily because the \ciioi \ versus \ciioitir \ parameter space produces robust constraints to the parameters for the largest number of pixels and best facilitates a comparison with the literature. However, the standard corrections we applied to these observable quantities (Sec.~\ref{sec:adjustments}) yield a disconcerting, counter-intuitive picture of a disc with $G_{0}$ remaining fairly constant whilst $n$ increases in the radial and vertical directions, trends which are not seen in similar studies of resolved galaxies (e.g. \citealp{lebouteiller2012}; \citealp{croxall2012}; \citealp{parkin2013}) and also appear contrary to some of the results of our empirical analysis (Sec.~\ref{sec:results}). For example, the increase of the $\nu F_{\nu}(70~\mu\mathrm{m})/\nu F_{\nu}(160~\mu\mathrm{m})$ FIR colour towards the center (Fig.~\ref{fig:tirpahcomp}) would suggest a stronger FUV field or higher density of the ISM in these regions than what we find from the PDR modelling. Given this is the first time this type of analysis is applied to an edge-on galaxy, we suspect these trends may arise from regional variations in the optical depth of the fine-structure lines not accounted for in our corrections.

The preferential method for investigating the effects of optical depth variations would be to constrain the optical depth of the \oi \ 63 line in each region/pixel, perhaps from an estimate of the extinction from the observed \oi 63/\oi 145 ratio (via the relations in e.g. Fig.~\ref{fig:oilines_av}), and use this to correct the observed line intensities. In practice, however, such a correction becomes highly speculative considering the uncertainties in the ill-constrained relationships between the \oi 63/\oi 145 ratio and extinction due to PDRs, the extinction and the optical depth, and the PDR geometry, etc. For example, the derived $A_V$ values are average values at the resolution of the images, but may be much higher on smaller physical scales, particularly in the dense cores from where the \oi ~emission originates. Even if we adopt the simplest possible assumptions, an accurate interpretation is not straightforward. Recognising these limitations, we instead take a more conservative approach and increase the \oi ~63~$\mu$m intensity by a different factor for each region defined in Fig.~\ref{fig:tirmapmask}; we adopt a factor of four (e.g. \citealp{stacey2010}) in the center and mid-plane regions (yellow and green pixels) where the optical depth is likely highest, and a factor of two in the outer regions (red and blue pixels), as the pixels towards the edge display on average the lowest $A_V$ and \oi 63/\oi 145 ratios suggesting lower optical depths.

\begin{figure}
\begin{center}
\includegraphics[width=0.99\columnwidth]{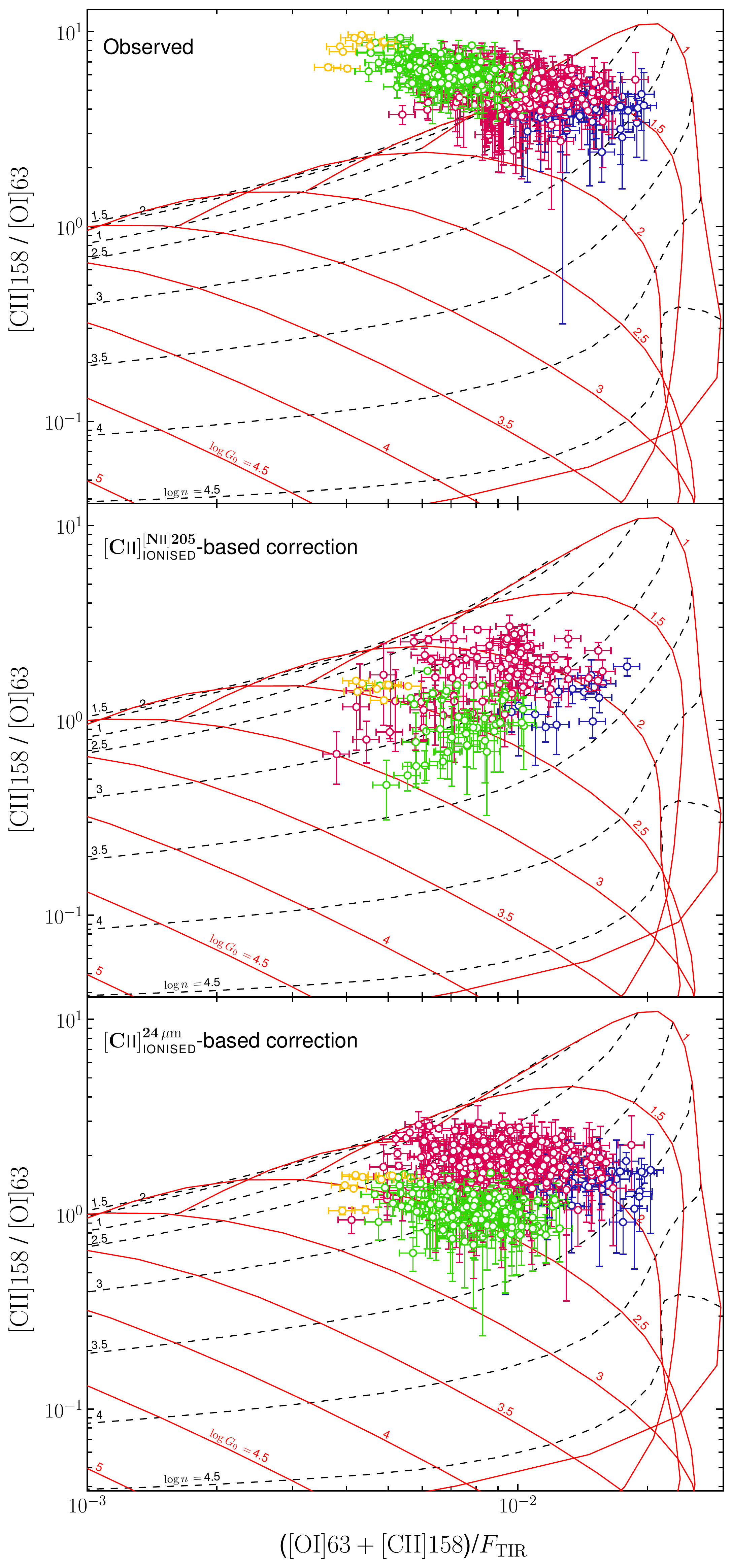}
\end{center}
\vspace{-0.4cm}
\caption[Diagnostic diagrams]{Diagnostic diagrams of the \ciioi \ ratio plotted against the \ciioitir ratio for NGC 891, similar to Fig.~\ref{fig:pdrdiag}. We superimpose our adjusted observations onto a grid of constant hydrogen nuclei density, log $n$ (black dashed lines), and FUV radiation field strength, log $G_{0}$ (red solid lines), determined from the PDR model of \citet{kaufman1999,kaufman2006}. Each data point represents one pixel, with colours as described in Fig.~\ref{fig:tirpahcomp}. We present our unadjusted observations (\textit{upper panel}) and the observations including the adjustments applied to the \cii \ and \tir \ emission as described in Sec.~\ref{sec:adjustments}, and compare our two approaches to remove the fraction of the \cii \ emission arising from ionised gas: the \ciiobsion -based (\textit{middle panel}) and \ciisynion -based corrections (\textit{lower panel}). See Fig.~\ref{fig:mapsdiagparams_av} for the corresponding maps.}\label{fig:pdrdiagav}
\end{figure}

\begin{figure*}
\begin{center}
\includegraphics[width=0.99\textwidth]{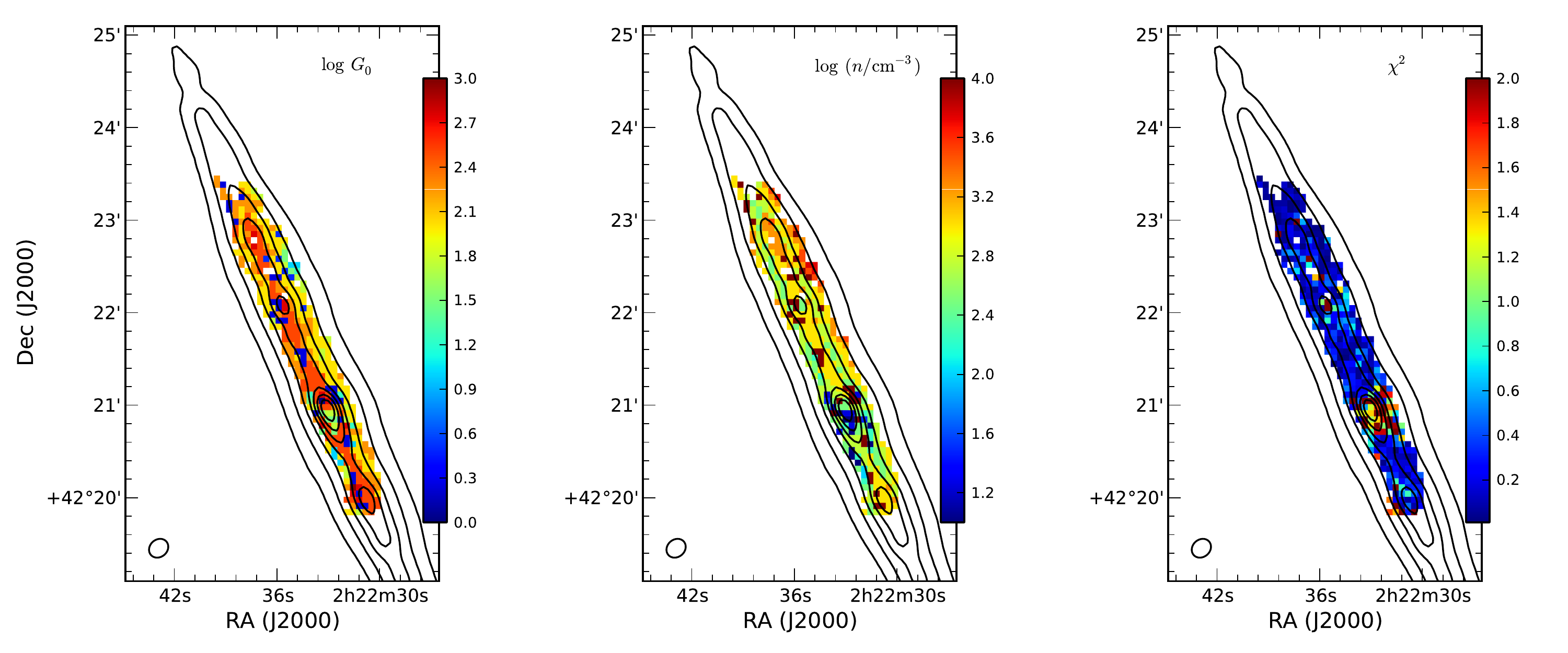}
\end{center}
\vspace{-0.3cm}
\caption[PACS far-infrared spectroscopic maps]{The maps of FUV radiation field strength, $G_{0}$ (\textit{left}), hydrogen nuclei density, $n$ (\textit{middle}), and $\chi^{2}$ (\textit{right}) determined from fitting the adjusted \cii , \oi ~63~$\mu$m, and \tir \ emission with the \citet{kaufman1999,kaufman2006} PDR model on a pixel-by-pixel basis. Here, the \oi~63~$\mu$m emission has been corrected using a varying factor to attempt to account for optical depth effects, as described in Sec.~\ref{sec:oi145pdr}. The maps are centred on $\alpha = 2^\mathrm{h}\ 22^\mathrm{m}\ 35\fs7$, $\delta = +42\degr\ 22\arcmin\ 05\farcs9$ (J2000.0) and are presented in the resolution and pixel size of the PACS 160~$\mu$m map. Contours from the $F_{\mathrm{TIR}}$ map (see Fig.~\ref{fig:tirmap}) are superimposed on each image as a visual aid with levels as listed in Fig.~\ref{fig:obsmaps3sig}.}\label{fig:mapsdiagparams_av}
\end{figure*}

Applying these corrections, we construct a diagnostic diagram of the \ciioi \ versus \ciioitir \ parameter space and compare the adjusted observations to the PDR model grid lines of constant log ($n/\mathrm{cm}^{-3}$) and log $G_{0}$ from the \citet{kaufman1999} diagnostic plots (see Fig.~\ref{fig:pdrdiagav}). The overall range of the parameter space inhabited by the pixels remains fairly unchanged, i.e. most of the PDRs in the disc have hydrogen densities between $1 < \log n/\mathrm{cm}^{-3} < 3.5$ and an incident FUV radiation field strength varying between $\log G_{0} \approx 1.7$ to 3. However, the trends seen in the previous analysis are almost reversed: central pixels now lie in a region of the parameter space corresponding to the PDR model grid lines of the strongest $\log G_{0}$, whereas pixels towards the outskirts of the disc now exhibit the weakest field strength. Fitting these observations with the PDR model on a pixel-by-pixel basis, again via the PDRT, confirms that the center and mid-plane regions typically have $2.5 < \log G_0 < 3.5$ compared to the FUV field strengths of $1.5 < \log G_0 < 2$ at the edge of the disc, whereas the density is typically $2.5 < \log n/\mathrm{cm}^{-3} < 3.5$ all along the disc (see Fig.~\ref{fig:mapsdiagparams_av}). However, the trends in the median values of each region remain unchanged (c.f. Table~\ref{tab:pdrprops}). As a sanity check, we find that these best-fitting $G_0$ and $n$ values predict \oi 145/\oi 63 ratios between 0.03 to 0.075 that are consistent with the observed \oi 145/\oi 63 ratios (see Fig.~\ref{fig:mapsdiags}, left panel) corrected by a factor four. Without such a correction, the observed \oi 145/\oi 63 ratios in central and mid-plane pixels fall outside of the parameter space described by the PDR model. The central pixels are still not well fit by the model ($\chi^{2} \gg 4$), despite increasing the \oi 63 correction factor to account for the higher optical depth in the center. There also appears to be more scatter in the parameter maps from pixels with very high-$n$, low-$G_0$ solutions. Whilst the origin of this scatter is unclear, it is possible we are now either under- or over-correcting the \cii \ and \oi \ emission in these regions.

Before we discuss these results, it is important to stress that whilst the adoption of a varying \oi \ correction factor has some clear physical motivation, our choice in the factors to apply to each region/pixel are very poorly constrained and more work is required to develop better corrections for the optical depth. We also note the fact that the counter-intuitive trend of increasing gas density towards the outer regions of the galaxy is not only seen when using the \oi ~63~$\mu$m line as a gas diagnostic, but also in the \oi ~145~$\mu$m line (although to a lesser extent, c.f. Fig.~\ref{fig:pdrdiag}). Although this seems to indicate that both lines are affected by optical depth effects in some regions of the galaxy, there also remains the possibility that different beam filling factors for the \cii \ and \oi ~lines could play a role. Given that most of the young star-forming regions are found in a disk with scale height 60-80 pc (\citealp{schechtmanrook2013}), one might envisage the dense PDRs in the same thin disk. Yet, moving away from the mid-plane, \cii ~emission may become increasingly associated with more diffuse \hi ~clouds, albeit fewer in number, and correcting for this difference in source size would make the \cii /\oi  ~ratios (from both \oi ~ lines) higher for larger distances from the mid-plane. At present, the main result from our analysis is that the trends in the density and FUV field strength across the disc appear to be highly sensitive to the regional variations in the optical depth, implying that care should be taken when applying such an analysis to observations of high inclination systems.

\section{Discussion}\label{sec:discussion}

In this section, we now compare our results to previous studies and discuss how observational errors and issues may affect our conclusions. We briefly note, however, that there are of course uncertainties in the PDR parameters associated with our choice of PDR model. \citet{rollig2007} performed a detailed comparison of PDR models to identify differences in the codes and examine their effects on the physical properties and chemical structures of the model clouds. One important feature of a PDR model is the adopted geometry; the plane-parallel geometry of the \citet{kaufman1999,kaufman2006} model is a first order approximation and, as shown here, a spherical model might be more appropriate. Whilst the benchmarking exercise demonstrated that resulting trends in physical parameters are consistent between the participating codes, they warn that discrepancies remain between observables computed with different codes - including the atomic fine-structure line intensities - and that these uncertainties should be kept in mind when comparing PDR model results to observations in order to constrain physical parameters, such as density, temperature and radiation field strength.

To summarise our main results from the comparison of our observations to the predictions of the \citet{kaufman1999,kaufman2006} PDR model, we find that, with the exception of the central region, the majority of the PDRs in NGC 891's disc have hydrogen nuclei with densities ranging from $1 < \log n < 3.5$ cm$^{-3}$ with a mean of $\log n/\mathrm{cm}^{-3} \sim 3$, and experience an incident FUV radiation field with a strength between $1.7 < \log G_0 < 3$ normalised to the \citet{habing1968} Field (see Table~\ref{tab:pdrprops}). Although similar results are found regardless of our adopted approach for adjusting the \cii \ emission to conform with the model requirements, very different trends are found across the disc, dependent on the correction to the \oi ~63~$\mu$m account for the effects of varying optical depth. Using ISO observations of the \cii ~158, \oi ~63 and 145~$\mu$m lines integrated along the galactic plane, \citet[][see their Fig. 10]{stacey2010} determined a best-fit FUV field strength of $G_{0} \sim 100$ and density $n \sim 3 \times 10^{3}$ cm$^{-3}$. Combined with observations of the pure rotational H$_{2}$ lines obtained with the \textit{Spitzer} Infrared Spectrograph, the closest model solution satisfying both the H$_{2}$ and FIR lines involves a common field strength of $G_{0} \sim 100-200$ albeit with unequal densities of $n \geq 6 \times 10^{4}$ and $0.4 \times 10^{4} < n < 1 \times 10^{4}$ cm$^{-3}$, respectively. Our results are consistent with the PDR parameters from the FIR lines, but we likely don't reach hydrogen densities as high as $n \sim 6 \times 10^{4}\,\mathrm{cm}^{-3}$ because the H$_{2}$ lines arise from deeper in the PDRs than the FIR lines and thus probe denser gas \citep*[see also][]{valentijn1999}. The gas properties we find in the disc of NGC 891 are also consistent with the previous surveys of global, integrated observations, such as the \citet{malhotra2001} ISO survey that found $2 \leq \log n/\mathrm{cm}^{-3} \leq 4.5$ and $2 \leq \log G_0 \leq 4.5$, and with targeted resolved studies of nearby objects (c.f. Table 9 in \citealp{parkin2013}; see also e.g. \citealp{lebouteiller2012}; \citealp{croxall2012}; \citealp{parkin2014}).

By following the methodology of \citet{parkin2013}, we can confidently make a direct comparison between the gas properties in NGC 891 and those found in the various regions of M51. The spiral arm and inter-arm regions in M51 both exhibit hydrogen densities and FUV radiation field strengths of $2.75 \leq \log n \leq 3$ cm$^{-3}$ and $2.25 \leq \log G_0 \leq 2.5$, respectively, despite the latter region have lower star formation rate surface densities compared to in the spiral arms, suggesting that the molecular clouds have similar properties but are more abundant in the arms than inter-arm regions \citep{parkin2013}. The $n$ and $G_{0}$ values we derive for most of NGC 891's disc at larger vertical distances are consistent with these values, supporting the body of evidence that this galaxy is a typical star-forming disc with spiral arms. However, whilst the majority of the disc in NGC 891 thus has very similar properties to the spiral arm and inter-arm regions in M51, the comparison and interpretation of the central and mid-plane regions of the edge-on galaxy are somewhat more complex than for a face-on disc. M51 has much higher ranges in the values of $n$ and $G_{0}$ for both the central ($3 \leq \log n/\mathrm{cm}^{-3} \leq 3.5$, $2.75 \leq \log G_0 \leq 3$) and nuclear ($3.75 \leq \log n/\mathrm{cm}^{-3} \leq 4$, $3.25 \leq \log G_0 \leq 3.75$) regions, which arise from the lower values of their observations in the \ciioi \ versus \ciioitir \ parameter space (see their Fig. 7), than compared with the center of NGC 891. When using the same adjustments to the observations as \citet{parkin2013}, we find the central and mid-plane pixels do not fall within the \ciioi \ versus \ciioitir \ parameter space described by the PDR model (see our Fig.~\ref{fig:pdrdiag}).

Such an offset between these values for the central regions likely arise due to several different factors affecting the \ciioi \ ratio in these two galaxies. M51 displays a lower \ciioi \ ratio in the center than the rest of the galaxy, due to a peak in the \oi ~63~$\mu$m line emission from the nucleus. There remains a possibility that the line emission may be contaminated by shock heating \citep{hollenbach1989} from M51's Seyfert type-2 nucleus \citep{ho1997} rather than star light, whereby we would expect a higher \ciioi \ ratio arising from the PDRs and, as a consequence, a shift in the derived PDR parameters to lower $n$ and $G_{0}$ values. In the case of NGC~891, there is some evidence to suggest the presence of a weak AGN; \citet{strickland2004} found a faint, hard (2--8~keV) X-ray source in \textit{Chandra} observations (not detected with \textit{XMM-Newton}; \citealp{temple2005}) towards the central radio continuum point source \citep{rupen1991}. We thus can't exclude AGN contamination to the line emission. In addition, we are not strictly observing just the center/nucleus (as in the case of a face-on galaxy like M51), as the line-of-sight towards the center will also include \cii \ emission originating from PDRs in the disc lying between our line-of-sight and the nucleus. Any conflation of the disc and nucleus along the line-of-sight may artificially increase the diagnostic \ciioi \ and \ciioitir \ ratios in the center, and thus drive the observations out of the parameter space described by the PDR model (e.g. Fig.~\ref{fig:mapsdiagparams}). Furthermore, the line-of-sight towards the center will pass through the densest regions and, since the \oi ~63~$\mu$m line becomes optically thick faster than the \cii ~158~$\mu$m line (e.g. \citealp{abel2007}), we may be significantly underestimating the amount of \oi ~63~$\mu$m line emission escaping away from our line of sight. The fact that the same trends in $n$ and $G_{0}$ across the disc are found when using the \oi ~145~$\mu$m line to constrain the PDR model, which we assume remains optically thin, suggests that the factor of two correction to the \oi ~63~$\mu$m line emission is appropriate for most of the disc along the plane and at increasing vertical heights above the plane, but also implies that perhaps an even higher correction factor is required in the center (e.g. \citealp{stacey2010}). Only by accounting for optical depth effects in the \oi ~63~$\mu$m line are we able to reproduce the overall trends found in M51 and bring the observations of the centers of NGC~891 and M51 into better agreement (compare our Fig.~\ref{fig:pdrdiagav} to Fig. 7 in \citealp{parkin2013}), implying that optical depth effects become increasingly important to consider when interpreting high inclination systems. Future studies should pursue more robust constraints on the measurement of the optical depth to accurately correct the \oi \ emission in central pixels (or, for example, all pixels with \oi 145 /\oi 63 $>$ 0.15 or $A_V$ $>$ 10 mag).

On the far north eastern side of the disc, we observe enhancements in the \cii , \oi \ 63~$\mu$m and \oiii \ line emission relative to the TIR contours, and, when we extend our analysis by exploiting the empirical correlation we find between the \nii ~205~$\mu$m and the 24 ~$\mu$m emission, this region consistently demonstrates relatively higher FUV field strengths, gas densities and PDR surface temperatures with respect to the rest of the disc. An important question remains, however, regarding whether this enhancement is genuinely physical, or merely an artefact of our method for estimating the ionised gas density in this part of the disc. For example, it is already known that NGC 891's 24~$\mu$m emission exhibits an enhancement in this region compared to the opposite location on the south-western side (e.g. \citealp{kamphuis2007}; \citealp{hughes2014}) and, from our method, this enhancement would lead to higher estimates of the \nii ~205~$\mu$m line emission, lower ionised gas densities (at fixed \nii ~122~$\mu$m flux densities, c.f. Fig.~\ref{fig:ionisedgasfraction}), lower estimates for the contribution of ionised gas to the \cii \ 158 emission, and hence a decrease in the \ciioi \ and \ciioitir \ ratios. Therefore, any enhancement in the 24~$\mu$m emission would consequently lead to higher $G_{0}$ values. Yet, if we consider that the NE side has more prominent and extended H$\alpha$ and UV emission than the SW side of the disc (\citealp{dettmar1990}; \citealp{rand1990}; \citealp{kamphuis2007}), possibly due to a higher SFR in the northern part of the disc than in the southern part \citep{rossa2004}, then we may expect this asymmetry in the SFR to manifest in the \nii ~205~$\mu$m line since it traces star formation (e.g. \citealp{zhao2013}; \citealp{wu2014}). In fact, the \nii ~205~$\mu$m flux density in the single 17$\arcsec$ pixel covering the NE region is 3.69$\times10^{-8}$ $\mathrm{W}\,\mathrm{m}^{-2}\,\mathrm{sr}^{-1}$, three times higher than the measured flux density of 1.14$\times10^{-8}$ $\mathrm{W}\,\mathrm{m}^{-2}\,\mathrm{sr}^{-1}$ in a single pixel on the diametrically opposite side of the disc (see Fig.~\ref{fig:obsmaps3sig}, lower right panel), which could tentatively hint at an asymmetry in the \nii ~205~$\mu$m line emission were we to neglect to consider the $\sim$7\% flux calibration errors on these measurements. We thus argue that perhaps the enhancement in the FUV radiation field strength is physical in nature.

In applying the PDR model of \citet{kaufman1999,kaufman2006}, we adjusted the observations by (i) correcting the \cii ~158~$\mu$m line emission to remove the contribution to the emission arising from diffuse ionised gas, (ii) increasing the \oi ~63~$\mu$m emission by a factor of two to account for photons emitted away from our line of sight, and (iii) reducing the TIR emission by a factor of two to account for the optically thin continuum flux emitting not just towards the observer but from both sides of the PDR slab (see Sec. \ref{sec:adjustments}). We assume these adjustments are correct to the first order for facilitating a proper comparison of our observations and the model. However, in addition to the possibility that the observed \oi ~63~$\mu$m line intensity should in fact be corrected for these geometric issues by as much as a factor of four for high optical depth in a spherical cloud geometry \citep{stacey2010}, it may also be necessary to further reduce the TIR emission to account for continuum emission from other non-PDR sources, such as e.g. \hii regions. Such additional corrections to all pixels would shift our \ciiobsion - and \ciisynion -based adjusted observations downwards and to the right in the \ciioi \ versus \ciioitir \ parameter space in the Fig.~\ref{fig:pdrdiag} diagnostic diagrams (and also in Fig.\ref{fig:pdrdiagav}), shifting $n$ and $G_0$ to higher densities and potentially lower FUV field strengths. We further caution that random shifts may also occur due to errors associated with the flux calibration, mismatching of the PSFs assumed for the convolution and rescaling the images to the resolution and pixel size of the \textit{Herschel} 160~$\mu$m image, and small offsets in the position angles of the various images \citep[see][]{hughes2014}.
            
Finally, we stress that, despite the huge advancements over previous FIR experiments in the quality of observations of the far-infrared fine-structure lines made possible by the \textit{Herschel} Space Observatory, enabling us to resolve features on sub-kiloparsec scales, one of the main limitations of this analysis was the relatively sparse coverage and low resolution of the SPIRE FTS observations of the \nii ~205~$\mu$m line. Our methods to determine the fraction of \cii ~158~$\mu$m emission arising from diffuse ionised gas using the direct measurements of the \nii ~205~$\mu$m emission at 17$\arcsec$ ($\sim$0.79~kpc) resolution and, alternatively, estimates of the \nii ~205~$\mu$m emission at 12$\arcsec$ ($\sim$0.56~kpc) resolution via the MIPS 24 ~$\mu$m data, introduced additional uncertainty into our analysis. Indeed, the empirical relationship we find between the \nii ~205~$\mu$m and the 24 ~$\mu$m emission should be the focus of greater study, preferably using observations of face-on or less inclined galaxies. These uncertainties, plus others discussed throughout this work, may be addressed by future far-infrared facilities, in particular the planned SPICA (Space Infrared Telescope for Cosmology and Astrophysics) mission \citep[e.g.][]{nakagawa2012} with the SAFARI instrument \citep[e.g.][]{roelfsema2012}, a far-infrared imaging FTS-spectrometer designed to cover the $\sim$34 to 210~$\mu$m waveband at unprecedented resolution using a cryogenically cooled ($<$6 K) $\sim$3.2 m space telescope.   
          
\renewcommand{\arraystretch}{1.2}          
            
\begin{sidewaystable*}
\begin{center}
\begin{minipage}{\textwidth}
\caption{Summary of best-fitting PDR model parameters found from each set of observed diagnostic lines and the various corrections considered in this work, where we state the range of values together with the corresponding median value (in brackets) determined for each of the regions in Fig.~\ref{fig:tirmapmask}.}
\label{tab:pdrprops}
\begin{center}
\begin{tabular}{l c c c c c c c c c c c}
\hline 
\hline
\multicolumn{1}{l}{Diagnostics} & \multicolumn{3}{c}{\cii ~ observed} & & \multicolumn{3}{c}{\cii ~corrected via \ciiobsion } & & \multicolumn{3}{c}{\cii ~corrected via \ciisynion} \\ \cline{2-4} \cline{6-8} \cline{10-12}
 &  $\log n/\mathrm{cm}^{3}$ &$\log G_0$& $T$/K & & $\log n/\mathrm{cm}^{3}$ &$\log G_0$& $T$/K & & $\log n/\mathrm{cm}^{3}$ &$\log G_0$& $T$/K\\
\hline
\oi ~63~$\mu$m  & \multicolumn{3}{l}{Observed} & & \multicolumn{3}{l}{Constant correction\footnoteremember{const}{Observed \oi ~63~$\mu$m flux increased by factor of 2, and the TIR emission reduced by a factor of 2 (see Sec.~\ref{sec:adjustments}).}} & & \multicolumn{3}{l}{Constant correction $\color{blue}^{a}$} \\
\hline
\multicolumn{12}{c}{ } \\
Center     & 3.75-6.25(6.25)  & 0.50-1.00(0.50)  & 27-33(30)  & & 1.50-2.00(2.00) & 1.50-1.75(1.75)  & 170-347(170) & &  1.50-4.25(2.00) & 1.00-1.75(1.75)  & 37-391(170)  \\
Mid-plane  & 1.00-6.25(3.75)  & 0.50-1.25(0.25)  & 27-282(37) & & 1.00-2.75(2.50) & 1.00-2.75(2.25)  & 170-598(256) & &  1.00-2.75(2.25) & 1.00-2.50(2.00)  & 170-598(256)  \\
Off-plane  & 1.00-4.00(2.00)  & 0.25-2.00(1.25)  & 27-391(136)& & 1.00-4.25(2.75) & 0.25-3.00(2.00)  & 42-598(202) &  & 1.00-4.25(2.75) & 0.25-2.75(2.00)  & 43-598(197)  \\
Outer disc & 2.00-3.75(2.75)  & 0.50-1.75(1.50)  & 36-138(96) & & 3.00-3.50(3.25) & 1.75-2.25(2.00)  & 97-197(163) &  & 3.00-4.25(3.25) & 0.50-2.25(2.00)  & 45-197(155)  \\
\multicolumn{12}{c}{ } \\
\hline
\oi ~63~$\mu$m & \multicolumn{3}{l}{Observed} & & \multicolumn{3}{l}{Varying correction\footnoteremember{vary}{Observed \oi ~63~$\mu$m flux increased by factor of 4 in center and mid-plane and a factor of 2 everywhere else, and the TIR emission reduced by a factor of 2 (see Sec.~\ref{sec:oi145pdr}).}} & & \multicolumn{3}{l}{Varying correction $\color{blue}^{b}$} \\
\hline
\multicolumn{12}{c}{ } \\
Center     &  3.75-6.25(6.25)  & 0.50-1.00(0.50)  & 27-33(30)  & & 1.00-2.75(1.75)  & 1.50-2.50(2.15)  & 220-598(401) & & 1.00-4.50(1.25) & 1.00-2.75(1.75) & 37-598(547)  \\
Mid-plane  &  1.00-6.25(3.75)  & 0.50-1.25(0.25)  & 27-282(37) & & 1.00-4.50(3.00)  & 0.25-3.00(2.50)  & 45-1080(197) & & 1.00-4.50(3.00) & 0.25-3.00(2.50) & 45-598(197)  \\
Off-plane  &  1.00-4.00(2.00)  & 0.25-2.00(1.25)  & 27-391(136)& & 2.75-4.50(3.25)  & 0.25-3.00(2.25)  & 45-320 (191) & & 2.50-4.25(3.25) & 0.25-3.00(2.25) & 45-307(191)  \\
Outer disc &  2.00-3.75(2.75)  & 0.50-1.75(1.50)  & 36-138(96) & & 3.50-4.25(3.75)  & 0.75-2.50(2.25)  & 50-191 (144) & & 3.50-4.25(4.00) & 0.50-2.50(1.75) & 45-191(80)   \\
\multicolumn{12}{c}{ } \\
\hline
\oi ~145~$\mu$m & \multicolumn{3}{l}{Observed} & & \multicolumn{3}{l}{Observed} & & \multicolumn{3}{l}{Observed}\\
\hline
\multicolumn{12}{c}{ } \\
Center     &  1.00-2.25(1.75) & 1.50-2.50(2.00)  &  256-1080(598) & & 1.00-2.75(2.00)& 1.75-3.00(2.50) & 322-1080(380)&  & 1.00-4.75(1.75)& 0.25-3.00(2.25) & 45-1470(823) \\
Mid-plane  &  1.00-4.50(2.50) & 0.25-3.00(2.00)  &  45-620(234)   & & 1.00-4.50(3.00)& 0.25-3.00(2.50) & 45-1080(197) &  & 1.00-4.50(2.50)& 0.25-3.00(2.25) & 45-1470(256) \\
Off-plane  &  1.00-4.50(3.25) & 0.25-3.00(2.25)  &  45-598(191)   & & 1.00-4.50(3.25)& 0.25-3.00(2.25) & 45-823(191)  &  & 1.00-4.50(3.25)& 0.25-3.00(2.25) & 45-1080(191) \\
Outer disc &  2.75-4.50(4.00) & 0.50-2.50(2.00)  &  45-220(112)   & & 3.00-4.25(4.00)& 0.50-2.50(1.75) & 45-197(80)    & &  2.75-4.50(4.00)& 0.50-2.50(2.00) & 45-220(112) \\
\multicolumn{12}{c}{ } \\
\hline
\end{tabular}
\end{center}
\end{minipage}
\end{center}
\end{sidewaystable*}

\section{Conclusions}\label{sec:conclusions}

We present \textit{Herschel} PACS and SPIRE FTS spectroscopy, focussing on the most important far-infrared cooling lines in NGC~891: \cii ~$\lambda$~158~$\mu$m,  \nii ~$\lambda\lambda$~122,~205~$\mu$m, \oi ~$\lambda\lambda$~63,~145~$\mu$m, and \oiii~$\lambda$~88~$\mu$m. We find that the photoelectric heating efficiency of the gas, traced via the (\cii + \oi ~$\lambda$~63~$\mu$m)/$F_{\mathrm{TIR}}$ ratio, varies from a mean \ciioitir \ of $3.5\times$10$^{-3}$ in the centre up to $8\times$10$^{-3}$ at increasing radial and vertical distances in the disc. We find a decrease in \ciioitir \ with increasing FIR colour, which corresponds to a decrease in the heating efficiency in the nucleus and inner plane regions relative to regions at higher radial and vertical distances along the disc, yet observe no similar variation in \ciioipah \ with increasing FIR colour. This result may suggest that in the central regions the gas heating becomes dominated by PAHs rather than dust grains.

We compare the observed flux of the FIR cooling lines and total IR emission with the predicted flux from a PDR model to determine the characteristics of the gas such as density, temperature and the incident far-ultraviolet radiation field, $G_{0}$, resolving details on physical scales of roughly 0.6~kpc. A pixel-by-pixel analysis reveals that, with the exception of the central region, the majority of the PDRs in NGC 891's disc have hydrogen nuclei with densities ranging from $1 < \log n/\mathrm{cm}^{-3} < 3.5$ with a mean of $\log n/\mathrm{cm}^{-3} \sim 3$, and experience an incident FUV radiation field with a strength between $1.7 < \log G_0 < 3$ normalised to the \citet{habing1968} Field. However, the variations in the $n$ and $G_0$ with increasing radius and vertical height were found to be highly sensitive to optical depth effects. Using a constant correction factor to the \oi ~63~$\mu$m line emission, we see an increase in the density with increasing radial distance and vertical height but less variation in the FUV radiation field strength, contrary to previous results. Whilst the $n$ and $G_{0}$ values we derive for most of NGC 891's disc are consistent with the gas properties found in PDRs in the spiral arms and inter-arm regions of M51, only by increasing this factor to account for optical depth effects in the \oi ~63~$\mu$m line are we able to reproduce the overall trends found in M51 and in similar studies of other nearby galaxies. We were, however, unable to account for the same trends found using the optically-thin \oi ~145~$\mu$m line as a gas diagnostic. These results imply that optical depth effects become increasingly important to consider when interpreting high inclination systems. 

We use an empirical linear relationship between the MIPS 24~$\mu$m data and the 205~$\mu$m data to predict the \nii ~205~$\mu$m line emission and hence increase the resolution and coverage of our estimate of the fraction of \cii ~158~$\mu$m emission arising from diffuse ionised gas. This alternative technique not only reproduces the aforementioned variations of $n$ and $G_{0}$ across the disc, but also estimates that the FUV radiation field in the far north-eastern side of the disc is on average slightly stronger ($\log G_{0} \sim 2.4$) than the rest of the disc albeit with a moderate hydrogen density ($\log n/\mathrm{cm}^{-3} \sim 2.5$). Whilst these enhancements in this region coincide with the above-average star formation rate surface densities and gas-to-dust ratios compared to the rest of the disc, a direct interpretation remains difficult due to uncertainties in the observations and PDR modelling. Up-coming far-infrared facilities, such as SPICA with the SAFARI instrument, will be necessary to investigate such variations in the gas heating and cooling mechanisms for much larger samples of galaxies.

\begin{acknowledgements}
We thank the referee for useful comments and suggestions which significantly improved the quality of this paper. TMH gratefully acknowledges the financial support from the Belgian Science Policy Office (BELSPO) in the frame of the PRODEX project C90370 (Herschel-PACS Guaranteed Time and Open Time Programs: Science Exploitation). IDL is a postdoctoral fellow of the Flemish Fund for Scientific Research (FWO-Vlaanderen). MB also acknowledges the financial support of the same Flemish Fund for Scientific Research. PACS has been developed by a consortium of institutes led by MPE (Germany) and including UVIE (Austria); KU Leuven, CSL, IMEC (Belgium); CEA, LAM (France); MPIA (Germany); INAF-IFSI/OAA/OAP/OAT, LENS, SISSA (Italy); IAC (Spain). This development has been supported by the funding agencies BMVIT (Austria), ESA-PRODEX (Belgium), CEA/CNES (France), DLR (Germany), ASI/INAF (Italy), and CICYT/MCYT (Spain). SPIRE has been developed by a consortium of institutes led by Cardiff University (UK) and including Univ. Lethbridge (Canada); NAOC (China); CEA, LAM (France); IFSI, Univ. Padua (Italy); IAC (Spain); Stockholm Observatory (Sweden); Imperial College London, RAL, UCL-MSSL, UKATC, Univ. Sussex (UK); and Caltech, JPL, NHSC, Univ. Colorado (USA). This development has been supported by national funding agencies: CSA (Canada); NAOC (China); CEA, CNES, CNRS (France); ASI (Italy); MCINN (Spain); SNSB (Sweden); STFC, UKSA (UK); and NASA (USA). This research has made use of the NASA/IPAC Extragalactic Database (NED) which is operated by the Jet Propulsion Laboratory, California Institute of Technology, under contract with the NASA (USA). This research made use of APLpy, an open-source plotting package for Python hosted at \url{http://aplpy.github.com}, and Montage, which is funded by the NASA (USA) Earth Science Technology Office, Computation Technologies Project, under Cooperative Agreement Number NCC5-626 between NASA and Caltech, and maintained by the NASA/IPAC Infrared Science Archive.
\end{acknowledgements}

\bibliography{aa24732-14}

\begin{thebibliography}{116}
\expandafter\ifx\csname natexlab\endcsname\relax\def\natexlab#1{#1}\fi

\bibitem[{{Abel} {et~al.}(2009){Abel}, {Dudley}, {Fischer}, {Satyapal}, \& {van
  Hoof}}]{abel2009}
{Abel}, N.~P., {Dudley}, C., {Fischer}, J., {Satyapal}, S., \& {van Hoof},
  P.~A.~M. 2009, \apj, 701, 1147

\bibitem[{{Abel} {et~al.}(2007){Abel}, {Sarma}, {Troland}, \&
  {Ferland}}]{abel2007}
{Abel}, N.~P., {Sarma}, A.~P., {Troland}, T.~H., \& {Ferland}, G.~J. 2007,
  \apj, 662, 1024

\bibitem[{{Aniano} {et~al.}(2011){Aniano}, {Draine}, {Gordon}, \&
  {Sandstrom}}]{aniano2011}
{Aniano}, G., {Draine}, B.~T., {Gordon}, K.~D., \& {Sandstrom}, K. 2011, \pasp,
  123, 1218

\bibitem[{{Bakes} \& {Tielens}(1994)}]{bakes1994}
{Bakes}, E.~L.~O. \& {Tielens}, A.~G.~G.~M. 1994, \apj, 427, 822

\bibitem[{{Beir{\~a}o} {et~al.}(2010){Beir{\~a}o}, {Armus}, {Appleton},
  {Smith}, {Croxall}, {Murphy}, {Dale}, {Helou}, {Kennicutt}, {Calzetti},
  {Bolatto}, {Brandl}, {Crocker}, {Draine}, {Dumas}, {Engelbracht}, {Gil de
  Paz}, {Gordon}, {Groves}, {Hao}, {Hinz}, {Hunt}, {Johnson}, {Koda}, {Krause},
  {Leroy}, {Meidt}, {Richer}, {Rix}, {Rahman}, {Roussel}, {Sandstrom},
  {Sauvage}, {Schinnerer}, {Skibba}, {Srinivasan}, {Walter}, {Warren},
  {Wilson}, {Wolfire}, \& {Zibetti}}]{beirao2010}
{Beir{\~a}o}, P., {Armus}, L., {Appleton}, P.~N., {et~al.} 2010, \aap, 518, L60

\bibitem[{{Beir{\~a}o} {et~al.}(2012){Beir{\~a}o}, {Armus}, {Helou},
  {Appleton}, {Smith}, {Croxall}, {Murphy}, {Dale}, {Draine}, {Wolfire},
  {Sandstrom}, {Aniano}, {Bolatto}, {Groves}, {Brandl}, {Schinnerer},
  {Crocker}, {Hinz}, {Rix}, {Kennicutt}, {Calzetti}, {Gil de Paz}, {Dumas},
  {Galametz}, {Gordon}, {Hao}, {Johnson}, {Koda}, {Krause}, {van der Laan},
  {Leroy}, {Li}, {Meidt}, {Meyer}, {Rahman}, {Roussel}, {Sauvage},
  {Srinivasan}, {Vigroux}, {Walter}, \& {Warren}}]{beirao2012}
{Beir{\~a}o}, P., {Armus}, L., {Helou}, G., {et~al.} 2012, \apj, 751, 144

\bibitem[{{Bendo} {et~al.}(2014){Bendo}, {Baes}, {Bianchi}, {Boquien},
  {Boselli}, {Cooray}, {Cortese}, {De Looze}, {di Serego Alighieri}, {Fritz},
  {Gentile}, {Hughes}, {Lu}, {Pappalardo}, {Smith}, {Spinoglio}, {Viaene}, \&
  {Vlahakis}}]{bendo2014}
{Bendo}, G.~J., {Baes}, M., {Bianchi}, S., {et~al.} 2014, ArXiv e-prints,
  astro-ph/1409.1815

\bibitem[{{Bendo} {et~al.}(2008){Bendo}, {Draine}, {Engelbracht}, {Helou},
  {Thornley}, {Bot}, {Buckalew}, {Calzetti}, {Dale}, {Hollenbach}, {Li}, \&
  {Moustakas}}]{bendo2008}
{Bendo}, G.~J., {Draine}, B.~T., {Engelbracht}, C.~W., {et~al.} 2008, \mnras,
  389, 629

\bibitem[{{Bendo} {et~al.}(2012){Bendo}, {Galliano}, \& {Madden}}]{bendo2012}
{Bendo}, G.~J., {Galliano}, F., \& {Madden}, S.~C. 2012, \mnras, 423, 197

\bibitem[{{Bianchi} \& {Xilouris}(2011)}]{bianchi2011}
{Bianchi}, S. \& {Xilouris}, E.~M. 2011, \aap, 531, L11

\bibitem[{{Blum} \& {Pradhan}(1992)}]{blum1992}
{Blum}, R.~D. \& {Pradhan}, A.~K. 1992, \apjs, 80, 425

\bibitem[{{Brauher} {et~al.}(2008){Brauher}, {Dale}, \& {Helou}}]{brauher2008}
{Brauher}, J.~R., {Dale}, D.~A., \& {Helou}, G. 2008, \apjs, 178, 280

\bibitem[{{Bregman}(1980)}]{bregman1980}
{Bregman}, J.~N. 1980, \apj, 236, 577

\bibitem[{{Burgdorf} {et~al.}(2007){Burgdorf}, {Ashby}, \&
  {Williams}}]{burgdorf2007}
{Burgdorf}, M., {Ashby}, M.~L.~N., \& {Williams}, R. 2007, \apj, 668, 918

\bibitem[{{Calzetti} {et~al.}(2007){Calzetti}, {Kennicutt}, {Engelbracht},
  {Leitherer}, {Draine}, {Kewley}, {Moustakas}, {Sosey}, {Dale}, {Gordon},
  {Helou}, {Hollenbach}, {Armus}, {Bendo}, {Bot}, {Buckalew}, {Jarrett}, {Li},
  {Meyer}, {Murphy}, {Prescott}, {Regan}, {Rieke}, {Roussel}, {Sheth}, {Smith},
  {Thornley}, \& {Walter}}]{calzetti2007}
{Calzetti}, D., {Kennicutt}, R.~C., {Engelbracht}, C.~W., {et~al.} 2007, \apj,
  666, 870

\bibitem[{{Calzetti} {et~al.}(2005){Calzetti}, {Kennicutt}, {Bianchi},
  {Thilker}, {Dale}, {Engelbracht}, {Leitherer}, {Meyer}, {Sosey}, {Mutchler},
  {Regan}, {Thornley}, {Armus}, {Bendo}, {Boissier}, {Boselli}, {Draine},
  {Gordon}, {Helou}, {Hollenbach}, {Kewley}, {Madore}, {Martin}, {Murphy},
  {Rieke}, {Rieke}, {Roussel}, {Sheth}, {Smith}, {Walter}, {White}, {Yi},
  {Scoville}, {Polletta}, \& {Lindler}}]{calzetti2005}
{Calzetti}, D., {Kennicutt}, Jr., R.~C., {Bianchi}, L., {et~al.} 2005, \apj,
  633, 871

\bibitem[{{Contursi} {et~al.}(2002){Contursi}, {Kaufman}, {Helou},
  {Hollenbach}, {Brauher}, {Stacey}, {Dale}, {Malhotra}, {Rubio}, {Rubin}, \&
  {Lord}}]{contursi2002}
{Contursi}, A., {Kaufman}, M.~J., {Helou}, G., {et~al.} 2002, \aj, 124, 751

\bibitem[{{Contursi} {et~al.}(2013){Contursi}, {Poglitsch}, {Gr{\'a}cia
  Carpio}, {Veilleux}, {Sturm}, {Fischer}, {Verma}, {Hailey-Dunsheath}, {Lutz},
  {Davies}, {Gonz{\'a}lez-Alfonso}, {Sternberg}, {Genzel}, \&
  {Tacconi}}]{contursi2013}
{Contursi}, A., {Poglitsch}, A., {Gr{\'a}cia Carpio}, J., {et~al.} 2013, \aap,
  549, A118

\bibitem[{{Cormier} {et~al.}(2010){Cormier}, {Madden}, {Hony}, {Contursi},
  {Poglitsch}, {Galliano}, {Sturm}, {Doublier}, {Feuchtgruber}, {Galametz},
  {Geis}, {de Jong}, {Okumura}, {Panuzzo}, \& {Sauvage}}]{cormier2010}
{Cormier}, D., {Madden}, S.~C., {Hony}, S., {et~al.} 2010, \aap, 518, L57

\bibitem[{{Crawford} {et~al.}(1985){Crawford}, {Genzel}, {Townes}, \&
  {Watson}}]{crawford1985}
{Crawford}, M.~K., {Genzel}, R., {Townes}, C.~H., \& {Watson}, D.~M. 1985,
  \apj, 291, 755

\bibitem[{{Croxall} {et~al.}(2012){Croxall}, {Smith}, {Wolfire}, {Roussel},
  {Sandstrom}, {Draine}, {Aniano}, {Dale}, {Armus}, {Beir{\~a}o}, {Helou},
  {Bolatto}, {Appleton}, {Brandl}, {Calzetti}, {Crocker}, {Galametz}, {Groves},
  {Hao}, {Hunt}, {Johnson}, {Kennicutt}, {Koda}, {Krause}, {Li}, {Meidt},
  {Murphy}, {Rahman}, {Rix}, {Sauvage}, {Schinnerer}, {Walter}, \&
  {Wilson}}]{croxall2012}
{Croxall}, K.~V., {Smith}, J.~D., {Wolfire}, M.~G., {et~al.} 2012, \apj, 747,
  81

\bibitem[{{Dale} {et~al.}(2001){Dale}, {Helou}, {Contursi}, {Silbermann}, \&
  {Kolhatkar}}]{dale2001}
{Dale}, D.~A., {Helou}, G., {Contursi}, A., {Silbermann}, N.~A., \&
  {Kolhatkar}, S. 2001, \apj, 549, 215

\bibitem[{{de Vaucouleurs} {et~al.}(1976){de Vaucouleurs}, {de Vaucouleurs}, \&
  {Corwin}}]{RC21976}
{de Vaucouleurs}, G., {de Vaucouleurs}, A., \& {Corwin}, J.~R. 1976, in Second
  reference catalogue of bright galaxies, 1976, Austin: University of Texas
  Press., 0

\bibitem[{{Dettmar}(1990)}]{dettmar1990}
{Dettmar}, R.-J. 1990, \aap, 232, L15

\bibitem[{{Draine} {et~al.}(2007){Draine}, {Dale}, {Bendo}, {Gordon}, {Smith},
  {Armus}, {Engelbracht}, {Helou}, {Kennicutt}, {Li}, {Roussel}, {Walter},
  {Calzetti}, {Moustakas}, {Murphy}, {Rieke}, {Bot}, {Hollenbach}, {Sheth}, \&
  {Teplitz}}]{draine2007}
{Draine}, B.~T., {Dale}, D.~A., {Bendo}, G., {et~al.} 2007, \apj, 663, 866

\bibitem[{{Draine} \& {Li}(2007)}]{draineli2007}
{Draine}, B.~T. \& {Li}, A. 2007, \apj, 657, 810

\bibitem[{{Engelbracht} {et~al.}(2007){Engelbracht}, {Blaylock}, {Su}, {Rho},
  {Rieke}, {Muzerolle}, {Padgett}, {Hines}, {Gordon}, {Fadda},
  {Noriega-Crespo}, {Kelly}, {Latter}, {Hinz}, {Misselt}, {Morrison},
  {Stansberry}, {Shupe}, {Stolovy}, {Wheaton}, {Young}, {Neugebauer},
  {Wachter}, {P{\'e}rez-Gonz{\'a}lez}, {Frayer}, \&
  {Marleau}}]{engelbracht2007}
{Engelbracht}, C.~W., {Blaylock}, M., {Su}, K.~Y.~L., {et~al.} 2007, \pasp,
  119, 994

\bibitem[{{Farrah} {et~al.}(2013){Farrah}, {Lebouteiller}, {Spoon},
  {Bernard-Salas}, {Pearson}, {Rigopoulou}, {Smith}, {Gonz{\'a}lez-Alfonso},
  {Clements}, {Efstathiou}, {Cormier}, {Afonso}, {Petty}, {Harris}, {Hurley},
  {Borys}, {Verma}, {Cooray}, \& {Salvatelli}}]{farrah2013}
{Farrah}, D., {Lebouteiller}, V., {Spoon}, H.~W.~W., {et~al.} 2013, \apj, 776,
  38

\bibitem[{{Fazio} {et~al.}(2004){Fazio}, {Hora}, {Allen}, {Ashby}, {Barmby},
  {Deutsch}, {Huang}, {Kleiner}, {Marengo}, {Megeath}, {Melnick}, {Pahre},
  {Patten}, {Polizotti}, {Smith}, {Taylor}, {Wang}, {Willner}, {Hoffmann},
  {Pipher}, {Forrest}, {McMurty}, {McCreight}, {McKelvey}, {McMurray}, {Koch},
  {Moseley}, {Arendt}, {Mentzell}, {Marx}, {Losch}, {Mayman}, {Eichhorn},
  {Krebs}, {Jhabvala}, {Gezari}, {Fixsen}, {Flores}, {Shakoorzadeh}, {Jungo},
  {Hakun}, {Workman}, {Karpati}, {Kichak}, {Whitley}, {Mann}, {Tollestrup},
  {Eisenhardt}, {Stern}, {Gorjian}, {Bhattacharya}, {Carey}, {Nelson},
  {Glaccum}, {Lacy}, {Lowrance}, {Laine}, {Reach}, {Stauffer}, {Surace},
  {Wilson}, {Wright}, {Hoffman}, {Domingo}, \& {Cohen}}]{fazio2004}
{Fazio}, G.~G., {Hora}, J.~L., {Allen}, L.~E., {et~al.} 2004, \apjs, 154, 10

\bibitem[{{Ferkinhoff} {et~al.}(2011){Ferkinhoff}, {Brisbin}, {Nikola},
  {Parshley}, {Stacey}, {Phillips}, {Falgarone}, {Benford}, {Staguhn}, \&
  {Tucker}}]{ferkinhoff2011}
{Ferkinhoff}, C., {Brisbin}, D., {Nikola}, T., {et~al.} 2011, \apjl, 740, L29

\bibitem[{{Galametz} {et~al.}(2012){Galametz}, {Kennicutt}, {Albrecht},
  {Aniano}, {Armus}, {Bertoldi}, {Calzetti}, {Crocker}, {Croxall}, {Dale},
  {Donovan Meyer}, {Draine}, {Engelbracht}, {Hinz}, {Roussel}, {Skibba},
  {Tabatabaei}, {Walter}, {Weiss}, {Wilson}, \& {Wolfire}}]{galametz2012}
{Galametz}, M., {Kennicutt}, R.~C., {Albrecht}, M., {et~al.} 2012, \mnras, 425,
  763

\bibitem[{{Galametz} {et~al.}(2013){Galametz}, {Kennicutt}, {Calzetti},
  {Aniano}, {Draine}, {Boquien}, {Brandl}, {Croxall}, {Dale}, {Engelbracht},
  {Gordon}, {Groves}, {Hao}, {Helou}, {Hinz}, {Hunt}, {Johnson}, {Li},
  {Murphy}, {Roussel}, {Sandstrom}, {Skibba}, \& {Tabatabaei}}]{galametz2013}
{Galametz}, M., {Kennicutt}, R.~C., {Calzetti}, D., {et~al.} 2013, \mnras, 431,
  1956

\bibitem[{{Galavis} {et~al.}(1997){Galavis}, {Mendoza}, \&
  {Zeippen}}]{galavis1997}
{Galavis}, M.~E., {Mendoza}, C., \& {Zeippen}, C.~J. 1997, \aaps, 123, 159

\bibitem[{{Galavis} {et~al.}(1998){Galavis}, {Mendoza}, \&
  {Zeippen}}]{galavis1998}
{Galavis}, M.~E., {Mendoza}, C., \& {Zeippen}, C.~J. 1998, \aaps, 131, 499

\bibitem[{{Gordon} {et~al.}(2008){Gordon}, {Engelbracht}, {Rieke}, {Misselt},
  {Smith}, \& {Kennicutt}}]{gordon2008}
{Gordon}, K.~D., {Engelbracht}, C.~W., {Rieke}, G.~H., {et~al.} 2008, \apj,
  682, 336

\bibitem[{{Gordon} {et~al.}(2005){Gordon}, {Rieke}, {Engelbracht}, {Muzerolle},
  {Stansberry}, {Misselt}, {Morrison}, {Cadien}, {Young}, {Dole}, {Kelly},
  {Alonso-Herrero}, {Egami}, {Su}, {Papovich}, {Smith}, {Hines}, {Rieke},
  {Blaylock}, {P{\'e}rez-Gonz{\'a}lez}, {Le Floc'h}, {Hinz}, {Latter},
  {Hesselroth}, {Frayer}, {Noriega-Crespo}, {Masci}, {Padgett}, {Smylie}, \&
  {Haegel}}]{gordon2005}
{Gordon}, K.~D., {Rieke}, G.~H., {Engelbracht}, C.~W., {et~al.} 2005, \pasp,
  117, 503

\bibitem[{{Gould} \& {Salpeter}(1963)}]{gould1963}
{Gould}, R.~J. \& {Salpeter}, E.~E. 1963, \apj, 138, 393

\bibitem[{{Graci{\'a}-Carpio} {et~al.}(2008){Graci{\'a}-Carpio},
  {Garc{\'{\i}}a-Burillo}, {Planesas}, {Fuente}, \& {Usero}}]{graciacarpio2008}
{Graci{\'a}-Carpio}, J., {Garc{\'{\i}}a-Burillo}, S., {Planesas}, P., {Fuente},
  A., \& {Usero}, A. 2008, \aap, 479, 703

\bibitem[{{Graci{\'a}-Carpio} {et~al.}(2011){Graci{\'a}-Carpio}, {Sturm},
  {Hailey-Dunsheath}, {Fischer}, {Contursi}, {Poglitsch}, {Genzel},
  {Gonz{\'a}lez-Alfonso}, {Sternberg}, {Verma}, {Christopher}, {Davies},
  {Feuchtgruber}, {de Jong}, {Lutz}, \& {Tacconi}}]{graciacarpio2011}
{Graci{\'a}-Carpio}, J., {Sturm}, E., {Hailey-Dunsheath}, S., {et~al.} 2011,
  \apjl, 728, L7

\bibitem[{{Griffin} {et~al.}(2010){Griffin}, {Abergel}, {Abreu}, {Ade},
  {Andr{\'e}}, {Augueres}, {Babbedge}, {Bae}, {Baillie}, {Baluteau}, {Barlow},
  {Bendo}, {Benielli}, {Bock}, {Bonhomme}, {Brisbin}, {Brockley-Blatt},
  {Caldwell}, {Cara}, {Castro-Rodriguez}, {Cerulli}, {Chanial}, {Chen},
  {Clark}, {Clements}, {Clerc}, {Coker}, {Communal}, {Conversi}, {Cox},
  {Crumb}, {Cunningham}, {Daly}, {Davis}, {de Antoni}, {Delderfield}, {Devin},
  {di Giorgio}, {Didschuns}, {Dohlen}, {Donati}, {Dowell}, {Dowell}, {Duband},
  {Dumaye}, {Emery}, {Ferlet}, {Ferrand}, {Fontignie}, {Fox}, {Franceschini},
  {Frerking}, {Fulton}, {Garcia}, {Gastaud}, {Gear}, {Glenn}, {Goizel},
  {Griffin}, {Grundy}, {Guest}, {Guillemet}, {Hargrave}, {Harwit}, {Hastings},
  {Hatziminaoglou}, {Herman}, {Hinde}, {Hristov}, {Huang}, {Imhof}, {Isaak},
  {Israelsson}, {Ivison}, {Jennings}, {Kiernan}, {King}, {Lange}, {Latter},
  {Laurent}, {Laurent}, {Leeks}, {Lellouch}, {Levenson}, {Li}, {Li},
  {Lilienthal}, {Lim}, {Liu}, {Lu}, {Madden}, {Mainetti}, {Marliani}, {McKay},
  {Mercier}, {Molinari}, {Morris}, {Moseley}, {Mulder}, {Mur}, {Naylor},
  {Nguyen}, {O'Halloran}, {Oliver}, {Olofsson}, {Olofsson}, {Orfei}, {Page},
  {Pain}, {Panuzzo}, {Papageorgiou}, {Parks}, {Parr-Burman}, {Pearce},
  {Pearson}, {P{\'e}rez-Fournon}, {Pinsard}, {Pisano}, {Podosek}, {Pohlen},
  {Polehampton}, {Pouliquen}, {Rigopoulou}, {Rizzo}, {Roseboom}, {Roussel},
  {Rowan-Robinson}, {Rownd}, {Saraceno}, {Sauvage}, {Savage}, {Savini},
  {Sawyer}, {Scharmberg}, {Schmitt}, {Schneider}, {Schulz}, {Schwartz},
  {Shafer}, {Shupe}, {Sibthorpe}, {Sidher}, {Smith}, {Smith}, {Smith},
  {Spencer}, {Stobie}, {Sudiwala}, {Sukhatme}, {Surace}, {Stevens}, {Swinyard},
  {Trichas}, {Tourette}, {Triou}, {Tseng}, {Tucker}, {Turner}, {Vaccari},
  {Valtchanov}, {Vigroux}, {Virique}, {Voellmer}, {Walker}, {Ward}, {Waskett},
  {Weilert}, {Wesson}, {White}, {Whitehouse}, {Wilson}, {Winter}, {Woodcraft},
  {Wright}, {Xu}, {Zavagno}, {Zemcov}, {Zhang}, \& {Zonca}}]{griffin2010}
{Griffin}, M.~J., {Abergel}, A., {Abreu}, A., {et~al.} 2010, \aap, 518, L3

\bibitem[{{Habing}(1968)}]{habing1968}
{Habing}, H.~J. 1968, \bain, 19, 421

\bibitem[{{Helou} {et~al.}(2004){Helou}, {Roussel}, {Appleton}, {Frayer},
  {Stolovy}, {Storrie-Lombardi}, {Hurt}, {Lowrance}, {Makovoz}, {Masci},
  {Surace}, {Gordon}, {Alonso-Herrero}, {Engelbracht}, {Misselt}, {Rieke},
  {Rieke}, {Willner}, {Pahre}, {Ashby}, {Fazio}, \& {Smith}}]{helou2004}
{Helou}, G., {Roussel}, H., {Appleton}, P., {et~al.} 2004, \apjs, 154, 253

\bibitem[{{Hildebrand}(1983)}]{hildebrand1983}
{Hildebrand}, R.~H. 1983, \qjras, 24, 267

\bibitem[{{Ho} {et~al.}(1997){Ho}, {Filippenko}, \& {Sargent}}]{ho1997}
{Ho}, L.~C., {Filippenko}, A.~V., \& {Sargent}, W.~L.~W. 1997, \apjs, 112, 315

\bibitem[{{Hollenbach} \& {McKee}(1989)}]{hollenbach1989}
{Hollenbach}, D. \& {McKee}, C.~F. 1989, \apj, 342, 306

\bibitem[{{Hollenbach} {et~al.}(1991){Hollenbach}, {Takahashi}, \&
  {Tielens}}]{hollenbach1991}
{Hollenbach}, D.~J., {Takahashi}, T., \& {Tielens}, A.~G.~G.~M. 1991, \apj,
  377, 192

\bibitem[{{Holwerda} {et~al.}(2012){Holwerda}, {Bianchi}, {B{\"o}ker},
  {Radburn-Smith}, {de Jong}, {Baes}, {van der Kruit}, {Xilouris}, {Gordon}, \&
  {Dalcanton}}]{holwerda2012}
{Holwerda}, B.~W., {Bianchi}, S., {B{\"o}ker}, T., {et~al.} 2012, \aap, 541, L5

\bibitem[{{Howk} \& {Savage}(1999)}]{howk1999}
{Howk}, J.~C. \& {Savage}, B.~D. 1999, \aj, 117, 2077

\bibitem[{{Hudson} \& {Bell}(2004)}]{hudson2004}
{Hudson}, C.~E. \& {Bell}, K.~L. 2004, \mnras, 348, 1275

\bibitem[{{Hughes} {et~al.}(2014){Hughes}, {Baes}, {Fritz}, {Smith}, {Parkin},
  {Gentile}, {Bendo}, {Wilson}, {Allaert}, {Bianchi}, {De Looze}, {Verstappen},
  {Viaene}, {Boquien}, {Boselli}, {Clements}, {Davies}, {Galametz}, {Madden},
  {R{\'e}my-Ruyer}, \& {Spinoglio}}]{hughes2014}
{Hughes}, T.~M., {Baes}, M., {Fritz}, J., {et~al.} 2014, \aap, 565, A4

\bibitem[{{Hunter} {et~al.}(2001){Hunter}, {Kaufman}, {Hollenbach}, {Rubin},
  {Malhotra}, {Dale}, {Brauher}, {Silbermann}, {Helou}, {Contursi}, \&
  {Lord}}]{hunter2001}
{Hunter}, D.~A., {Kaufman}, M., {Hollenbach}, D.~J., {et~al.} 2001, \apj, 553,
  121

\bibitem[{{Kamphuis} {et~al.}(2007){Kamphuis}, {Holwerda}, {Allen}, {Peletier},
  \& {van der Kruit}}]{kamphuis2007}
{Kamphuis}, P., {Holwerda}, B.~W., {Allen}, R.~J., {Peletier}, R.~F., \& {van
  der Kruit}, P.~C. 2007, \aap, 471, L1

\bibitem[{{Karczewski} {et~al.}(2013){Karczewski}, {Barlow}, {Page}, {Kuin},
  {Ferreras}, {Baes}, {Bendo}, {Boselli}, {Cooray}, {Cormier}, {De Looze},
  {Galametz}, {Galliano}, {Lebouteiller}, {Madden}, {Pohlen}, {R{\'e}my-Ruyer},
  {Smith}, \& {Spinoglio}}]{karczewski2013}
{Karczewski}, O.~{\L}., {Barlow}, M.~J., {Page}, M.~J., {et~al.} 2013, \mnras,
  431, 2493

\bibitem[{{Kaufman} {et~al.}(2006){Kaufman}, {Wolfire}, \&
  {Hollenbach}}]{kaufman2006}
{Kaufman}, M.~J., {Wolfire}, M.~G., \& {Hollenbach}, D.~J. 2006, \apj, 644, 283

\bibitem[{{Kaufman} {et~al.}(1999){Kaufman}, {Wolfire}, {Hollenbach}, \&
  {Luhman}}]{kaufman1999}
{Kaufman}, M.~J., {Wolfire}, M.~G., {Hollenbach}, D.~J., \& {Luhman}, M.~L.
  1999, \apj, 527, 795

\bibitem[{{Kennicutt} \& {Evans}(2012)}]{kennicutt2012}
{Kennicutt}, R.~C. \& {Evans}, N.~J. 2012, \araa, 50, 531

\bibitem[{{Kramer} {et~al.}(2013){Kramer}, {Abreu-Vicente},
  {Garc{\'{\i}}a-Burillo}, {Rela{\~n}o}, {Aalto}, {Boquien}, {Braine},
  {Buchbender}, {Gratier}, {Israel}, {Nikola}, {R{\"o}llig}, {Verley}, {van der
  Werf}, \& {Xilouris}}]{kramer2013}
{Kramer}, C., {Abreu-Vicente}, J., {Garc{\'{\i}}a-Burillo}, S., {et~al.} 2013,
  \aap, 553, A114

\bibitem[{{Kramer} {et~al.}(2005){Kramer}, {Mookerjea}, {Bayet},
  {Garcia-Burillo}, {Gerin}, {Israel}, {Stutzki}, \& {Wouterloot}}]{kramer2005}
{Kramer}, C., {Mookerjea}, B., {Bayet}, E., {et~al.} 2005, \aap, 441, 961

\bibitem[{{Kreckel} {et~al.}(2013){Kreckel}, {Groves}, {Schinnerer}, {Johnson},
  {Aniano}, {Calzetti}, {Croxall}, {Draine}, {Gordon}, {Crocker}, {Dale},
  {Hunt}, {Kennicutt}, {Meidt}, {Smith}, \& {Tabatabaei}}]{kreckel2013}
{Kreckel}, K., {Groves}, B., {Schinnerer}, E., {et~al.} 2013, \apj, 771, 62

\bibitem[{{Lebouteiller} {et~al.}(2007){Lebouteiller}, {Brandl},
  {Bernard-Salas}, {Devost}, \& {Houck}}]{lebouteiller2007}
{Lebouteiller}, V., {Brandl}, B., {Bernard-Salas}, J., {Devost}, D., \&
  {Houck}, J.~R. 2007, \apj, 665, 390

\bibitem[{{Lebouteiller} {et~al.}(2012){Lebouteiller}, {Cormier}, {Madden},
  {Galliano}, {Indebetouw}, {Abel}, {Sauvage}, {Hony}, {Contursi}, {Poglitsch},
  {R{\'e}my}, {Sturm}, \& {Wu}}]{lebouteiller2012}
{Lebouteiller}, V., {Cormier}, D., {Madden}, S.~C., {et~al.} 2012, \aap, 548,
  A91

\bibitem[{{Liseau} {et~al.}(2006){Liseau}, {Justtanont}, \&
  {Tielens}}]{liseau2006}
{Liseau}, R., {Justtanont}, K., \& {Tielens}, A.~G.~G.~M. 2006, \aap, 446, 561

\bibitem[{{Luhman} {et~al.}(1998){Luhman}, {Satyapal}, {Fischer}, {Wolfire},
  {Cox}, {Lord}, {Smith}, {Stacey}, \& {Unger}}]{luhman1998}
{Luhman}, M.~L., {Satyapal}, S., {Fischer}, J., {et~al.} 1998, \apjl, 504, L11

\bibitem[{{Luhman} {et~al.}(2003){Luhman}, {Satyapal}, {Fischer}, {Wolfire},
  {Sturm}, {Dudley}, {Lutz}, \& {Genzel}}]{luhman2003}
{Luhman}, M.~L., {Satyapal}, S., {Fischer}, J., {et~al.} 2003, \apj, 594, 758

\bibitem[{{Madden} {et~al.}(1993){Madden}, {Geis}, {Genzel}, {Herrmann},
  {Jackson}, {Poglitsch}, {Stacey}, \& {Townes}}]{madden1993}
{Madden}, S.~C., {Geis}, N., {Genzel}, R., {et~al.} 1993, \apj, 407, 579

\bibitem[{{Madden} {et~al.}(1994){Madden}, {Geis}, {Genzel}, {Herrmann},
  {Poglitsch}, {Stacey}, \& {Townes}}]{madden1994}
{Madden}, S.~C., {Geis}, N., {Genzel}, R., {et~al.} 1994, Infrared Physics and
  Technology, 35, 311

\bibitem[{{Makovoz} \& {Khan}(2005)}]{makovoz2005}
{Makovoz}, D. \& {Khan}, I. 2005, in Astronomical Society of the Pacific
  Conference Series, Vol. 347, Astronomical Data Analysis Software and Systems
  XIV, ed. P.~{Shopbell}, M.~{Britton}, \& R.~{Ebert}, 81

\bibitem[{{Malhotra} {et~al.}(2001){Malhotra}, {Kaufman}, {Hollenbach},
  {Helou}, {Rubin}, {Brauher}, {Dale}, {Lu}, {Lord}, {Stacey}, {Contursi},
  {Hunter}, \& {Dinerstein}}]{malhotra2001}
{Malhotra}, S., {Kaufman}, M.~J., {Hollenbach}, D., {et~al.} 2001, \apj, 561,
  766

\bibitem[{{Marble} {et~al.}(2010){Marble}, {Engelbracht}, {van Zee}, {Dale},
  {Smith}, {Gordon}, {Wu}, {Lee}, {Kennicutt}, {Skillman}, {Johnson}, {Block},
  {Calzetti}, {Cohen}, {Lee}, \& {Schuster}}]{marble2010}
{Marble}, A.~R., {Engelbracht}, C.~W., {van Zee}, L., {et~al.} 2010, \apj, 715,
  506

\bibitem[{{Mochizuki}(2004)}]{mochizuki2004}
{Mochizuki}, K. 2004, Journal of Korean Astronomical Society, 37, 193

\bibitem[{{Mookerjea} {et~al.}(2011){Mookerjea}, {Kramer}, {Buchbender},
  {Boquien}, {Verley}, {Rela{\~n}o}, {Quintana-Lacaci}, {Aalto}, {Braine},
  {Calzetti}, {Combes}, {Garcia-Burillo}, {Gratier}, {Henkel}, {Israel},
  {Lord}, {Nikola}, {R{\"o}llig}, {Stacey}, {Tabatabaei}, {van der Tak}, \&
  {van der Werf}}]{mookerjea2011}
{Mookerjea}, B., {Kramer}, C., {Buchbender}, C., {et~al.} 2011, \aap, 532, A152

\bibitem[{{Nakagawa} {et~al.}(2012){Nakagawa}, {Matsuhara}, \&
  {Kawakatsu}}]{nakagawa2012}
{Nakagawa}, T., {Matsuhara}, H., \& {Kawakatsu}, Y. 2012, in Society of
  Photo-Optical Instrumentation Engineers (SPIE) Conference Series, Vol. 8442,
  Society of Photo-Optical Instrumentation Engineers (SPIE) Conference Series

\bibitem[{{Norman} \& {Ikeuchi}(1989)}]{norman1989}
{Norman}, C.~A. \& {Ikeuchi}, S. 1989, \apj, 345, 372

\bibitem[{{Nozawa} \& {Kozasa}(2013)}]{nozawa2013}
{Nozawa}, T. \& {Kozasa}, T. 2013, \apj, 776, 24

\bibitem[{{Oberst} {et~al.}(2011){Oberst}, {Parshley}, {Nikola}, {Stacey},
  {L{\"o}hr}, {Lane}, {Stark}, \& {Kamenetzky}}]{oberst2011}
{Oberst}, T.~E., {Parshley}, S.~C., {Nikola}, T., {et~al.} 2011, \apj, 739, 100

\bibitem[{{Oberst} {et~al.}(2006){Oberst}, {Parshley}, {Stacey}, {Nikola},
  {L{\"o}hr}, {Harnett}, {Tothill}, {Lane}, {Stark}, \& {Tucker}}]{oberst2006}
{Oberst}, T.~E., {Parshley}, S.~C., {Stacey}, G.~J., {et~al.} 2006, \apjl, 652,
  L125

\bibitem[{{Ott}(2010)}]{ott2010}
{Ott}, S. 2010, in Astronomical Society of the Pacific Conference Series, Vol.
  434, Astronomical Data Analysis Software and Systems XIX, ed. Y.~{Mizumoto},
  K.-I. {Morita}, \& M.~{Ohishi}, 139

\bibitem[{{Parkin} {et~al.}(2013){Parkin}, {Wilson}, {Schirm}, {Baes},
  {Boquien}, {Boselli}, {Cooray}, {Cormier}, {Foyle}, {Karczewski},
  {Lebouteiller}, {de Looze}, {Madden}, {Roussel}, {Sauvage}, \&
  {Spinoglio}}]{parkin2013}
{Parkin}, T.~J., {Wilson}, C.~D., {Schirm}, M.~R.~P., {et~al.} 2013, \apj, 776,
  65

\bibitem[{{Parkin} {et~al.}(2014){Parkin}, {Wilson}, {Schirm}, {Baes},
  {Boquien}, {Boselli}, {Cormier}, {Galametz}, {Karczewski}, {Lebouteiller},
  {De Looze}, {Madden}, {Roussel}, {Smith}, \& {Spinoglio}}]{parkin2014}
{Parkin}, T.~J., {Wilson}, C.~D., {Schirm}, M.~R.~P., {et~al.} 2014, \apj, 787,
  16

\bibitem[{{Pilbratt} {et~al.}(2010){Pilbratt}, {Riedinger}, {Passvogel},
  {Crone}, {Doyle}, {Gageur}, {Heras}, {Jewell}, {Metcalfe}, {Ott}, \&
  {Schmidt}}]{pilbratt2010}
{Pilbratt}, G.~L., {Riedinger}, J.~R., {Passvogel}, T., {et~al.} 2010, \aap,
  518, L1

\bibitem[{{Poglitsch} {et~al.}(2010){Poglitsch}, {Waelkens}, {Geis},
  {Feuchtgruber}, {Vandenbussche}, {Rodriguez}, {Krause}, {Renotte}, {van
  Hoof}, {Saraceno}, {Cepa}, {Kerschbaum}, {Agn{\`e}se}, {Ali}, {Altieri},
  {Andreani}, {Augueres}, {Balog}, {Barl}, {Bauer}, {Belbachir}, {Benedettini},
  {Billot}, {Boulade}, {Bischof}, {Blommaert}, {Callut}, {Cara}, {Cerulli},
  {Cesarsky}, {Contursi}, {Creten}, {De Meester}, {Doublier}, {Doumayrou},
  {Duband}, {Exter}, {Genzel}, {Gillis}, {Gr{\"o}zinger}, {Henning},
  {Herreros}, {Huygen}, {Inguscio}, {Jakob}, {Jamar}, {Jean}, {de Jong},
  {Katterloher}, {Kiss}, {Klaas}, {Lemke}, {Lutz}, {Madden}, {Marquet},
  {Martignac}, {Mazy}, {Merken}, {Montfort}, {Morbidelli}, {M{\"u}ller},
  {Nielbock}, {Okumura}, {Orfei}, {Ottensamer}, {Pezzuto}, {Popesso},
  {Putzeys}, {Regibo}, {Reveret}, {Royer}, {Sauvage}, {Schreiber}, {Stegmaier},
  {Schmitt}, {Schubert}, {Sturm}, {Thiel}, {Tofani}, {Vavrek}, {Wetzstein},
  {Wieprecht}, \& {Wiezorrek}}]{poglitsch2010}
{Poglitsch}, A., {Waelkens}, C., {Geis}, N., {et~al.} 2010, \aap, 518, L2

\bibitem[{{Pound} \& {Wolfire}(2008)}]{pound2008}
{Pound}, M.~W. \& {Wolfire}, M.~G. 2008, in Astronomical Society of the Pacific
  Conference Series, Vol. 394, Astronomical Data Analysis Software and Systems
  XVII, ed. R.~W. {Argyle}, P.~S. {Bunclark}, \& J.~R. {Lewis}, 654

\bibitem[{{Rand} {et~al.}(1990){Rand}, {Kulkarni}, \& {Hester}}]{rand1990}
{Rand}, R.~J., {Kulkarni}, S.~R., \& {Hester}, J.~J. 1990, \apjl, 352, L1

\bibitem[{{Rand} {et~al.}(2008){Rand}, {Wood}, \& {Benjamin}}]{rand2008}
{Rand}, R.~J., {Wood}, K., \& {Benjamin}, R.~A. 2008, \apj, 680, 263

\bibitem[{{Rand} {et~al.}(2011){Rand}, {Wood}, {Benjamin}, \&
  {Meidt}}]{rand2011}
{Rand}, R.~J., {Wood}, K., {Benjamin}, R.~A., \& {Meidt}, S.~E. 2011, \apj,
  728, 163

\bibitem[{{Riechers} {et~al.}(2013){Riechers}, {Bradford}, {Clements},
  {Dowell}, {P{\'e}rez-Fournon}, {Ivison}, {Bridge}, {Conley}, {Fu}, {Vieira},
  {Wardlow}, {Calanog}, {Cooray}, {Hurley}, {Neri}, {Kamenetzky}, {Aguirre},
  {Altieri}, {Arumugam}, {Benford}, {B{\'e}thermin}, {Bock}, {Burgarella},
  {Cabrera-Lavers}, {Chapman}, {Cox}, {Dunlop}, {Earle}, {Farrah}, {Ferrero},
  {Franceschini}, {Gavazzi}, {Glenn}, {Solares}, {Gurwell}, {Halpern},
  {Hatziminaoglou}, {Hyde}, {Ibar}, {Kov{\'a}cs}, {Krips}, {Lupu}, {Maloney},
  {Martinez-Navajas}, {Matsuhara}, {Murphy}, {Naylor}, {Nguyen}, {Oliver},
  {Omont}, {Page}, {Petitpas}, {Rangwala}, {Roseboom}, {Scott}, {Smith},
  {Staguhn}, {Streblyanska}, {Thomson}, {Valtchanov}, {Viero}, {Wang},
  {Zemcov}, \& {Zmuidzinas}}]{riechers2013}
{Riechers}, D.~A., {Bradford}, C.~M., {Clements}, D.~L., {et~al.} 2013, \nat,
  496, 329

\bibitem[{{Rieke} {et~al.}(2004){Rieke}, {Young}, {Engelbracht}, {Kelly},
  {Low}, {Haller}, {Beeman}, {Gordon}, {Stansberry}, {Misselt}, {Cadien},
  {Morrison}, {Rivlis}, {Latter}, {Noriega-Crespo}, {Padgett}, {Stapelfeldt},
  {Hines}, {Egami}, {Muzerolle}, {Alonso-Herrero}, {Blaylock}, {Dole}, {Hinz},
  {Le Floc'h}, {Papovich}, {P{\'e}rez-Gonz{\'a}lez}, {Smith}, {Su}, {Bennett},
  {Frayer}, {Henderson}, {Lu}, {Masci}, {Pesenson}, {Rebull}, {Rho}, {Keene},
  {Stolovy}, {Wachter}, {Wheaton}, {Werner}, \& {Richards}}]{rieke2004}
{Rieke}, G.~H., {Young}, E.~T., {Engelbracht}, C.~W., {et~al.} 2004, \apjs,
  154, 25

\bibitem[{{Rigopoulou} {et~al.}(2014){Rigopoulou}, {Hopwood}, {Magdis},
  {Thatte}, {Swinyard}, {Farrah}, {Huang}, {Alonso-Herrero}, {Bock},
  {Clements}, {Cooray}, {Griffin}, {Oliver}, {Pearson}, {Riechers}, {Scott},
  {Smith}, {Vaccari}, {Valtchanov}, \& {Wang}}]{rigopoulou2014}
{Rigopoulou}, D., {Hopwood}, R., {Magdis}, G.~E., {et~al.} 2014, \apjl, 781,
  L15

\bibitem[{{Roelfsema} {et~al.}(2012){Roelfsema}, {Giard}, {Najarro},
  {Wafelbakker}, {Jellema}, {Jackson}, {Swinyard}, {Audard}, {Doi}, {Griffin},
  {Helmich}, {Kerschbaum}, {Meyer}, {Naylor}, {Nielsen}, {Olofsson},
  {Poglitsch}, {Spinoglio}, {Vandenbussche}, {Isaak}, \&
  {Goicoechea}}]{roelfsema2012}
{Roelfsema}, P., {Giard}, M., {Najarro}, F., {et~al.} 2012, in Society of
  Photo-Optical Instrumentation Engineers (SPIE) Conference Series, Vol. 8442,
  Society of Photo-Optical Instrumentation Engineers (SPIE) Conference Series

\bibitem[{{R{\"o}llig} {et~al.}(2007){R{\"o}llig}, {Abel}, {Bell}, {Bensch},
  {Black}, {Ferland}, {Jonkheid}, {Kamp}, {Kaufman}, {Le Bourlot}, {Le Petit},
  {Meijerink}, {Morata}, {Ossenkopf}, {Roueff}, {Shaw}, {Spaans}, {Sternberg},
  {Stutzki}, {Thi}, {van Dishoeck}, {van Hoof}, {Viti}, \&
  {Wolfire}}]{rollig2007}
{R{\"o}llig}, M., {Abel}, N.~P., {Bell}, T., {et~al.} 2007, \aap, 467, 187

\bibitem[{{Rossa} {et~al.}(2004){Rossa}, {Dettmar}, {Walterbos}, \&
  {Norman}}]{rossa2004}
{Rossa}, J., {Dettmar}, R.-J., {Walterbos}, R.~A.~M., \& {Norman}, C.~A. 2004,
  \aj, 128, 674

\bibitem[{{Rubin}(1985)}]{rubin1985}
{Rubin}, R.~H. 1985, \apjs, 57, 349

\bibitem[{{Rupen}(1991)}]{rupen1991}
{Rupen}, M.~P. 1991, \aj, 102, 48

\bibitem[{{Savage} \& {Sembach}(1996)}]{savage1996}
{Savage}, B.~D. \& {Sembach}, K.~R. 1996, \araa, 34, 279

\bibitem[{{Schechtman-Rook} \& {Bershady}(2013)}]{schechtmanrook2013}
{Schechtman-Rook}, A. \& {Bershady}, M.~A. 2013, \apj, 773, 45

\bibitem[{{Schirm} {et~al.}(2014){Schirm}, {Wilson}, {Parkin}, {Kamenetzky},
  {Glenn}, {Rangwala}, {Spinoglio}, {Pereira-Santaella}, {Baes}, {Barlow},
  {Clements}, {Cooray}, {De Looze}, {Karczewski}, {Madden}, {R{\'e}my-Ruyer},
  \& {Wu}}]{schirm2014}
{Schirm}, M.~R.~P., {Wilson}, C.~D., {Parkin}, T.~J., {et~al.} 2014, \apj, 781,
  101

\bibitem[{{Shapiro} \& {Field}(1976)}]{shapiro1976}
{Shapiro}, P.~R. \& {Field}, G.~B. 1976, \apj, 205, 762

\bibitem[{{Smith} {et~al.}(2007){Smith}, {Draine}, {Dale}, {Moustakas},
  {Kennicutt}, {Helou}, {Armus}, {Roussel}, {Sheth}, {Bendo}, {Buckalew},
  {Calzetti}, {Engelbracht}, {Gordon}, {Hollenbach}, {Li}, {Malhotra},
  {Murphy}, \& {Walter}}]{smith2007}
{Smith}, J.~D.~T., {Draine}, B.~T., {Dale}, D.~A., {et~al.} 2007, \apj, 656,
  770

\bibitem[{{Stacey} {et~al.}(2010){Stacey}, {Charmandaris}, {Boulanger}, {Wu},
  {Combes}, {Higdon}, {Smith}, \& {Nikola}}]{stacey2010}
{Stacey}, G.~J., {Charmandaris}, V., {Boulanger}, F., {et~al.} 2010, \apj, 721,
  59

\bibitem[{{Stacey} {et~al.}(1991){Stacey}, {Geis}, {Genzel}, {Lugten},
  {Poglitsch}, {Sternberg}, \& {Townes}}]{stacey1991}
{Stacey}, G.~J., {Geis}, N., {Genzel}, R., {et~al.} 1991, \apj, 373, 423

\bibitem[{{Stacey} {et~al.}(1983){Stacey}, {Smyers}, {Kurtz}, \&
  {Harwit}}]{stacey1983}
{Stacey}, G.~J., {Smyers}, S.~D., {Kurtz}, N.~T., \& {Harwit}, M. 1983, \apjl,
  268, L99

\bibitem[{{Stacey} {et~al.}(1985){Stacey}, {Viscuso}, {Fuller}, \&
  {Kurtz}}]{stacey1985}
{Stacey}, G.~J., {Viscuso}, P.~J., {Fuller}, C.~E., \& {Kurtz}, N.~T. 1985,
  \apj, 289, 803

\bibitem[{{Strickland} {et~al.}(2004){Strickland}, {Heckman}, {Colbert},
  {Hoopes}, \& {Weaver}}]{strickland2004}
{Strickland}, D.~K., {Heckman}, T.~M., {Colbert}, E.~J.~M., {Hoopes}, C.~G., \&
  {Weaver}, K.~A. 2004, \apj, 606, 829

\bibitem[{{Temple} {et~al.}(2005){Temple}, {Raychaudhury}, \&
  {Stevens}}]{temple2005}
{Temple}, R.~F., {Raychaudhury}, S., \& {Stevens}, I.~R. 2005, \mnras, 362, 581

\bibitem[{{Thompson} {et~al.}(2004){Thompson}, {Howk}, \&
  {Savage}}]{thompson2004}
{Thompson}, T.~W.~J., {Howk}, J.~C., \& {Savage}, B.~D. 2004, \aj, 128, 662

\bibitem[{{Tielens} \& {Hollenbach}(1985)}]{tielens1985}
{Tielens}, A.~G.~G.~M. \& {Hollenbach}, D. 1985, \apj, 291, 722

\bibitem[{{Vacca} {et~al.}(1996){Vacca}, {Garmany}, \& {Shull}}]{vacca1996}
{Vacca}, W.~D., {Garmany}, C.~D., \& {Shull}, J.~M. 1996, \apj, 460, 914

\bibitem[{{Valentijn} \& {van der Werf}(1999)}]{valentijn1999}
{Valentijn}, E.~A. \& {van der Werf}, P.~P. 1999, \apjl, 522, L29

\bibitem[{{Verstappen} {et~al.}(2013){Verstappen}, {Fritz}, {Baes}, {Smith},
  {Allaert}, {Bianchi}, {Blommaert}, {De Geyter}, {De Looze}, {Gentile},
  {Gordon}, {Holwerda}, {Viaene}, \& {Xilouris}}]{verstappen2013}
{Verstappen}, J., {Fritz}, J., {Baes}, M., {et~al.} 2013, \aap, 556, A54

\bibitem[{{Watson}(1972)}]{watson1972}
{Watson}, W.~D. 1972, \apj, 176, 103

\bibitem[{{Whaley} {et~al.}(2009){Whaley}, {Irwin}, {Madden}, {Galliano}, \&
  {Bendo}}]{whaley2009}
{Whaley}, C.~H., {Irwin}, J.~A., {Madden}, S.~C., {Galliano}, F., \& {Bendo},
  G.~J. 2009, \mnras, 395, 97

\bibitem[{{Wolfire} {et~al.}(1995){Wolfire}, {Hollenbach}, {McKee}, {Tielens},
  \& {Bakes}}]{wolfire1995}
{Wolfire}, M.~G., {Hollenbach}, D., {McKee}, C.~F., {Tielens}, A.~G.~G.~M., \&
  {Bakes}, E.~L.~O. 1995, \apj, 443, 152

\bibitem[{{Wolfire} {et~al.}(1990){Wolfire}, {Tielens}, \&
  {Hollenbach}}]{wolfire1990}
{Wolfire}, M.~G., {Tielens}, A.~G.~G.~M., \& {Hollenbach}, D. 1990, \apj, 358,
  116

\bibitem[{{Wu} {et~al.}(2014){Wu}, {Madden}, \& {et al.}}]{wu2014}
{Wu}, R., {Madden}, S.~C., \& {et al.} 2014, \aap,\ submitted

\bibitem[{{Xilouris} {et~al.}(1998){Xilouris}, {Alton}, {Davies}, {Kylafis},
  {Papamastorakis}, \& {Trewhella}}]{xilouris1998}
{Xilouris}, E.~M., {Alton}, P.~B., {Davies}, J.~I., {et~al.} 1998, \aap, 331,
  894

\bibitem[{{Zhao} {et~al.}(2013){Zhao}, {Lu}, {Xu}, {Gao}, {Lord}, {Howell},
  {Isaak}, {Charmandaris}, {Diaz-Santos}, {Appleton}, {Evans}, {Iwasawa},
  {Leech}, {Mazzarella}, {Petric}, {Sanders}, {Schulz}, {Surace}, \& {van der
  Werf}}]{zhao2013}
{Zhao}, Y., {Lu}, N., {Xu}, C.~K., {et~al.} 2013, \apjl, 765, L13

\end{thebibliography}

\end{document}